\documentclass[prd,aps,
showpacs,
nofootinbib,
floatfix,
superscriptaddress]{revtex4}
\usepackage{graphicx}
\usepackage{epsfig}
\usepackage{rotating}
\usepackage{amssymb}
\usepackage{axodraw}
\usepackage{dsfont}
\usepackage{psfrag}
\usepackage{amsmath,euscript,array,mathrsfs}

\newcommand{\SP}[1]{\begin{equation}\begin{split} #1
\end{split}\end{equation}}

\newcommand{\beq}{\begin{equation}}
\newcommand{\eeq}{\end{equation}}
\newcommand{\beqs}{\begin{eqnarray}}
\newcommand{\eeqs}{\end{eqnarray}}
\newcommand{\lsim}{\mathrel{\raisebox{-
.6ex}{$\stackrel{\textstyle<}{\sim}$}}}
\newcommand{\gsim}{\mathrel{\raisebox{-
.6ex}{$\stackrel{\textstyle>}{\sim}$}}}
\newcommand{\Tr}{{\rm Tr}}

\def\dslash{\raisebox{1pt}{$\slash$} \hspace{-0pt} \partial}
\def\Dslash{\raisebox{1pt}{$\slash$} \hspace{-2pt} D}
\def\hbar{\hspace{0pt}\raisebox{1pt}{$-$} \hspace{-7pt} h}

\def\di{\mbox{d}}
\def\r{\rho}

\newcommand{\be}{\begin{equation}}
\newcommand{\ee}{\end{equation}}
\newcommand{\bea}{\begin{eqnarray}}
\newcommand{\eea}{\end{eqnarray}}
\newcommand{\nn}{\nonumber}

\def\co{{\cal O}}

\def\lbldef#1#2{\expandafter\gdef\csname #1\endcsname {#2}}

\def\href#1#2{#2}

\newcommand{\ber}{\begin{eqnarray}}
\newcommand{\eer}{\end{eqnarray}}

\newcommand{\beqar}{\begin{eqnarray}}

\newcommand{\eeqar}{\end{eqnarray}}

\newcommand{\dsl}
  {\kern.06em\hbox{\raise.15ex\hbox{$/$}\kern-.56em\hbox{$\partial$}}}

\newcommand{\eeqarr}{\end{eqnarray}}
\newcommand{\ZZ}{{\rm \kern 0.275em Z \kern -0.92em Z}\;}

\def\CC{{\mathchoice
{\rm C\mkern-8mu\vrule height1.45ex depth-.05ex
width.05em\mkern9mu\kern-.05em}
{\rm C\mkern-8mu\vrule height1.45ex depth-.05ex
width.05em\mkern9mu\kern-.05em}
{\rm C\mkern-8mu\vrule height1ex depth-.07ex
width.035em\mkern9mu\kern-.035em}
{\rm C\mkern-8mu\vrule height.65ex depth-.1ex
width.025em\mkern8mu\kern-.025em}}}

\def\RR{{\rm I\kern-1.6pt {\rm R}}}

\def\ZZ{{\rm Z}\kern-3.8pt {\rm Z} \kern2pt}
\def\IB{\relax{\rm I\kern-.18em B}}
\def\ID{\relax{\rm I\kern-.18em D}}
\def\II{\relax{\rm I\kern-.18em I}}
\def\IP{\relax{\rm I\kern-.18em P}}

\newcommand{\bear}{\begin{eqnarray}}
\newcommand{\eear}{\end{eqnarray}}

\def\to{\rightarrow}

\def\to{\rightarrow}

\def\a{\alpha}

  \def\w{\omega}
\def\r{\rho}                                     
\def\t{\tau}

\def\lab{\label}
\def\6{\partial}

\def\bea{\begin{eqnarray}}
\def\eea{\end{eqnarray}}

\def\beqx{\begin{displaymath}}
\def\eeqx{\end{displaymath}}

\newcommand{\bmat}{\left(\begin{array}}
\newcommand{\emat}{\end{array}\right)}

\def\a{\alpha}

\def\r{\rho}

\def\t{\tau}

\def\co{{\cal O}}

\def\bo{{\raise-.3ex\hbox{\large$\Box$}}}               
\def\gtap{\raisebox{-.4ex}{\rlap{$\sim$}} \raisebox{.4ex}{$>$}}   
\def\face{{\raise.2ex\hbox{$\displaystyle \bigodot$}\mskip-2.2mu \llap {$\ddot
        \smile$}}}                                   
\def\>{\rangle}                                      
\def\<{\langle}                                      

\def\slash#1{\rlap{\hbox{$\mskip 1 mu /$}}#1}        
\def\leftrightarrowfill{$\mathsurround=0pt \mathord\leftarrow \mkern-6mu
        \cleaders\hbox{$\mkern-2mu \mathord- \mkern-2mu$}\hfill
        \mkern-6mu \mathord\rightarrow$}        
\def\dvec#1{\vbox{\ialign{##\crcr
        \leftrightarrowfill\crcr\noalign{\kern-1pt\nointerlineskip}
        $\hfil\displaystyle{#1}\hfil$\crcr}}}           
\def\Tr{{\rm Tr \,}}                                    

\def\-{\hphantom{-}}

\begin{document}
\title{Lectures on walking technicolor, holography and gauge/gravity dualities.
}

\author{ Maurizio Piai}
\affiliation{Swansea University, Physics Department,
Singleton Park, Swansea, SA2 8PP, Wales, UK}



\begin{abstract}
Dynamical Electro-weak symmetry breaking is an appealing, strongly-coupled
alternative to the weakly-coupled models based on an elementary scalar field developing 
a vacuum expectation value.
In the first two sections of this set of lectures, I  summarize  the arguments, 
based on low-energy phenomenology,
supporting  walking technicolor as a realistic
realization of  this idea.
This pedagogical introduction to
walking technicolor, and more generally to the physics of 
extensions of the standard model,
makes extensive use of effective field theory arguments,
symmetries and counting rules.
The strongly-coupled nature of the underlying interactions,
and the peculiar quasi-conformal behavior of the theory,
requires to use non-perturbative methods in order to 
address many fundamental questions within this framework.
The recent development  of gauge/gravity dualities provides an ideal
set of such non-perturbative instruments.
The remaining  two sections  illustrate the potential of these techniques with two 
technical examples, one within the
bottom-up phenomenological approach to holography in five-dimensions, the other 
within a more systematic top-down construction derived from ten-dimensional  type-IIB supergravity.

\end{abstract}


\maketitle

\tableofcontents

\newpage
\section*{Introduction and outline}

This set  of notes on walking technicolor 
is based in part on sets of lectures delivered to string theory students
in the context of their Ph.~D.  programs at  Swansea University and at 
the University of Barcelona.
As such, the content is intended for readers who are familiar with the 
ideas of gauge/gravity dualities~\cite{AdSCFT}, but not with technicolor (TC)~\cite{TC}, walking technicolor (WTC)~\cite{WTC}, extended technicolor (ETC)~\cite{ETC}
 and more in general  with  the details of the physics of the standard model (SM) and the 
 problematics arising when dealing with its extensions.
  
This is not a complete and exhaustive review of the subject of technicolor, which can be found elsewhere,
but rather a collection of exercises and arguments aimed at singling out a subset of  problems that emerge
in the context of physics beyond the standard model (BSM), and that are particularly 
important when, 
as is the case for  WTC,
the BSM new physics  is strongly coupled.
This set of problems 
admits a natural formulation (if not yet a natural solution) in the context of  gauge/gravity dualities,
which is the main message these lectures want to convey.
The  reader who wants to have a complete overview of the subject  of strongly-coupled dynamical electro-weak symmetry breaking (EWSB)
should complement the reading of these notes with other reviews, which develop 
in more details many aspects and ideas that are only glanced at in the present set of lectures~\cite{reviews}.

The first two parts of these notes are very general, and should be easily accessible to 
anybody who has a working knowledge of quantum field theory and a basic knowledge of the 
standard model.
For pedagogical reasons, before talking about  WTC,
I start by reminding the reader 
about some classical results that are at the core of 
our modern understanding of the Minimal version of the Standard Model (MSM)~\footnote{I will use the
MSM acronym to mean the version of the standard model with one Higgs scalar doublet, ultimately responsible for EWSB, in such a way as to distinguish it
from the acronym SM, by which I mean a generic theory with the fermions and gauge bosons of the Standard Model, without committing to the details
of the symmetry-breaking sector itself.},
and of its weakly-coupled extensions.
These include the Coleman-Weinberg results for the loop generated effective Higgs potential~\cite{CW},
a discussion of the big and little hierarchy problems,
the definition of the electro-weak chiral Lagrangian~\cite{EWCL} and of the oblique precision parameters~\cite{PT,Barbieri} (and their
use in precision electro-weak physics), 
the GIM mechanism and the resulting suppression of flavor-changing neutral current (FCNC) processes~\cite{GIM}, 
the Froggatt-Nielsen mechanism for  the generation of fermion mass hierarchy~\cite{FN}, the see-saw mechanism~\cite{seesaw}, and
the analysis of perturbative unitarity of longitudinally-polarized gauge-boson elastic-scattering amplitudes~\cite{PU}.
As such, this first half of the material should also provide a  useful introduction 
 for any reader who is interested in BSM physics in general.

The first section of the paper contains some generic introductory
material, and part of the discussion is  only semi-quantitative
(more precise statements will be developed in later sections).
It starts from a reminder about the basic building blocks of the SM,
the introduction of the concept of naturalness,
and a discussion of the big hierarchy problem.
I also provide an example of technicolor model,
for pedagogical reasons chosen to be peculiarly simple~\cite{FS}. 
Unconventionally, in comparing three popular scenarios for solving the big hierarchy problem (supersymmetry,
technicolor and warped extra-dimensions), rather than the specific differences I highlight the 
similarities between these three approaches. In particular the fact that at low-energies they all
yield to the little hierarchy problem.
In this way, it should be clear to the reader that the tension between the 
absence of fine-tuning (for which a new physics scale $\Lambda^{\ast}_1 \lsim 1$ TeV would be a natural expectation),
and the absence of indirect effects of new physics in precision measurements (which would naturally require a larger new physics scale 
$\Lambda^{\ast}_2\gsim 5$ TeV),
is a rather general problem of all BSM scenarios, not just a problem of strongly-coupled extensions of the SM.
After all, the very fact that, thanks to the work of our colleagues in many successful 
experimental programs, we have now an unprecedented body of high-precision information about 
nature, per se implies that not every  generic new physics model can be allowed by the data.
Hence the presence of what appears to be a  ${\cal O}(\Lambda_1^{\ast\,2}/\Lambda_2^{\ast\,2})\sim 5\%$ fine-tuning in the 
low-energy effective Lagrangian extending the SM is not necessarily a surprising and  bad thing, 
but rather seems to suggest to us that something very special is going to be discovered at the LHC,
hence explaining this very special low-energy  results.
What  singles out the strongly coupled scenarios as particularly problematic 
is the strong coupling itself, which severely limits the possibility of 
exploring the parameter-space of models, in the search of regions that are compatible with the current data,
in the same way as is done in the  weakly-coupled context, and hence explaining  the origin of the aforementioned 5\% tuning
in terms of specific, special detail properties of the new physics.

The second section contains a more systematic and precise pedagogical discussion, and, besides
a preliminary introduction of the idea of walking dynamics, consists 
of four main subsections. 
In the first subsection, precision electro-weak physics is discussed, starting from the effective field theory (EFT)
treatment, introducing the oblique parameters, and summarizing some classical perturbative results. 
In comparing to strongly-coupled scenarios, I introduce the rules of naive dimensional analysis (NDA)~\cite{NDA},
and hidden local symmetry (HLS)~\cite{HLS}.
The second subsection is devoted to the physics of flavor. Starting again from the weakly coupled 
case, the GIM mechanism is introduced, and the basic logic behind the construction of 
weakly-coupled models of flavor generation explained.
ETC and WTC are then introduced and their specific physical implications 
are explained, and contrasted with what happens in the weakly-coupled case.
I do not discuss CP violation, the treatment of which to large extent follows the same logic
as the physics of FCNC, up to subtleties in dealing with dipole moments that,
while important phenomenologically, 
are not very important for the present purposes.
The third subsection is a brief reminder about a classical result in the MSM: the scattering amplitude
of longitudinally polarized gauge bosons is shown to be manifestly unitary in perturbation theory,
at the diagrammatic level, provided all the couplings (including the Higgs self-coupling) are perturbative.
This is obviously not the case in TC, where showing explicitly that the scattering amplitudes never grow beyond the unitarity bounds
would require a full non-perturbative analysis: many ideas have been proposed addressing this problem,
which is actively studied, 
but a definite answer does not exist, and hence I will limit the discussion to the well-understood perturbative case.
In the fourth subsection I summarize, by means of examples, a few very general expectations about the spectrum of 
TC, focusing on those bound states (techni-pions, techni-rhos and techni-dilaton) the physics of which
is controlled by strong symmetry arguments.
I conclude with a brief subsection containing a wish list, to some extent personal and incomplete. 
This is a list of specific questions about 
strongly-coupled EWSB, all of which emerge from the previous discussion,
all of which have crucial phenomenological implications, and answering  to any of which requires a 
very high level of understanding of non-perturbative physics.  Gauge/gravity dualities 
represent a remarkable opportunity for addressing some of these questions in a new context, which is
complementary to what is known from traditional techniques.

The second half of this paper contains a set of examples of applications
of the ideas and techniques derived from the context of gauge/gravity dualities,
in order to address some of the questions in the wish list.
This is not done systematically, rather I explain in some details two very special such applications,
one in the context of effective five-dimensional theories (bottom-up approach),
and one in the context of ten-dimensional supergravity and string theory (top-down approach).
This may  be understood as a proof of principle, showing that indeed 
some of the questions in the wish list can be addressed within specific holographic models,
and, furthermost, that the technology necessary for this type of studies  to be carried out does exist and  
is well understood.
It must be stressed that many more examples exist in the literature, that are not even mentioned in 
these lectures. Since the early papers on the AdS/CFT correspondence~\cite{AdSCFT} and 
by Randall and Sundrum~\cite{RS}, 
a huge body of literature on these and related subjects appeared, reviewing which 
goes very far beyond the aims of these lecture notes.

The third section is devoted to  one specific five-dimensional effective theory that most resembles what might be the dual of a 
four-dimensional walking technicolor theory~\cite{AdSTC,AdSTC2}.
I  show how calculability is strongly improved within this 
approach, in respect to what happens in four-dimensional EFT and HLS approaches,
yielding useful phenomenological correlations 
between observable quantities such as the oblique precision parameters and the spectrum and couplings of techni-rho mesons.
I  also show that some fundamental questions cannot be answered within this approach, but require more information
about the underlying dynamical properties of the models.

In the fourth section, a more ambitious approach is  discussed~\cite{NPP,ENP,NPR}. 
The proposal is to study the properties of walking dynamics in isolation, separating them from the
specific embedding of walking into a detailed model of electro-weak symmetry breaking and fermion mass generation.
This is  done by constructing a 10-dimensional string-theory model such that, in the supergravity limit,
 walking, or some feature that resembles it,  emerges dynamically,
and for which the technology exists allowing to perform a set of calculations that can yield a quantitative answer to
some of the questions about walking dynamics that are left open by bottom-up approaches.
This is a rather novel research program, and hence only few such calculations exist. I will limit myself to
one specific model, which has its own advantages and disadvantages. 
A few other possible models and other interesting calculations are suggested, but the main aim of this section
is to encourage the reader  to use his own experience, knowledge and creativity to devise 
alternative scenarios, and use them to address some of the problems listed earlier on.

\newpage
\section{Electro-weak symmetry breaking, within and beyond the standard model.}

\subsection{The Minimal Standard Model vs. Technicolor.}

The standard model (SM) is a quantum field theory in which
 weak, strong and electromagnetic interactions
 are described by a gauge theory with group $G_{SM}=SU(3)_c\times SU(2)_L\times U(1)_Y$.
All the known fundamental matter particles are described by  
(chiral) fermions transforming in some representation of $G_{SM}$ (see table~\ref{Fig:MSM}).
\begin{table}
\begin{tabular}{|c|c|c|c|c|c|}
\hline\hline
&  $J$ & $SU(3)$ & $SU(2)_L$ & $U(1)_Y$&$N$\cr
\hline
$B^{\mu}$& $1$ & $1$ & $1$ & $0$&$1$\cr
$W^{\mu}$& $1$ & $1$ & $3$ & $0$&$1$\cr
$G^{\mu}$& $1$ & $8$ & $1$ & $0$&$1$\cr
\hline
$q_L=\left(\begin{array}{c}u\cr d\end{array}\right)_L$&$1/2$&$3$&$2$&$1/6$&$3$\cr
$u_R$&$1/2$&$3$&$1$&$2/3$&$3$\cr
$d_R$&$1/2$&$3$&$1$&$-1/3$&$3$\cr
$\ell_L=\left(\begin{array}{c}\nu\cr e\end{array}\right)_L$&$1/2$&$1$&$2$&$-1/2$&$3$\cr
$e_R$&$1/2$&$1$&$1$&$-1$&$3$\cr
$\nu_R$&$1/2$&$1$&$1$&$0$&$3?$\cr
\hline
$H$&$0$&$1$&$2$&$+1/2$&$1$\cr
\hline\hline
\end{tabular}
\caption{Field content of the Minimal Standard Model. All the fermions are represented by 
a Left-Handed or a Right-Handed chiral field, with different quantum numbers. 
Specified is also the number $N$ of different species having the same quantum numbers. 
The 3 copies of fermions are referred  to in the text as {\it families}.
The neutrino singlet $\nu_R$ is a hypothetical particle, such as is the Higgs $H$.}
\label{Fig:MSM}
\end{table}
If the gauge symmetry were unbroken, all of the resulting particles would be massless.
However, the electro-weak gauge symmetry is spontaneously 
broken as $SU(2)_L\times U(1)_Y \rightarrow U(1)_{\rm e.m.}$,
by some condensate that provides a non-trivial structure for the vacuum of the theory.
The unbroken $U(1)_{\rm e.m.}$ is the gauge symmetry of electromagnetism.
Hence all the fermions, together with the $W$ and $Z$ bosons,
acquire a mass, proportional to the symmetry-breaking condensate.
The proportionality constant is the coupling  of the field to the condensate,
and depends on the specific field.
All of this is well established, and can be found nowadays even in popular science books.

It is also known that none of the SM fermions, nor any of the SM gauge interactions,
are responsible for electro-weak symmetry breaking itself~\footnote{The chiral condensate of QCD does, in fact,
spontaneously break the SM gauge symmetry. But the scale $f_{\pi}$ is so small that this fact can be ignored for most phenomenological purposes.}. 
Some new fields and new interactions
must be present, in order to implement in the theory a mechanism yielding spontaneous electro-weak symmetry breaking.
Identifying these new fields and, equally important, new interactions is the main goal of 
large hadron collider (LHC) program at CERN. In the case of the minimal version of the standard model
(MSM), the sector responsible for electro-weak symmetry breaking is constituted by one scalar field,
transforming as $(1,2,1/2)$ under $G_{SM}$. This is assumed to have a weakly coupled description
in terms of a scalar potential with a negative quadratic interaction and a quartic interaction,
the latter being the only free parameter to be measured at LHC. The classical minimum of the potential
is non-trivial, the Higgs field acquires a vacuum expectation value (VEV) 
and hence  yields electro-weak symmetry breaking.
The complete Lagrangian of the MSM contains five types of terms, which depend on the fields and on their covariant derivatives.
We will discuss in more detail some of these terms later on, for the time being it suffices to write them out schematically
(suppressing all the indexes).
\begin{itemize}
\item Gauge-boson Lagrangian
\beqs
{\cal L}_1 &=& -\frac{1}{2}\Tr F_{\mu\nu}F^{\mu\nu}\,+\,\cdots\,.
\eeqs
\item Fermion kinetic terms
\beqs
{\cal L}_{1/2} &=& \bar{\psi}\,i\,\Dslash\, \psi\,+\,\cdots\,.
\eeqs
\item Scalar kinetic term
\beqs
{\cal L}_0 &=& \left(D_{\mu} H\right)^{\dagger}D^{\mu}H\,.
\eeqs
\item Yukawa couplings
\beqs
\label{Eq:Yukawa}
{\cal L}_y &=& -y\,\bar{\psi_L}\,H\,\psi_R\,+\,\cdots\,.
\eeqs
\item Scalar potential
\beqs
\label{Eq:Higgs}
{\cal L}_{\cal V} &=& -{\cal V}\,=\,-\mu^2\,H^{\dagger}H\,-\,\lambda\left(H^{\dagger}H\right)^2\,.
\eeqs
\end{itemize}

The MSM is a very successful model. It agrees to very high accuracy with a number of precision electro-weak tests,
carried on in particular at SLAC, LEP, and TeVatron. It also successfully reproduces most of the 
flavor physics studied experimentally in rare processes involving B-mesons, kaons, pions,
muons. One important reason for this success is that  the MSM automatically implements
the GIM mechanism, so that all sources of flavor changing processes and of CP-violation are
encoded in the CKM matrix, hence naturally suppressing flavor changing neutral current (FCNC) processes. 
It is also a remarkably satisfactory model from a theoretical point of view,
in the sense that it automatically yields results that are 
very non-trivial. For example, thanks to the cancellation of all gauge anomalies,  and to the perturbative nature 
of all the couplings, including the Higgs potential, it is possible to show 
that unitarity  of certain scattering amplitudes is manifest at the diagrammatic level up to very high scales.

Yet, there is no experimental evidence supporting the idea that symmetry breaking be due 
to the weakly-coupled Higgs sector of the MSM, with LEP and TeVatron yielding exclusion regions and bounds on the mass,
but no unambiguous positive discovery signal.
  It is hence important to consider alternatives.
In particular, there is no real reason why the sector responsible for electro-weak symmetry
be weakly coupled. Actually, it is a remarkable fact that in the two most celebrated
examples of spontaneous symmetry breaking (the theory of superconductors, and 
the description of chiral symmetry breaking in QCD) the condensate does not arise from an
elementary scalar developing a VEV, but rather a strong interaction produces the formation
of a composite condensate.

Besides this simple fact, a set of  serious reasons for concern arises when studying the UV behavior of the 
Higgs sector itself. The Higgs potential does not satisfy the 
naturalness condition: switching off any of the couplings in the Higgs potential does not result in enhancing the 
symmetry of the system. We will discuss this in more details 
later. As a result, both the quadratic and quartic operator receive additive renormalization 
from (divergent) loop diagrams involving all the couplings of the model. In a sense, the dynamics encoded
in the potential is not fundamental, and as a result it is affected by fine-tuning problems.
Finally, the beta-function of the quartic coupling of the Higgs and of the Yukawa couplings is not 
asymptotically free (which is related to the triviality problem). 
All of this suggests that the Higgs sector might provide a very good description of low-energy physics,
but needs to be UV completed.

Technicolor models are a radical alternative to the Higgs of the MSM.
There is no weakly-coupled scalar, but rather
a completely new non-abelian gauge interaction is assumed to be present,
morally analog to QCD.
At low-energies (of the order of the electro-weak scale), the interaction becomes strong enough to trigger confinement
and chiral symmetry breaking, via the formation of a condensate of fermions.
As a  simple example, consider for instance a $SU(N_{T})$ gauge theory, in which fermions $Q_L$, $U_R$, $D_R$,
$\ell_L$, $E_R$ and $N_R$ with the same quantum numbers 
as a family of SM fermions transform on the fundamental
of  $SU(N_T)$~\footnote{Here and in several other points of these first two sections I chose to refer to this 
specific model mainly for its striking simplicity and elegance. The reader however should not be mislead into thinking 
that this is always what a TC model has to look like, but should simply think of TC as a strongly-coupled theory in which
a dynamically generated condensate induces EWSB. For example, lots of attention has been given in recent years to
small-$N_T$ models with techni-quarks in 2-index representations of the gauge group~\cite{Sannino2}, 
rather than on the fundamental representation,
and nothing prevents from much more exotic scenarios to be  considered.}.
At the electro-weak scale, condensates form:
\beqs
\langle \bar{Q}_L  U_R\rangle&=&
\langle \bar{D}_R  Q_L\rangle
\,=\,\langle \bar{E}_R  L_L\rangle
\,=\,\langle \bar{L}_L  N_R\rangle\,\sim\,{\cal O}(4\pi v_W^3)\,.
\eeqs
This condensate breaks electro-weak symmetry in exactly the same way as the Higgs 
of the MSM, because each of the bi-linears entering the condensates has the same 
quantum numbers as $H$.

The properties of models of this type are the opposite of those of the MSM Higgs sector.
All the couplings are natural, there is no fine-tuning nor triviality  problem, all the scales are generated
dynamically and naturally stabilized.
However, calculability (and hence predictivity) becomes problematic, because of the strong coupling.
Back-of-the-envolope estimates based on naive dimensional analysis (NDA) suggest
that it is difficult to reconcile these models with electro-weak precision measurements.
Unitarity of scattering amplitudes is not manifest at the diagrammatic level, but relies on uncalculable contributions from 
the strong dynamics. And, contrary to what happens in the MSM, the generation of the mass
of the SM fermions cannot proceed via marginal (Yukawa) couplings, but requires the introduction
of a whole new sector of interactions, via what is referred to as extended technicolor (ETC).
As a result, there is reason of concern regarding the contribution of this new ETC sector to FCNC processes,
because an analog to the GIM mechanism is not automatically present.

One very plausible way to alleviate the shortcomings of technicolor, while preserving its good features,
is provided by walking technicolor. The basic idea is that if the new strongly coupled theory is not some variation of 
QCD, but rather it has a very different dynamics, such that it be strongly coupled over a large range of energies 
above the electro-weak scale, then large anomalous dimensions are likely to be generated non-perturbatively.
If so, the counting rules of NDA do not hold, and  large enough masses for the SM fermions 
(the top quark is particularly problematic) might be generated dynamically,
without necessarily running into troubles with precision electro-weak parameters and FCNC processes.
The dynamical justification for this scenario is the assumption that an approximate fixed point governs the physics
in the IR, just above the electro-weak scale, hence slowing down  the running of the couplings in the strongly-coupled regime, 
whence the name {\it walking}.
While this adds to the calculability problems of traditional approaches to technicolor, it also provides
an ideal arena in which the techniques of the gauge/string duality can be applied. Indeed, in the walking regime 
the theory is very strongly coupled, and approximately conformal.

\subsection{Naturalness,  Hierarchy Problem(s) and their proposed solutions.}

All known interactions can be reduced, at the microscopic level, to three
fundamental ones: electro-weak, strong and gravitational interactions.
The strength of these interactions can be characterized in terms of a typical 
{\it scale}, that can be identified, respectively, with three dimensionful coupling constants:
\beqs
v_{w}\,\equiv\,\frac{1}{\sqrt{\sqrt{2}G_F}}&\simeq& 246\,{\rm GeV}\,,\\
f_{\pi}&\simeq& 93\,{\rm MeV}\,,\\
M_P\,\equiv\,\frac{1}{\sqrt{G_N}}&\simeq& 1.22\,\times\, 10^{19}\,{\rm GeV}\,,
\eeqs
where $G_F$ is the Fermi constant,  $f_{\pi}$ is the pion decay constant,
and $G_N$ the Newton constant. 

At present, the SM does not incorporate a universally accepted theory of quantum gravity.
However, it is conceivable that if such a formulation exists, the SM should be thought of 
as containing the leading-order terms  a low-energy effective field theory (EFT) description of all fundamental interactions.
In this spirit, quantum gravity should amount to two main effects (at low-energies): the renormalization of 
the SM Lagrangian ${\cal L}_{SM}$ itself, and the generation of higher-order corrections, encoded in higher-dimensional operators 
to be added to ${\cal L}_{SM}$.
Both effects would  be controlled by the typical scale of gravitational interactions $M_P$. 
Here is one way of seeing the emergence of a technical problem, the {\it big hierarchy problem}.
On the one hand, the fact that 
\beqs
v_w^2/M_P^2 \simeq 10^{-34}
\eeqs
justifies phenomenologically the fact that higher-order terms
can be safely neglected, because strongly suppressed. On the other hand, the very fact that these two scales be 
so far apart, in a context in which quantum corrections are present, needs to be explained. 
In the absence of such an explanation, one would expect  either that the electro-weak scale and the Planck scale 
should be of the same order of magnitude, $v_w \sim M_P$, or  that their hierarchy is due to accidental 
cancellations between interactions of very different nature, 
coming from physics taking place above and below the Planck scale,
hence requiring an implausible amount of fine-tuning.

This formulation of the big hierarchy problem shows explicitly that it has a very general origin. 
The only explicit assumptions used in order to highlight the problem itself 
being the existence of a quantum theory of gravity (or more general of any new interaction
beyond the SM), and of 
a sensible low-energy EFT description, characterized by the Planck scale $M_P$ 
(or by a new physics scale $\Lambda \gg v_W$). 
There is another hidden assumption in all of the above: 
that the space-time symmetries be 
described by the four-dimensional Lorentz (or Poincar\'e) group.
A vast literature exists on possible solutions to the big hierarchy problem, that propose to modify the
fundamental theory is very different ways. It is remarkable that 
the three main classes of such solutions (based upon supersymmetry, extra-dimension and technicolor, respectively)
all rely on the idea of extending the group of space-time symmetries.

The reason why this is a good idea has to do with the relation between the concept of {\it naturalness}
and renormalization in a general field theory. Using a definition often attributed to 't Hooft:
\begin{center}
a coupling is {\it natural} if, in the limit in which it vanishes, the theory has 
an enlarged symmetry.
\end{center}
In  field theory, quantum corrections  renormalize the 
bare Lagrangian in two possible ways: some couplings renormalize multiplicatively,
while others renormalize additively (due to operator mixing). 

Couplings that are not natural may in general renormalize additively.
Hence, they may be affected by fine-tuning problems. Setting to a very small value
a parameter that is renormalized additively does not ensure that quantum corrections will
preserve its smallness. A simple illustration of this is given within the MSM
by the Higgs potential ${\cal V}$ in Eq.~(\ref{Eq:Higgs}),
with neither  $\mu^2$ nor $\lambda$ being natural parameter~\footnote{It should be noticed that
setting $\mu^2=0$ would render the MSM Lagrangian scale-invariant at the classical level. However,
such invariance is not preserved by quantum correction, due to diagrams involving the quartic, gauge and Yukawa couplings,
and even allowing for fine-tuning in the quadratically divergent part of quantum corrections one ends
up with the appearance of a scale in the 1-loop generated Higgs potential, which ultimately might lead to 
electro-weak symmetry breaking itself~\cite{CW}.}.
As a consequence, nothing prevents them from receiving additive renormalization.
A way to compute the (divergent) part of the (perturbative) 1-loop correction makes use of the results
from Coleman and Weinberg~\cite{CW} for the quantum effective action:
\beqs
\label{Eq:CW}
\delta {\cal V}&=&\frac{\Lambda^2}{32\pi^2}{\cal ST}r\, {\cal M}^2\,+\,\frac{1}{64\pi^2}{\cal ST}r \left({\cal M}^4 \ln \left(\frac{{\cal M}^2}{\Lambda^2}\right)\right)\,,
\eeqs
where $\Lambda$ is the UV cut-off, ${\cal M}^2$ is the mass matrix of all fields as a function of the classical external scalar fields,
and ${\cal ST}r$ is the supertrace, a trace in which fermion and boson degrees of freedom enter with opposite signs.
For instance, in the MSM the coefficient $\mu^2$ of the quadratic operator  receives, among others,
 (quadratically divergent)  positive contributions from loops of gauge bosons
 and negative from loops of the top:
\beqs
\delta \mu^2 &\simeq& c_g g^2 \Lambda^2\,-\,c_y y_t^2\Lambda^2\,+\cdots\,,
\eeqs
with $c_g$ and $c_t$ numerical coefficients with mild 1-loop suppressions,
$g$ the gauge coupling and $y_t$ the top Yukawa coupling.
If the cut-off is at the Planck scale $\Lambda\sim M_P$, 
in order for the resulting quadratic coupling to be 
at the electroweak scale, a very fine-tuned cancellation would be needed between this loop correction
and an appropriately chosen counter-term.

Conversely, a natural coupling (barring the possibility of anomalies)  
renormalizes multiplicatively: in the limit in which the coupling is set to zero,
the theory has an extended symmetry, that must be preserved also by quantum corrections.
The fact that quantum corrections are themselves proportional to the natural coupling
ensures that this symmetry property be automatically satisfied.
As such, if one could find a way to render natural the parameter in the Lagrangian setting the electro-weak 
scale, this would not receive dangerously large additive renormalization via Eq.~(\ref{Eq:CW}),
and the coexistence of the widely separated electro-weak scale $v_W$ and Planck scale $M_P$ 
would not require fine-tuning, making the EFT useful and predictive.
This is what is done in the most popular extensions of the standard model.

A  most popular class of solutions proposes to  enlarge the symmetry by incorporating some form of supersymmetry,
i.~e. by effectively adding fermionic directions to the space-time (superspace formulation),
and assuming that perturbation theory be a good tool up to the Planck scale.
The standard example is the Minimal Supersymmetric Standard Model (MSSM)~\cite{susy}.
The basic idea is to enlarge the spectrum by adding a super-partner with different 
spin and statistics to each of the MSM fields,  and to adjust the couplings so that the new Lagrangian 
is supersymmetric. Details about the construction require to introduce two Higgs doublets instead of one.
The technical way in which supersymmetry solves the hierarchy problem is that,
when computing quantum corrections to the relevant operator controlling the electro-weak scale,
loops involving bosons and fermions in the same supersymmetry multiplets
cancel against each other.
In other words, in a ${\cal N}=1$ supersymmetric theory in a generic vacuum 
(and under a few  rather soft conditions such as the absence of mixed gauge-gravity anomalies)
\beqs
{\cal ST}r\, {\cal M}^2&=&0\,.
\eeqs
 As a result, the quadratically divergent
corrections to the supersymmetric 
generalization of Eq.~(\ref{Eq:Higgs})
 drop,  in favor of a milder logarithmic divergence.
In the MSSM, supersymmetry is broken explicitly, by adding to the supersymmetric Lagrangian a restricted set of
mass terms and couplings the effect of which are soft: they preserve the logarithmic 
dependence of the divergence on the cut-off scale. 
Interestingly, because the Higgs potential, in the supersymmetric limit, does not admit an electro-weak symmetry breaking
minimum, the latter is controlled by the scale of supersymmetry breaking itself $\Lambda^{\ast}$, hence linking 
electro-weak scale and supersymmetry-breaking scale.

A second class of solutions to the big hierarchy problem involves the introduction of
space-like extra-dimensions. There is a vast number of variations on this idea.
One most appealing such realization is the Randall-Sundrum scenario~\cite{RS}, possibly assisted by some
implementation of the Goldberger-Wise mechanism~\cite{GW}.
The basic idea is to introduce a fifth dimension $z$, and assume that the space-time has the geometry of 
a slice of AdS space
\beqs
\di s^2&=&\frac{L^2}{z^2}\left(\eta_{\mu\nu}\di x^{\mu}\di x^{\nu}\,-\,\di z^2\right)\,, 
\eeqs 
with  $L$ the curvature scale, and with two 4-dimensional  boundaries so that $L_0<z<L_1$.
The radial direction is related to the scale of the 4-dimensional theory living on a 
probe localized at $z$, with $z\rightarrow L_0$ corresponding to the UV and 
 $z\rightarrow L_1$ corresponding to the IR.
 While the 4-dimensional Planck scale $M_P$ is related to the 
 five-dimensional fundamental scale $M_5$ as
 \beqs
 M_P^2&\simeq&\frac{M_5^3L^3}{L_0^2}\,,
 \eeqs
 and one chooses (up to $O(1)$ factors) $L_0\sim L \sim 1/M_5 \sim 1/M_P$, 
 the typical scale of the physics taking place in the IR is suppressed as
 \beqs
 \Lambda^{\ast\,2}&\sim &\frac{L_0^2}{L_1^2} M_{P}^2\,.
 \eeqs
 In the Goldberger-Wise mechanism, the hierarchy between the position of the two boundaries
 is determined dynamically to be
 \beqs
 \frac{L_0}{L_1} &\propto& \left(\frac{v_1}{v_0}\right)^{1/\epsilon}\,,
 \eeqs
where  $v_{0,1}={\cal O}(M_P)$ are the boundary values of some dynamical bulk scalar,
and $\epsilon$ is a natural parameter, related to the coefficient of the 
only relevant coupling of the five-dimensional scalar (i.~e. its five-dimensional mass).
If the physics of electro-weak symmetry breaking  is to take place near the IR boundary, 
its scale is going to be exponentially suppressed in respect to the Planck scale, hence 
solving the big hierarchy problem, provided a modest hierarchy (for instance $v_1/v_0\sim 1/10$)
is chosen, together with a large value of the exponent (for instance $\epsilon \sim 1/20$).
For discussions about the holographic interpretation of this picture see also~\cite{revRS}.

A simple way of introducing the idea behind the third class of 
solutions to the hierarchy problem (technicolor) is to go back to QCD,
observing that in what we said up to now, we never mentioned a strong-interaction
fine-tuning problem related to the fact that $f_{\pi}\ll M_P$. 
For a good reason: there is no such a problem!
The Lagrangian of a QCD-like theory with $N_f$ massless fermions transforming on the fundamental representation of $SU(N_c)$ is
\beqs
\label{Eq:QCD}
{\cal L}&=&-\frac{1}{2}\Tr F^2\,+\,i\bar{\psi} \Dslash \psi\,,
\eeqs
where the covariant derivative is
\beqs
D \psi &=& \partial \psi - i g A \psi\,.
\eeqs
The beta-function computed at 1-loop is
\beqs
\beta(g)\,&=&-\frac{g^3}{(4\pi)^2}\left(\frac{11}{3}N_c\,-\,\frac{2}{3} N_f\right)\,=-\frac{g^3}{(4\pi)^2}b_0\,,
\eeqs
and accordingly
\beqs
\label{Eq:qcd}
g^2(\mu)&=&\frac{g^2}{1+\frac{g^2}{(4\pi)^2}b_0\ln \frac{\mu^2}{\Lambda^2}}\,,
\eeqs
where $\mu$ is the renormalization scale, and $g$ is defined at some reference scale $\Lambda$.
The expression for $g^2(\mu)$ diverges for $\mu^2=\Lambda^{\ast\,2}$ defined by
\beqs
\Lambda^{\ast\,2}&\equiv&\Lambda^2 e^{-\frac{(4\pi)^2} {g^2 b_0}}\,.
\eeqs
If $\Lambda \sim M_P$, and for choices of $N_c$ and $N_f$ such that $b_0 \sim O(1)$,
this means that $\Lambda^{\ast}$ defined in this way is exponentially suppressed, provided the theory be
asymptotically free, and the coupling $g$ measured at the scale $\Lambda$ be perturbatively small.
Because $\Lambda^{\ast}$ is the scale at which the 
theory becomes strongly coupled, then one has $f_{\pi} \sim O(\Lambda^{\ast}) \ll M_P$.

The symmetry principle behind the mechanism that makes $f_{\pi}$ small compared to the Planck scale
can be read from Eq.~(\ref{Eq:QCD}). At the classical level, this Lagrangian is scale invariant, because all the operators 
are exactly marginal.
Quantum effects break conformal invariance, the gauge coupling representing a marginally relevant deformation of the massless free theory
living at the UV fixed-point.
However, the breaking is controlled by the coupling itself: setting the coupling $g$ (as defined at a given scale $\Lambda$) 
 to very tiny values, implies that $\Lambda^{\ast}/\Lambda\rightarrow 0$, because in the $g=0$ limit the theory reduces back to a trivial conformal theory of non-interacting, massless gluons and quarks.
Again, a space-time symmetry is present, broken only by a natural parameter (the gauge coupling) which renormalizes 
multiplicatively (all renormalizations of the gauge coupling are proportional to the gauge coupling itself).
As a result, no fine-tuning is needed in order to parametrically separate the scale of the strong dynamics from the Planck scale.

The basic mechanism behind  technicolor is the same: it relies on removing the Higgs sector from the MSM, and replacing it with a new strong sector
with a new gauge symmetry (technicolor) acting on new degrees of freedom (techni-fermions). The electro-weak gauge group 
is defined by gauging a subgroup of the global symmetry of such fermions (techni-flavor).
Once the theory becomes strongly coupled at scale $\Lambda^{\ast}$, a chiral condensate will form dynamically, generating $v_W$ 
and spontaneously breaking the electroweak symmetry,
in analogy to what happens in QCD.
Appropriately choosing the new gauge coupling, it is possible to arrange for
$\Lambda^{\ast} \ll M_P$, without any fine-tuning, because
in the $g\rightarrow 0$ limit one recovers a conformal theory, with space-time symmetry described by $SO(4,2)$.

Summarizing, the three examples discussed share the same basic idea of enlarging the space-time symmetry.
While the way in which this is done is very different,  it yields very similar results, turning the dangerous 
quadratic divergences relating the electro-weak and Planck scales into logarithms, and linking the electro-weak scale
to a new scale $\Lambda^{\ast}$ characterizing the breaking of the enlarged space-time symmetry itself.
Namely, below $\Lambda^{\ast}$, supersymmetry is lost, the theory is four-dimensional, and scale-invariance is lost, respectively, in the three cases.
Provided $\Lambda^{\ast}\lsim 1$ TeV, no fine-tuning is present in the electro-weak symmetry breaking sector.
The scale $\Lambda^{\ast}$ is also the scale at and above which a new physics sector 
(fields and interactions) appears. It controls the masses of the superpartners of the SM fields, 
the masses of the KK modes coming from the compact extra-dimension, and the masses of the resonances and composite states
in technicolor (technimesons, technibaryons \dots).
But $\Lambda^{\ast}$  also controls {\it indirect} contributions to low-energy precision physics,
which we will discuss at length in these lectures.

\subsubsection{Little hierarchy problem.}

Whichever your favorite solution to the big hierarchy problem is, provided it is based on the idea
of enlarging the space-time symmetries as in the three cases discussed earlier, 
the net result is that
 there are now four physical scales, 
 characterizing four distinct interactions: $f_{\pi}$, $v_W$, $M_P$ and $\Lambda^{\ast}$.
The more the two scales $v_W$ and $\Lambda^{\ast}$ are close to each other,
the more satisfactory is the solution to the hierarchy problem.

 Below $\Lambda^{\ast}$, the process of integrating out all of the sector responsible for 
 the breaking of the enlarged symmetry (and for the arising of $\Lambda^{\ast}$ itself) means that 
corrections to the low-energy description of electro-weak and strong interactions 
will be induced, controlled by $\Lambda^{\ast}$.
In other words, the EFT whose leading order terms are given by the standard model
will have corrections suppressed by inverse powers of $\Lambda^{\ast}$, besides
those suppressed by the Planck scale $M_P$.
These corrections are tightly bounded experimentally. The precision electro-weak tests
 can be summarized in the precision parameters of the electro-weak chiral Lagrangian 
(or equivalently in the oblique parameters such as $S$ and$T$), and they indicate that
 the physics of the electro-weak gauge bosons can receive correction
 coming from higher-dimensional operators  that are at most at the per mille level.
 Naively, this suggests that $\Lambda^{\ast} \gsim 5$ TeV.
 The physics of flavor changing neutral currents, tested in great details 
 by studying rare processes involving kaons, $B$ mesons, charmed mesons, pions
 and charged leptons, also indicates that, unless $\Lambda^{\ast}$ has nothing to do with 
 flavor, $\Lambda^{\ast}$ must be large.

In the specific case of the MSSM, analogous source of tension can be shown to have a more subtle origin.
Some of the  scalars from the Higgs multiplets complete, in the supersymmetric limit,
the heavy vector multiplets containing the gauge bosons and the gauginos.
In this limit the mass of such scalars is closely linked to the mass of the electro-weak gauge bosons.
In particular, for the mass of the lightest Higgs scalar  in the MSSM 
the bound $m_h^2 < M_Z^2$ holds in complete generality at the tree-level~\footnote{Technically, this originates from the fact that the quartic couplings 
in the Higgs potential are not free parameters, but actual gauge couplings derived from the
$D$-terms within the supersymmetric  gauge theory.}.
Loop effects involving (finite) diagrams affected by supersymmetry breaking 
partially remove this bound, and one can show for example that the corrections to the physical mass of the 
 lightest scalar particle in the spectrum, very  schematically look like~\cite{HiggsMSSM}
\beqs
\label{Eq:susyhiggs}
\delta m_h^2&\simeq& c_h v_W^2 \ln \frac{\Lambda^{\ast\,2}}{v_W^{2}},
\eeqs
where $c_h$ is again 1-loop suppressed.
Because $c_h$ is perturbatively small (loop-suppressed),
evading the experimental bounds from direct searches $m_h\gsim 114$ GeV 
requires, again, that $\Lambda^{\ast}$ should be somewhere in the few-TeV range
(or at least the mass of one of the scalar partners of the top should). The logarithmic dependence also
means that if the lower bound on the Higgs mass raises, evading it would require 
a very significant amount of fine-tuning.

Concluding, solving the big hierarchy problem requires linking the electro-weak scale
to a new scale $\Lambda^{\ast}$ which is kept {\it naturally} smaller  than the Planck scale,
and at which some new enlarged space-time symmetry be broken.
Absence of any fine-tuning would be achieved if $\Lambda^{\ast} \lsim 1$ TeV.
However, indirect searches strongly suggest that $\Lambda^{\ast} \gsim 5$ TeV,
both in weakly and strongly coupled scenarios,
substantially above the scale dictated by $v_W$.
This tension goes under the name of {\it little hierarchy problem}.
There are several ways of addressing the little hierarchy problem, which may require the introduction
of entirely new fields and symmetries (such as in the case of composite Higgs~\cite{compositeHiggs} models and Little Higgs models~\cite{LH}),
hence again naturally separating $v_{W}\ll \Lambda^{\ast}$.
Or one might  modify the solution to the big hierarchy problem in such a way as to implement some 
discrete symmetry, in such a way as to enforce cancellations and suppressing unwanted 
processes.  The literature on the subject is vast, and these proposals are very interesting phenomenologically, 
both for the LHC and for astrophysics and cosmology
(for examples, models with parities might play an important role in the physics of cold dark matter), but are not the 
subject of these lectures.

It must be stressed that numerically the little hierarchy problem is, indeed, little.
In a generic low-energy theory it amounts to a fine-tuning at most ${\cal O}(1\, {\rm TeV}^2/(5\, {\rm TeV})^2)\sim 5\%$.
The arguments presented here
are strongly affected by model-dependent coefficients that in this discussion have been taken to
be $O(1)$, but that in special models might just occur to be small enough to evade the experimental bounds.
Particularly in the case of technicolor, which naively is the one example in which the little hierarchy problem is
most worrisome, the strongly coupled nature of the theory at the $\Lambda^{\ast}$ scale makes the calculation of such 
coefficients quite delicate.
Accurate and efficient ways of computing such observables are needed. 
 Part of these lectures will be devoted to this specific topic,
 in the context of walking technicolor (which itself involves the introduction of a very special symmetry),
  and we will see that this possibility is not excluded.
  While doing so, the arguments presented in this short subsection will be made more rigorous and quantitative,
  and concrete examples will be developed.

\subsubsection{Final remarks}

Before we move away from these general introductory remarks, and start studying more concrete
examples and perform more specific calculations, a final set of comments are due to the reader.
First of all, one might argue that hierarchy problems as defined in the previous section
are not severe failures of the theory that require 
its complete rewriting: 
they do not indicate a fatal contradiction between theory and experiment.
After all, once properly renormalized, the MSM and its parameters can be used to reproduce correctly
a vast amount of experimental data, with no clear indication of a contradiction 
(barring the fact the the Higgs particle has not been observed yet,
and hence might not exist in nature,
which is a different problem).
In general, the emergence  of a hierarchy problem
can be looked at as the emergence of an
opportunity for research. A fine-tuned parameter 
is an indication given to us by nature of a possible avenue 
for discovery of actually new, interesting physical phenomena, by highlighting a sector of the theory 
which we do not really understand at the fundamental level, and which is probably incomplete. 
It would be regrettable not to pursue this avenue.
This is one reason why so much attention is paid by model-builders to
the hierarchy problems. Besides the electro-weak hierarchy problem(s) summarized here, similar 
attitude is applied towards other hierarchy (fine-tuning) problems, such as the strong-CP problem in QCD, and, most important, 
the problem of the cosmological constant. 

Second, it must be stressed that the smallness of a parameter does not necessarily 
constitute a hierarchy problem.
At least, not in the sense of naturalness used here.
For instance, one might legitimately ask why there exists such 
a large hierarchy between the numerical values of the masses of the SM fermions, 
comparing for instance the neutrinos to the electron, and to the top quark.
However, all of these masses are controlled by parameters that are {\it natural},
and hence their values are stable against radiative corrections, without the technical problem of 
fine-tuning. 
It would certainly be very interesting and useful to have a theory of flavor, in which these masses and their hierarchical 
structures are generated dynamically from first principles, and a vast 
literature exists on the subject, with beautiful and very interesting results having 
been collected  over the years. Yet, the argument in favor of such a research program is 
somewhat less compelling, at least within weakly-coupled theories.
However, this kind of research program becomes compelling in strongly-coupled extensions of the standard model,
 for reasons that will be explained later in these notes.

Finally, the ultimate reason why fine-tuned theories are unsatisfactory is
deeply connected with the concept of effective field theory. All progress in physics
up to recent times can be thought of in terms of a chain of EFTs, and the vast majority of physical 
phenomena that we believe we understand are actually understood in terms of their EFT description.
EFTs, as opposed to would-be fundamental theories, are constructed on the basis of very few and simple general
symmetry rules. These rules cannot, by themselves, single out a Lagrangian (or Action) 
which contains only a finite number of 
local interactions. In general, an EFT contains an infinite number of such interactions, and an infinite number 
of free parameters. In the absence of anything else, these theories cannot be calculable and predictive,
because no matter how many independent experimental quantities are measured and used as input, 
no unique prediction can be formulated.
What makes EFTs work (and they do in real situations), is that one can implement in the EFT also
counting rules, ensuring that within the regime of validity of the EFT, and requiring a given accuracy from the calculations
to be performed, only a finite number of such parameters are actually important. The (infinite) others are, on the basis of the
counting rules, expected to give contributions that, while uncalculable, are safely smaller than the required accuracy.
Hence, the EFT Lagrangian can be truncated to contain only a finite number of terms.

The  problem comes from the fact that
the counting rules do not yield actual predictions for the values of the 
(unknown) parameters, but provide only order-of-magnitude estimates.
Such estimates are hence useful only provided no fine-tuning turns out to be necessary.
In the case where one of the parameters that are kept in the truncated EFT turns out to be severely fine-tuned, such a parameter
is explicitly violating the counting rules. It becomes hence impossible to justify the truncation itself,
and predictivity is lost.
Hence, the hierarchy problem(s) in the standard model 
highlight a contradiction in the procedure 
that would allow to use it as (the leading order part of) an EFT description.
The history of successes of EFTs itself provides a very compelling reason to look
for solutions to the hierarchy problem, beyond the SM.

\newpage
\section{From weak to strong coupling: walking technicolor.}

We have already anticipated that technicolor provides a solution to the 
big hierarchy problem based on the same mechanism at work in QCD, 
namely at the Planck scale the theory is very close to a trivial fixed point
of the renormalization group equations. The coupling of a marginally relevant operator (the gauge coupling)
drives the theory away from this fixed point towards the IR. The theory becomes 
strongly coupled at a scale $\Lambda^{\ast}$. An exponential hierarchy $\Lambda^{\ast}\ll M_P$ 
is natural, because controlled by the (perturbative) value of the gauge coupling in the UV.
Finally, the strong coupling triggers the formation of a condensate that induces EWSB, hence linking $\Lambda^{\ast}$ and $v_W$.
We also anticipated that the phenomenological viability of such 
a scenario is heavily questioned by the fact that the little hierarchy problem 
is exacerbated by the strong coupling at the scale $\Lambda^{\ast}$, particularly in a QCD-like theory,
and that walking technicolor is one proposal  that might soften  such a problem.
Let us be more specific. 

\begin{figure}[htpb]
\includegraphics[width=5.8cm]{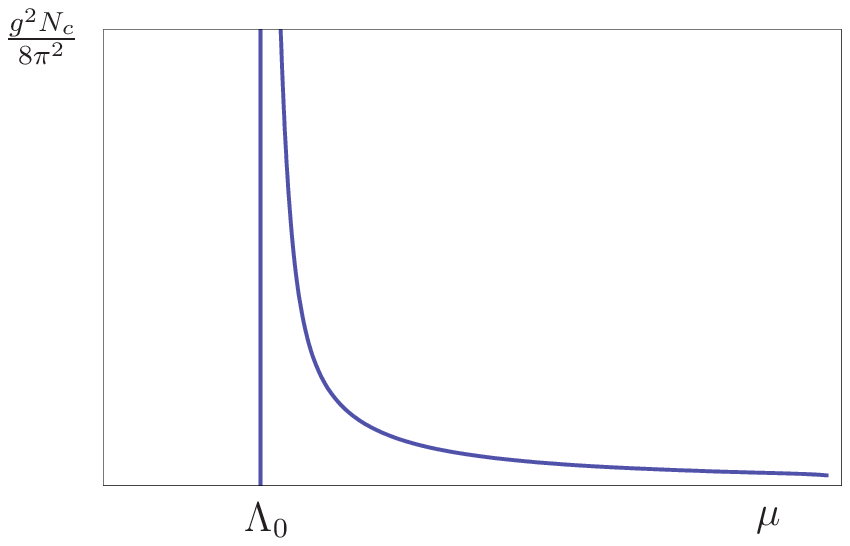}
\includegraphics[width=5.8cm]{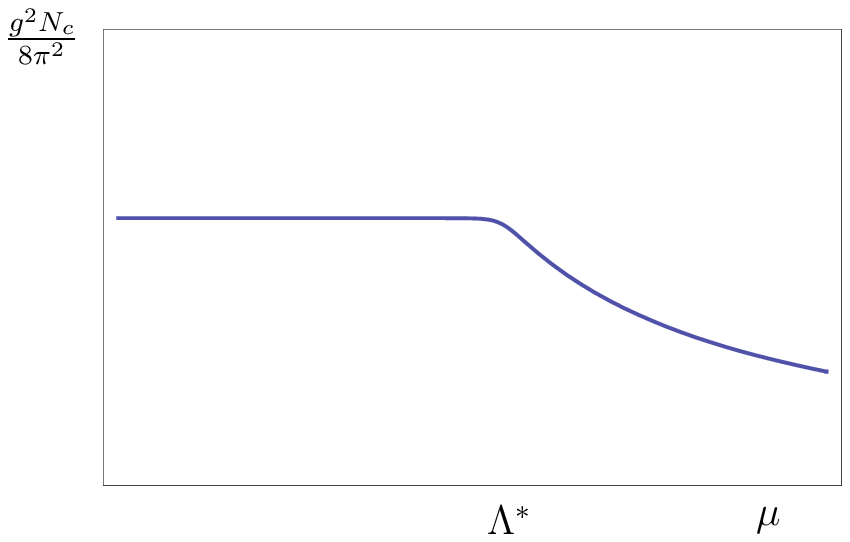}
\includegraphics[width=5.8cm]{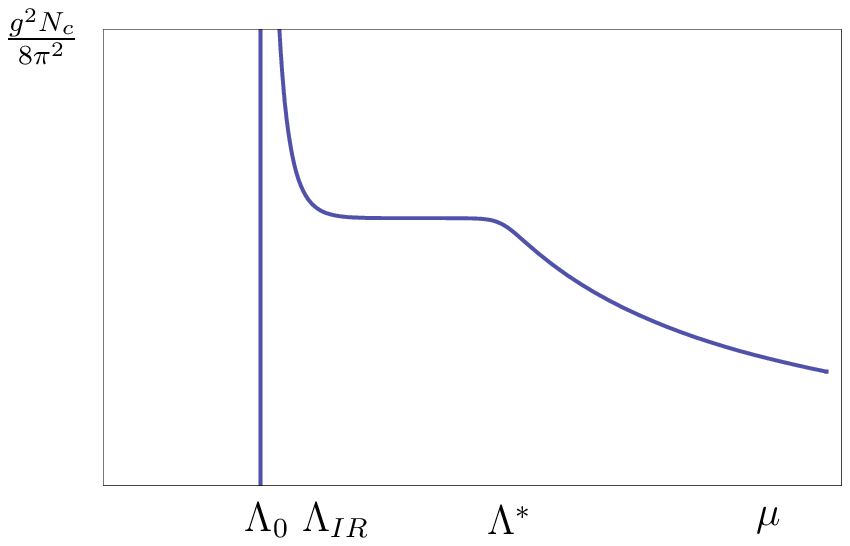}
\caption{Schematic depiction of the evolution of the gauge coupling as a function of the renormalization
scale in three different asymptotically free theories:  a QCD-like theory, a theory with an IR fixed point, and a walking theory (left to right).}
\label{Fig:running}
\end{figure}

Walking technicolor is based on a gauge theory
in which the running of the gauge coupling is completely different from what expected
in a QCD-like theory. Fig.~\ref{Fig:running} illustrates the basic idea.
On the left panel is the QCD-like running as obtained for instance from Eq.~(\ref{Eq:qcd}). 
One dynamical scale $\Lambda_0=\Lambda^{\ast}$
exists, identified by the divergence of the running coupling $g^2N_c/(8\pi^2)$.
The middle panel of  Fig.~\ref{Fig:running} presents a theory that
in the IR flows to a non-trivial fixed point.
Again, one dynamical scale exists $\Lambda^{\ast}$, separating the two energy regimes
at which the theory is well approximated by the UV and IR fixed points, respectively.
There exists two well-established examples of such behavior.
One can be seen to exist by looking at the 2-loop perturbative
$\beta$-function of a $SU(N_c)$ theory with $N_f$ vectorial fermions on the fundamental representation~\cite{BZ}.
If $N_f$ is very close to the limit beyond which asymptotic freedom is lost, the $\beta$-function is suppressed by anomalously small
coefficients. One then finds that a fixed point exists in the IR, and that provided this is perturbative (so that
neglecting higher-loop contributions is justified) the theory flows into a weakly-coupled non-abelian Coulomb phase.

Another example emerges for the supersymmetric version of 
the same gauge theory, in which case the beta function is~\cite{NSVZ}
\beqs
\beta(g)&=&-\frac{g^3}{16\pi^2}\left(\frac{3N_c-N_f-N_f\gamma(g^2)}{1-N_c\frac{g^2}{8\pi^2}}\right)\,,
\eeqs
where the anomalous dimension of the mass is
\beqs
\gamma(g^2)&=&\frac{g^2}{8\pi^2}\frac{N_c^2-1}{N_c}\,.
\eeqs
Again, when $N_f$ is very close to $3N_c$, a fixed point in the spirit of~\cite{BZ} exists.
However, because of supersymmetry the theory possesses a Seiberg duality~\cite{Seiberg},
which in turns means that such a fixed point exists in the whole conformal window
\beqs
\frac{3}{2} N_c \,<\, N_f\,<\, 3N_c\,.
\eeqs
The interesting fact is that towards the lower end of the conformal window,
the fixed point is strongly coupled. Again thanks to supersymmetry,
one can compute the anomalous dimension $\gamma$ at the fixed point:
\beqs
\gamma&=&-1+\frac{3N_c}{N_f}\,,
\eeqs
which agrees with the intuition by vanishing when $N_f\rightarrow 3N_c$,
but towards the other boundary of the conformal window yields
\beqs
\lim_{N_f\rightarrow 3N_c/2} \gamma&=&1\,,
\eeqs
with the dimension of the (super-)quark bilinear  $d(\tilde{Q}Q)=2-\gamma$ 
changing continuously in the range $1<d(\tilde{Q}Q)<2$ as a function of $N_f$, by assuming values that are ${\cal O}(1)$
and hence completely non-perturbative.

The dynamics of a walking theory is a combination of these confining 
and conformal theories.
In the IR, below the scale $\Lambda^{\ast}$, the
theory  is approaching an approximate fixed point, governed by a strongly-coupled
conformal field theory, described by operators that have large anomalous dimensions. 
The fixed point is approximate in the sense that some dynamical feature prevents the
theory from actually reaching such a fixed point. Deep into the IR, the theory actually confines and yields a 
symmetry breaking condensate at the electroweak scale $v_W$.
The gauge coupling evolves as in the right panel of  Fig.~\ref{Fig:running}, in a way that is somewhat an hybrid
of the previous two cases.

Notice that in the walking case there are at least three dynamical scales. A scale $\Lambda^{\ast}$
analogous to the conformal case, at which the theory enters a quasi-conformal phase,
a scale $\Lambda_0$ deep in the IR in which the gauge coupling diverges in the same sense as in the 
QCD-like case. And an additional intermediate scale $\Lambda_{IR}$, 
at which the theory stops walking, and enters the fast running leading ultimately to confinement at $\Lambda_0$.
The word {\it walking} refers to the fact that the coupling is approximately constant
in the region $\Lambda_{IR}<\mu<\Lambda^{\ast}$, where the beta function is 
parametrically small in comparison to the coupling itself.
Somewhere between the two lower scales (which might coincide) 
the symmetry breaking scale $v_W$ should be dynamically generated too.

A second feature, of utmost phenomenological relevance, is that because the physics in the walking region is governed
by the strongly-coupled  approximate IR fixed-point, its operators acquire in general 
very large, non-perturbative anomalous dimensions, which drastically change the 
power-counting rules expected from NDA and implemented in the low-energy effective description of electro-weak physics
in the standard model.

In principle, one would like to a have a non-perturbative, non-supersymmetric theory
with the properties mentioned above. We have no rigorous proof that such a thing exists,
but  it is usually believed that some critical $N_f^c$ exists above which
the theory flows into an IR fixed point, and below which it confines and condensates form.
A walking theory is supposed to be given by the limit in which $N_f$ is close to this critical value.
In this lecture we will be more precise about  the phenomenological implications of the previous statements, and 
discuss what are the challenges of this proposal, 
in direct comparison with what happens in perturbative theories such as the MSM.

\subsection{Electro-weak chiral Lagrangian and   oblique precision parameters.}

A convenient way of parameterizing the low-energy effects of the strongly-coupled sector responsible for 
electro-weak symmetry breaking is provided by the electro-weak chiral Lagrangian~\cite{EWCL}. 
The basic idea is that at very low energy the only relevant degrees of freedom are
given by the Goldstone bosons and the electro-weak gauge bosons~\footnote{In these lectures, it is assumed that the SM 
fermions do not carry quantum numbers of the technicolor sector, and are essentially treated as spectators.}. An effective description is then obtained by writing (infinite numbers of) 
possible local interactions involving pions and gauge bosons, compatibly with the symmetries of the problem,
and organizing the expansion as a  powers-series.

One can think of the resulting effective action as a systematic approximation of the quantum effective action 
that would be obtained by integrating out all 
the heavy degrees of freedom in the theory (techni-mesons, techni-baryons, techni-glueballs \dots). In principle, this is 
a non-local function of the momentum $q^2$. Provided $q^2$ is much smaller than a scale $\Lambda^{\ast\,2}$
related to the lightest among the masses of all the degrees of freedom
that are integrated out, one can expand in powers of $q^2/\Lambda^{\ast\,2}$, and truncate at a given order in the expansion.
The truncation reduces the number of independent couplings to be finite. 
At the same time, the truncation determines the level of accuracy of the results of calculations of physical observables,
which is itself controlled by a power of $q^2/\Lambda^{\ast\,2}$.

\begin{figure}[htpb]
\begin{picture}(0,100)(100,0)
\GCirc(70,26){10}{0}
\Line(80,26)(120,26)
\GCirc(130,26){10}{0.4}
\Text(100,38)[c]{ \Large $S$}
\Text(70,6)[c]{ $SU(2)_L$}
\Text(130,6)[c]{ $U(1)_Y$}
\Text(70,46)[c]{ $SU(2)_L$}
\Text(130,46)[c]{ $SU(2)_R$}
\end{picture}
\caption{Diagram of the 2-site model, the electro-weak chiral Lagrangian.}
\label{Fig:2sites}
\end{figure}
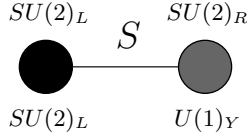

The electro-weak chiral Lagrangian~\cite{EWCL} is defined from the two-sites model in Fig.~\ref{Fig:2sites}. 
One defines a  $SU(2)_{{L}}\times SU(2)_{{R}}$ global symmetry,
gauges the $SU(2)_L\times U(1)_Y$ subgroup, with $U(1)_Y$ generated by $T^3$,
and introduces the dimensionless
\beqs
S&=&e^{\frac{2 i \pi}{{f}}}\,,
\eeqs
with $\pi=\pi^a T^a$, $T^a=\tau^a/2$ and $\tau^a$ the Pauli matrices,  
so that $S$ transforms as a bi-fundamental under $SU(2)_L\times SU(2)_R$. 
The decay constant is the electro-weak VEV,  ${f}=v_W$.
The pion fields $\pi^a$ are the Goldstone bosons of the spontaneous breaking $SU(2)_L\times SU(2)_R\rightarrow SU(2)_V$.
In unitary gauge $S=1$, and the pions become the longitudinal components of the electro-weak gauge bosons.
The covariant derivative is
\beqs
DS &=& \partial S\,+\,i\left({g}{W} S - {g}^{\prime}S{B}\right)\,,
\eeqs
with $W=W^aT^a$ the weakly-coupled gauge bosons of $SU(2)_L$,
$B=B^3T^3$ the gauge boson associated with the diagonal $U(1)_Y\subset SU(2)_R$,
and $g$ and $g^{\prime}$ the (weak) gauge couplings.
In terms of the field strength $W_{\mu\nu}$ and $B_{\mu\nu}$,
of the covariant derivative $D_{\mu}S$, of the SM currents $J_{W,B}$ (which contain the SM fermions) and of the custodial-symmetry breaking 
operator $t= S T^{3} S^{\dagger}$,
 the Lagrangian reads
\beqs
\label{Eq:Chi}
{\cal L}_{\chi} & = &  
{\cal L}_{J}
+ {\cal L}_{g}
+ {\cal L}_{2,0}
+ {\cal L}_{2,2}
+ {\cal L}_{4,0}
+ {\cal L}_{4,1}
+ {\cal L}_{4,2}
+{\cal L}_{4,4}\cdots
\eeqs
where
\beqs
{\cal L}_{J}&=&
+2 {g}\,\Tr{W}\,J_W\,-\,2 {g}^{\prime}\Tr{B}\,J_B\,,\nonumber\\
{\cal L}_{g}&=&
-\,\frac{1}{2}\Tr {W}_{\mu\nu}{W}^{\mu\nu}
\,-\,\frac{1}{2}\Tr {B}_{\mu\nu}{B}^{\mu\nu}\nonumber\,,\\
 {\cal L}_{2,0}&=&
 \frac{ {f}^2}{4}\Tr |DS|^2\,,\nonumber\\
  {\cal L}_{2,2}&=&
  \alpha_0g^2 {f}^2[\Tr(t D_\mu S S^{\dagger})]^2 \,,\\
 {\cal L}_{4,0}&=&
\alpha_1 {g} {g}^{\prime}\,\Tr\,S B_{\mu\nu} S^{\dagger} W^{\mu\nu}\,\nonumber\\
&&+\,i\alpha_2 {g}^{\prime}\,\Tr\left( B_{\mu\nu}\left[S^{\dagger}D^{\mu}S\,,\,S^{\dagger}D^{\nu}S\right]\right)\nonumber
\\ && \,+\,i\alpha_3 {g}\,\Tr\left( W_{\mu\nu}\left[D^{\mu}SS^{\dagger}\,,\,D^{\nu}SS^{\dagger}\right]\right)\nonumber\\
&&+\alpha_4\,\Tr\left(D_{\mu}SS^{\dagger}D_{\nu}SS^{\dagger}\right)\,\Tr\left(D^{\mu}SS^{\dagger}D^{\nu}SS^{\dagger}\right)\nonumber
\\ &&\,+\,\alpha_5\,\Tr\left(D_{\mu}SS^{\dagger}D^{\mu}SS^{\dagger}\right)^2\,.\nonumber\\
 {\cal L}_{4,1}&=&
  2\alpha_{11} {g}\epsilon^{\mu\nu\rho\lambda}Tr(t V_\mu)Tr(V_\nu W_{\rho\lambda})\nn \,,\\
 {\cal L}_{4,2}&=&
 4 \alpha_6\Tr\left(D_{\mu}SS^{\dagger}D_{\nu}SS^{\dagger}\right)
 \Tr(t D^\mu S S^{\dagger})\Tr(t D^\nu S S^{\dagger}) \nn 
             \\  && + 4 \alpha_7\Tr\left(D_{\mu}SS^{\dagger}D^{\mu}SS^{\dagger}\right)
             \Tr(t D_\nu S S^{\dagger})\Tr(t D^\nu S S^{\dagger}) \nn 
    \\             &&+  \alpha_{8} {g}^2 [\Tr(t W_{\mu\nu})]^2 \nn 
      \\        && +{2}i \alpha_9\Tr(t W_{\mu\nu})\Tr\left(t \left[D^{\mu}SS^{\dagger}\,,\,D^{\nu}SS^{\dagger}\right]\right) \nn\,,\\
 {\cal L}_{4,4}     &=&
8\alpha_{10}[ \Tr(t D_\mu S S^{\dagger})\Tr(t D_\nu S S^{\dagger}) ]^2\,.\nn
\eeqs

The notation introduced here is somewhat unconventional, and demands for an explanation.
In the terms ${\cal L}_{i,j}$, the index $i$ indicates the dimension of the operators,
obtained by counting $[S]=[E]^0$, $[\partial]=[D]=[E]^1$ and $[W_{\mu\nu}]=[B_{\mu\nu}]=[E]^2$.
The index $j$ counts the explicit appearances of $t$, the operator introducing  explicit breaking of custodial symmetry.
In practice, there are three expansions being performed, and used in classifying corrections beyond ${\cal L}_{2,0}$. 
One is the aforementioned expansion in derivatives ($q^2/\Lambda^{\ast\,2}\ll 1$). 
But there is also an expansion in a custodial-symmetry breaking parameter $\delta/\Lambda^{\ast}$, 
implicit within the definition of many of the $\alpha_i$.
And finally we are making use of the perturbative expansion in $g^2/(4\pi^2),g^{\prime\,2}/(4\pi^2)\ll 1$.
The way in which the Lagrangian is written is such that the small parameters are all 
implicit in the definition of the coefficients $\alpha_i$, which are hence expected to be small.

In order to understand the origin of all these couplings, and how the counting works,
let us start by assuming that custodial symmetry be exact. 
The only terms surviving in this limit are
the ones in ${\cal L}_{2,0}$ and ${\cal L}_{4,0}$~\footnote{Some insight might be obtained from the comparison to the chiral Lagrangian, and the classification by Gasser and Leutwyler~\cite{GL}
of the coefficients appearing at the $O(q^4)$.
$\alpha_1$ corresponds to $L_{10}$, $\alpha_2$ and $\alpha_3$ to $L_9$, $\alpha_4$ to $L_2$ and $\alpha_5$ to $L_1$.}.
One could start from ${\cal L}_{2,0}$ alone. The fact that the symmetry is non-linearly realized is reflected in the fact that
$S$ contains only the three degrees of freedom of the pion (the constraint $SS^{\dagger}=1$ is applied).
 Expanding in powers of $\pi/{f}$, at the leading-order one gets the 
(canonically normalized) kinetic terms for the pions, while  the next-to-leading order yields a (derivative) quartic coupling proportional to $1/{f}^2$
\beqs
{\cal L}_{2,0}&=&\Tr (\partial \pi)^2\,+\,\frac{1}{3{f}^2}\Tr [\pi,\partial \pi] [\pi,\partial \pi]\,+\,\cdots\,,
\eeqs
which, for instance, yields the leading-order contribution to the $\pi\pi\rightarrow \pi\pi$ elastic scattering amplitude.
This coupling can also be used to construct 1-loop diagrams.
Because the EFT is non-renormalizable, the divergences of the 1-loop diagrams 
cannot be easily removed: there are logarithmically divergent amplitudes proportional to $q^4$.
In order to cure these divergences, one has to introduce  new couplings
in the Lagrangian, which are those in ${\cal L}_{4,0}$.
This means that one should think of the expansion in $q^2/\Lambda^{\ast\,2}$ as an expansion in $\hbar$.
Going to higher loops will involve introducing new couplings with higher number of derivatives.
Most important, in comparing with experimental data, one has to be careful to use 
the Lagrangian ${\cal L}_{\chi}$ appropriately: if one wants to extract a leading-order value for, say $\alpha_1$,
 the comparison should be done by including 1-loop diagrams built with the couplings extracted from ${\cal L}_{2,0}$. 

Analogously, one can think of custodial-symmetry breaking effects as arising from loop diagrams involving 
fundamental degrees of freedom (techni-quarks) in the underlying theory, 
whose masses and couplings violate custodial symmetry.
And hence we associate also $\delta^2/\Lambda^{\ast\,2}$
with ${\cal O}(\hbar)$ corrections.
The power-counting is hence determined by assuming that
\beqs
\frac{q^2}{\Lambda^{\ast\,2}}&\sim&\frac{\delta^2}{\Lambda^{\ast\,2}}\,\sim\,\frac{g^2}{4\pi^2}\,\sim\,{\cal O}(\hbar)\,.
\eeqs

Looking back at ${\cal L}_{\chi}$ one can ask what is the physical meaning of the couplings
in terms of the physics of electro-weak gauge bosons.
One immediately sees that $\alpha_0$, $\alpha_1$ and $\alpha_8$ produce corrections to the 
gauge boson propagators, while $\alpha_2$, $\alpha_3$, $\alpha_{11}$ and $\alpha_9$  correct the cubic
couplings and $\alpha_4$, $\alpha_5$, $\alpha_6$, $\alpha_7$ and $\alpha_{10}$ the quartic couplings.
Precision studies of the electro-weak interactions bound mostly 
the corrections to the vacuum polarizations. 
Following~\cite{Barbieri}, the oblique parameters can be defined starting 
from the transverse vacuum polarization amplitudes of the gauge bosons  of the standard model,
expanding as
\beqs
\Pi(q^2)&=&\Pi(0)+q^2\Pi^{\prime}(0)+\frac{1}{2}(q^2)^2\Pi^{\prime\prime}(0)+\cdots\,,
\eeqs
and defining, for $\Pi^{\prime}_{WW}(0)=1=\Pi^{\prime}_{BB}(0)$:
\beqs
\label{Eq:Shat}
\hat{S}&=&\frac{g}{g^{\prime}}\Pi_{W^3B}^{\prime}(0)\,,\\
\hat{T}&=&\frac{1}{M_W^2}\left(\Pi_{W^3W^3}(0)-\Pi_{W^+W^-}(0)\right)\,,\\
\hat{U}&=&\left(-\Pi_{W^3W^3}^{\prime}(0)+\Pi_{W^+W^-}^{\prime}(0)\right)\,,\\
W&=&\frac{M_W^2}{2}\Pi_{W^3W^3}^{\prime\prime}(0)\,,\\
X&=&\frac{M_W^2}{2}\Pi_{W^3B}^{\prime\prime}(0)\,,\\
Y&=&\frac{M_W^2}{2}\Pi_{BB}^{\prime\prime}(0)\,.
\eeqs
The parameters $\alpha_0$, $\alpha_1$ and $\alpha_8$
are hence directly related to $S$, $T$ and $U$ of~\cite{PT}~\footnote{Notice that in the definition of $S$, $T$ and  $U$ the ${\cal O}(q^4)$ corrections are neglected, and hence the identification of $\hat{S}$ and $\alpha_1$ with $S$ is valid only up to ${\cal O}(q^4)$ corrections.}, and to the $\hat{S}$, $\hat{T}$ and $\hat{U}$ in~\cite{Barbieri}:
\beqs
\hat{S}&=&-{g}^2\alpha_1\,=\,\frac{\alpha}{4\sin^2\theta_W}S\,.\\
\hat{T}&=&2{g}^2\alpha_0\,=\,\alpha T  \,,\\
\hat{U} &=&{g}^2 \alpha_8 \,=\, - \frac{\alpha}{4\sin^2\theta_W}U  \label{a-STU}\,.
\eeqs 
Bounds on the other $\alpha_i$ exist, but are much less stringent, because much less data
is available for the cubic and quartic vertexes.
The parameters $X$, $W$ and $Y$ are related to terms ${\cal O}(q^6)$ in ${\cal L}_{\chi}$.

The strategy for  extracting bounds on new physics from the experimental data requires to 
compute measurable amplitudes from the MSM at 1-loop level, and include ${\cal L}_{2,2}$ and ${\cal L}_{4,0} $
(equivalently, introducing the corrections
from $S$ and  $T$ in the propagators) at the tree level.
In this way, all UV and IR divergences cancel, and a consistent set of amplitudes computed at ${\cal O}(\hbar)$
can be compared to the data (the reader should remember that at the tree-level the MSM itself does not provide a satisfactory fit
of all the electro-weak precision observables, for which the inclusion of 1-loop corrections is necessary).
One then performs a fit to the data using as free parameters the values of the electroweak MSM parameters
$g$, $g^{\prime}$ and $v_W$, of  the masses of top $m_t$ and Higgs $m_h$,
and the precision parameters $\hat{S}$ and $\hat{T}$. A list of the observables used
can be found for instance in~\cite{Barbieri,PDG}. 

Here comes one subtlety, which is going to be important also later: in computing the MSM
1-loop contributions to the physical observables, an explicit logarithmic dependence on the 
mass of the Higgs $m_h$ is present.
Borrowing the results from~\cite{PT}:
\beqs
\hat{S}&\simeq&\frac{\alpha}{48\pi\sin^2\theta_W}\ln\frac{m_h^2}{m_h^2({\rm ref})}\,,
\eeqs
with $m_h^2({\rm ref})$ some reference value for the Higgs mass.
The result is that the MSM provides a good fit to all the existing data, for values of the 
Higgs mass $m_h \lsim 1$ TeV,
provided all the precision parameters such as $\hat{S}$ and $\hat{T}$  are at most
of order few $10^{-3}$.
One may arbitrarily decide to do the fit while setting all the oblique parameters to zero.
In doing so, one finds a  bound on the mass of the Higgs to be $m_h\lsim 200$ GeV ($95\%$ c.~l.)\cite{Wells},
because of the logarithmic dependence on $m_h$ of the 1-loop corrections.
This bound amounts to assuming that no new physics be present at any scale.

The comparison of the above procedure to a generic extension 
of the SM can be done rigorously only provided the only light degrees of freedom include all the 
SM fermions and gauge bosons, together with one Higgs scalar, while all new physics can be integrated out and reabsorbed in the 
higher-order coefficients in ${\cal L}_{\chi}$.
This means that some caution must be used when comparing to models which predict the presence of
new light degrees of freedom, for which
the 1-loop perturbative part of the calculations should be redone. 

But also highlights a difficulty
with higgsless models, in which strictly speaking  the precision parameters 
diverge, and are hence not calculable. A physical cut-off should be kept in the calculation, entering logarithmically 
in place of the logarithmic dependence on the mass of the Higgs.
In practice, phenomenological studies usually assume that Higgsless models (including TC) 
be well described by the limit  of the MSM case in which $m_h\simeq 1 $ TeV.
This is motivated mostly by the pragmatic observation that, because the dependence on the mass of the
Higgs is logarithmic, by varying the Higgs mass between $m_h=115$ GeV and $m_h=800$ GeV
(which covers most of the regime over which the Higgs can be treated perturbatively),
the bound on $\hat{S}$ changes only by $\Delta \hat{S} \simeq 10^{-3}$, which is smaller that the 
error bar on $\hat{S}$ itself. 
Imposing, for instance,  $|\hat{S}|\lsim 3\,\times 10^{-3}$ on the calculable part of the new physics contribution is hence
going to yield sensible quantitative constraints.
Notice also that the result of the fit of the data yields some ellipsoid in the space of the complete set of precision parameters,
and the resulting bounds on $\hat{T}$ and $\hat{S}$ in particular are correlated. Because of the intrinsic uncertainty
in the treatment of strongly-coupled systems, in particular due to what we just said about the role of scalars,
 in the following I will use the bounds as uncorrelated, with the understanding that these are somewhat conservative bounds:
 models that end up outside the bounds are very unlikely to be fixed by improved calculations, and are hence excluded,
 while for models that fit into the bounds, a detailed strong-coupling calculations would be needed to 
 decide if they can or cannot fit the data. 

The non-decoupling of the Higgs degrees of freedom is one of the many subtle, 
phenomenologically important, aspects of the MSM, which sources theoretical problems
in its extensions. In the case of precision electro-weak tests, this 
feature still allows for useful quantitative information to be extracted, thanks to  the mild logarithmic dependence on the Higgs mass. 
In short, generic new physics contributions are larger than the uncertainty introduced by the Higgs-mass dependence,
and hence useful (and quite restrictive) constraints can be extracted.
We will see that this is not the case
for other observables, such as elastic-scattering amplitudes of longitudinally-polarized gauge bosons,
in which the non-decoupling manifests itself with a strong, quadratic dependence on $m_h$. In Higgless theories
one is going to face quadratically divergent contributions to these scattering amplitudes, that are removed by completely uncalculable
contributions from the scalar sector, hence limiting greatly the amount of information about new physics that can be extracted.
We will be  more precise on this topic in a later subsection.

Before going any further, the reader should be reminded of the role played
by precision studies in the discovery of the top, which is a nice illustration of how powerful 
simple power-counting is, and how important precision studies are.
The electro-weak chiral Lagrangian highlights a very special feature: custodial symmetry is broken by
an operator of dimension-2, in ${\cal L}_{2,2}$. 
Very interestingly, in the SM there is one very large, explicit source of breaking of $SU(2)_R$,
namely the mass of the top is much larger that the mass of the bottom. Dimensional analysis suggests
hence that $\hat{T}\propto m_t^2$ (or, equivalently, $\delta \propto m_t$).
(We will see later that indeed this is the result of perturbative calculations with a heavy fermion.)
Which means that the precision tests of the SM are very sensitive to the precise value of the top mass.
And indeed, a clear indication that the top is heavy, in the $170$ GeV region,
 was obtained by analyzing LEP data before the actual TeVatron discovery.
 As we already saw, similar arguments yield much more limited information about the Higgs mass, 
 that enters only logarithmically in the precision tests.

\subsubsection{Perturbative estimates of $\hat{S}$ and $\hat{T}$.}

Following Peskin and Takeuchi, we can start from computing the contribution to 
the precision parameters from a new family of weakly-coupled, heavy fermions.
This perturbative calculation will be useful in showing us how the corrections are generated,
and we will later discuss what to expect in the non-perturbative case.

Consider a fermion left-handed doublet $(N,E)_L$, with hypercharge $Y$,
and right-handed $N_R$, $L_R$ such that the physical states have masses,
respectively, $m$ and $m+\delta$. Assuming that $m \gg M_Z$, 
one can compute the correction to the vacuum polarizations induced at 1-loop 
from diagrams involving loops of such fermions, with SM gauge bosons on the external lines.
Assuming also,  for simplicity,  that $\delta \ll m$, the expressions simplify to
\beqs
\hat{S}&\simeq&\frac{\alpha}{4\sin^2\theta_W}\,\frac{1}{6\pi}\,,\\
\hat{T}&\simeq&\frac{(g^2+g^{\prime\,2})}{48\pi^2}\frac{\delta^2}{M_Z^2}\,,\\
\hat{U}&\simeq&-\frac{\alpha}{4\sin^2\theta_W}\frac{2}{15\pi}\frac{\delta^2}{m^2}\,.
\eeqs
First of all, this confirms that $\hat{U}$ is doubly suppressed, as anticipated in speaking about the  electro-weak
chiral Lagrangian, and hence can be ignored in phenomenological analysis.
Second, it explicitly shows that $\hat{T}$ is proportional to the amount of custodial-symmetry breaking in the new physics
sector, and is hence negligibly small provided $\delta/M_Z\ll 1$. For these reasons, we focus on  the $\hat{S}$ parameter.
 Finally, if one has $N_D$ different 
fermions, their contributions sum,
\beqs
\hat{S} &\simeq&0.0004 N_D\,,
\eeqs
which, taking as a reference the bound $\hat{S} \lsim 0.003$, implies $N_D \lsim 8$.

If one could borrow this result and use it as an estimate of 
the TC contribution, this would imply a strong bound on the possible choice of field content  of the theory.
For example, reconsidering the $SU(N_T)$ vectorial gauge theory introduced earlier on, 
with $N_f$ families of techni-quarks,
each family including fermions on the fundamental of $SU(N_T)$ that carry the quantum numbers
of a family of SM fermions, one obtains
\beqs
\hat{S}_{\rm perturbative} &\simeq& 0.0004\,\times\,(3+1)\,N_f\,N_T\,,
\eeqs
which effectively restricts to one point of parameter space with $N_T=2$ and $N_f=1$.
And still, this single point would lie on the $3\sigma$ upper bound.

The use of this perturbative estimate is, of course, very hard to motivate: by definition technicolor is strongly coupled.
The numerical coefficients have to be taken for what they are.
The robust messages that can be extracted from this exercise are that some form of custodial symmetry better  be present, 
to suppress $\hat{T}$,
that $\hat{S}$ is potentially problematic, and that the precision parameters are expected to grow with the number of 
degrees of freedom $N_D$ coupled to the $SU(2)_L$ gauge bosons.

\subsubsection{Non-perturbative  estimates of $\hat{S}$: NDA and hidden local symmetry.}

One might want to get a feeling for how big the coefficients of ${\cal L}_{\chi}$ are
expected to be, in particular $\alpha_1$ (equivalent to $\hat{S}$), which is the most severe  test of dynamical electro-weak symmetry breaking.
A set of very simple rules for obtaining such an estimate exists, and goes under the name of Naive Dimensional Analysis (NDA).
A summary and discussion of such rules can be found in~\cite{NDA}.
Here are the rules, applied to ${\cal L}_{\chi}$, restricted by setting to zero all the custodial-symmetry 
violating terms. First of all, one has to write out all possible local operators 
compatible with the symmetries, and include them in the effective Lagrangian. An estimate of the coefficients
is then given by the following three rules:
\begin{itemize}
\item include a factor of ${f}^2\Lambda^{\ast\,2}$ overall,
\item include a factor of $1/{f}$ for any occurrence of strongly coupled fields (such as the pions $\pi$),
\item include appropriate factors of $\Lambda^{\ast}$ to get the dimension to 4\,.
\end{itemize}
Hence, the coefficient of the terms in ${\cal L}_{2,0}$ is ${f}^2$,
and for ${\cal L}_{4,0}$ all the coefficients are $\alpha_i = {\cal O} ({f}^2/\Lambda^{\ast\,2})$.
The scale $\Lambda^{\ast}$ is associated with the mass of the lightest degree of freedom that has been integrated out of the theory,
which typically is the mass of the technirho meson $M_{\rho}$. Hence, one expects
\beqs
\label{Eq:Sestimate}
\hat{S}&\sim&g^2\frac{{f}^2}{M_{\rho}^2}\,.
\eeqs
Identifying ${f}$ with the electro-weak scale, taken literally this result would mean that 
$M_{\rho} \,\gtap\, 3$ TeV. We will see that much more sophisticated techniques agree 
to an impressive degree with this NDA estimate.
It must be stresses that additional symmetry arguments provide systematic ways of including exceptions to this rules,
as is illustrated by the fact that on the basis of this counting only one would conclude that $\hat{T}={\cal O}(1)$,
while promoting $SU(2)_R$ to an approximate symmetry makes $\hat{T} = {\cal O}(\delta^2/\Lambda^{\ast\,2})\ll 1$.

The relation between the $\hat{S}$ parameter and the physics of spin-1 resonances 
in a strongly-coupled theory
can be made more precise and model-independent
with the use of dispersion relations to rewrite Eq.~(\ref{Eq:Shat}) by saturating on the 
vector resonances ($\rho$)  and axial resonances ($a_1$), which results in:
\beqs
\hat{S}&\simeq&\frac{\alpha}{4\sin^2\theta_W}\,4\pi\sum_n\left(\frac{f^2_{\rho\,n}}{M^2_{\rho\,n}}-\frac{f^2_{a_1\,n}}{M^2_{a_1\,n}}\right)\,,
\eeqs
where the index $n=1,2,\cdots$ identifies the tower of resonances with mass $M^2_{n}$ and decay constant $f^2_{n}$.
In the limit in which $SU(2)\times SU(2)$ were unbroken, axial and vector states would be exactly degenerate and connected by 
a symmetry, and hence $\hat{S}=0$. The pion decay constant measures how broken this symmetry is. 
This expression is useful because it shown that a suppression of $\hat{S}$ not necessarily implies taking large values of $M_{\rho}$,
but might be achieved by suppressing the decay constants, or by ensuring a cancellation between vectorial and axial 
contributions. 

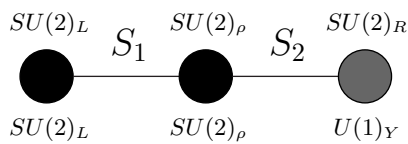
\begin{figure}[htpb]
\begin{picture}(0,100)(130,0)
\GCirc(70,26){10}{0}
\Line(80,26)(120,26)
\GCirc(130,26){10}{0}
\Line(140,26)(180,26)
\GCirc(190,26){10}{0.4}
\Text(100,38)[c]{ \Large $S_1$}
\Text(160,38)[c]{ \Large $S_2$}
\Text(70,6)[c]{ $SU(2)_L$}
\Text(130,6)[c]{ $SU(2)_{\rho}$}
\Text(70,46)[c]{ $SU(2)_L$}
\Text(130,46)[c]{ $SU(2)_{\rho}$}
\Text(190,46)[c]{ $SU(2)_R$}
\Text(190,6)[c]{ $U(1)_Y$}
\end{picture}
\caption{Diagram of the 3-site model.}
\label{Fig:3sites}
\end{figure}

One way of drawing a closer relation between $\hat{S}$ and the 
heavy composite states, in particular the lightest spin-1 meson, the
techni-rho, is provided by the inclusion
of the latter  in the low energy effective theory. This is done with the help of 
so called hidden local symmetry (HLS)~\cite{HLS}.
The basic trick is to extend the global symmetry of the chiral Lagrangian (truncated to the leading order),
by introducing an additional, spontaneously-broken gauged $SU(2)$ as in Fig.~\ref{Fig:3sites}. The $SU(2)_{\rho}$ in the middle site
is gauged with strength $g_{\rho}$, and there are now two sigma-model fields, so that in unitary gauge the new gauge bosons are massive.

The important terms in the Lagrangian are, at the leading-order,
\beqs
{\cal L}_{3s}&=&-\,\frac{1}{2}\Tr {W}_{\mu\nu}{W}^{\mu\nu}
\,-\,\frac{1}{2}\Tr {B}_{\mu\nu}{B}^{\mu\nu}
\,-\,\frac{1}{2}\Tr {\rho}_{\mu\nu}{\rho}^{\mu\nu}\,+\,
 \frac{{f}^2}{4}\Tr |DS_1|^2+ \frac{{f}^2}{4}\Tr |DS_2|^2\,,
\eeqs
with 
\beqs
S_{1,2}&=&e^{\frac{2i\pi_{1,2}}{f}}\,,
\eeqs
and the covariant derivatives
\beqs
DS_1 &=& \partial S_1\,+\,i\left(g{W} S_1 - g_{\rho}S_1{\rho}\right)\,,\\
DS_2 &=& \partial S_2\,+\,i\left(g_{\rho}{\rho} S_2 - g^{\prime}S_2{B}\right)\,.
\eeqs
In unitary gauge, $S_1=1=S_2$, and the mass of the neutral gauge bosons is,
in the $(W^3,\rho^3,B)$ basis:
\beqs
{\cal M}^2&=&
\frac{1}{4}f^2
\left(\begin{array}{ccc}
g^2 & -g g_{\rho} & 0\cr
-g g_{\rho} &2g_{\rho}^2 & - g^{\prime} g_{\rho}\cr
0 &  - g^{\prime} g_{\rho} &  g^{\prime\,2} 
\end{array}\right)\,.
\eeqs
Assuming that $g,g^{\prime} \ll g_{\rho}$, one can integrate out the heavy techni-rho, of mass
\beqs
M_{\rho}^2 = \frac{1}{8}(g^2+g^{\prime\,2}+4g_{\rho}^2)f^2 &\gg& M_Z^2 = \frac{1}{8}(g^2+g^{\prime\,2})f^2\,,
\eeqs
and finally obtain 
\beqs
\hat{S}&=&\frac{1}{4}\frac{g^2}{g_{\rho}^2}\,=\,\frac{1}{4}g^2\frac{f_{\pi}^2}{M_{\rho}^2}\,=\,\frac{M_W^2}{M_{\rho}^2}
\eeqs
where $f_{\pi}^2=f^2/2=v_W^2$. This estimate implies that the bound on $\hat{S}$ requires $M_{\rho}\gsim 1.5$ TeV,
which is a factor of $2$ milder that what derived from the NDA estimate Eq.~(\ref{Eq:Sestimate}), but consistent with it.

Some lesson and some comment on this little exercise.
One might regard the agreement between the equations we derived as a success.
They actually indicate a surprising degree of robustness of the NDA estimates.
However, one has to use some caution, for a number of reasons. First of all, the bound we just derived
implies that $g_{\rho}\gsim 6.5$, which means that the very idea of studying perturbatively the 
spectrum, couplings and correlators from ${\cal L}_{3s}$ does not work. 
In practice, this Lagrangian would be useful only in the regime in which 
$g_{\rho}$ be perturbative, in which the bounds on $\hat{S}$ are certainly exceeded.
Even worse: in writing ${\cal L}_{3s}$ we did not apply the basic rule of EFT, requiring to write,
order-by-order, all the possible terms allowed by symmetries, because we omitted
\beqs
\label{Eq:kappa}
 {\cal L}_{\kappa}&=&\kappa \frac{{f}^2}{4}\Tr |D(S_1S_2)|^2\,.
\eeqs 
Aside from the fact that this term is not  nearest-neighbour, it is
a perfectly legitimate part of the leading-order Lagrangian.
Its presence has two main effects, due to the fact that it changes both the kinetic terms of the pions and
the mass terms of the gauge bosons.
First of all, it splits the pion and techni-rho decay constants
\beqs
f_{\rho}^2=\frac{f^2}{2}&\neq&f_{\pi}^2=(1+2\kappa)\frac{f^2}{2}\,,
\eeqs 
notice that positivity of the kinetic terms requires $\kappa>-1/2$.
Furthermore, because the spectrum is given by $M_Z^2\simeq(1+2k)(g^2+g^{\prime\,2})f^2/8=(g^2+g^{\prime\,2})f_{\pi}^2/4$ and 
$M_{\rho}^2\simeq g_{\rho}^2f^2/2$,
it modifies the $\hat{S}$ parameter
\beqs
\hat{S}&=&\frac{1}{4}\frac{g^2}{g_{\rho}^2}\,=\,\frac{1}{8}g^2\frac{f^2}{M_{\rho}^2}\,=\,\frac{M_W^2}{M_{\rho}^2}\frac{1}{1+2\kappa}\,.
\label{Eq:newShat}
\eeqs
Finally, it changes the cubic coupling between $\rho$ and pions:
\beqs
g_{\rho\pi\pi}&=&\frac{g_{\rho}}{2+4k}\,.
\eeqs
Notice that, in real-world QCD, which can  be described with the same formalism up to minor modifications,
 $f_{\rho}\simeq 150$ MeV and $f_{\pi}\simeq 93$ MeV,
indicating that even in order  to describe QCD  $\kappa$ must be included and is ${\cal O}(1)$, not small.
In this specific case, the fact that $f_{\pi}$ is smaller that $f_{\rho}$ implies negative values of $\kappa$, and hence an enhancement of $\hat{S}$
in comparison with what we said earlier on. This is what actually happens if one 
estimates the oblique corrections by scaling up the QCD experimental results~\cite{PT}.

Before going any further, another important comment about these non-nearest-neighbour couplings.
By looking at Eq.~(\ref{Eq:newShat}), one might be tempted to conclude that 
this provides a perfectly natural way out of the little hierarchy problem, just by dialing $\kappa$ to a somewhat largish,
positive coefficient. 
This train of thought is quite dangerous, and yields to big problems.
The reason is that once non-nearest-neighbour interactions are allowed, and particularly 
if $\kappa$ is large, nothing prevents form writing much more dangerous terms, one of which
is $\Tr S_1S_2 B_{\mu\nu}S_2^{\dagger}S_1^{\dagger}W^{\mu\nu}$, which one recognizes to be 
a direct contribution to $\hat{S}$. This is a subleading term to be added to ${\cal L}_{3s}$, and as such one would
naively think of it as suppressed, but if $\kappa$ is taken large, then also the coefficient of this new term 
should be large, and this would become the dominant (uncalculable) contribution to $\hat{S}$.
 In other words, playing the game of dialing up the coefficients of
interactions such as Eq.~(\ref{Eq:kappa}) yields in the end a useless theory, in which $\hat{S}$
 is a free parameter.

In general, terms such as $ {\cal L}_{\kappa}$ are generated when integrating out other heavier states from the theory.
The one term in Eq.~(\ref{Eq:kappa}) is just one example. These terms and their proliferation with the inclusion of more sites
and links represent a very heavy limitation in the predictive power of the hidden local symmetry approach.
Also, one must question how appropriate it really is to include in the effective theory the rho mesons, but not other states, such as
their axial counterparts, the $a_1$.
One can use the  technique exemplified above to include many spin-1 resonances, along the lines of~\cite{HLS}.
The resulting model, in the limit of infinite number of sites,  is the deconstructed 
version of a fifth dimension, provided no non-local terms are included. 

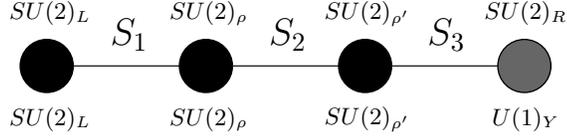
\begin{figure}[htpb]
\begin{picture}(0,100)(150,0)
\GCirc(70,26){10}{0}
\Line(80,26)(120,26)
\GCirc(130,26){10}{0}
\Line(140,26)(180,26)
\GCirc(190,26){10}{0}
\Line(200,26)(240,26)
\GCirc(250,26){10}{0.4}
\Text(100,38)[c]{ \Large $S_1$}
\Text(160,38)[c]{ \Large $S_2$}
\Text(220,38)[c]{ \Large $S_3$}
\Text(70,6)[c]{ $SU(2)_L$}
\Text(130,6)[c]{ $SU(2)_{\rho}$}
\Text(190,6)[c]{ $SU(2)_{\rho^{\prime}}$}
\Text(70,46)[c]{ $SU(2)_L$}
\Text(130,46)[c]{ $SU(2)_{\rho}$}
\Text(190,46)[c]{ $SU(2)_{\rho^{\prime}}$}
\Text(250,46)[c]{ $SU(2)_R$}
\Text(250,6)[c]{ $U(1)_Y$}
\end{picture}
\caption{Diagram of the 4-sites model.}
\label{Fig:4sites}
\end{figure}

Another simple exercise will help understanding what is the physical meaning of the $\hat{S}$ parameter.
In this case, we extend the 3-sites model to a 4-sites model as in Fig.~\ref{Fig:4sites}, hence incorporating in the EFT both the
vectorial (techni-rho) meson and its axial partner (techni-$a_1$).
The important terms in the Lagrangian are
\beqs
{\cal L}_{4s}&=&-\,\frac{1}{2}\Tr {W}_{\mu\nu}{W}^{\mu\nu}
\,-\,\frac{1}{2}\Tr {B}_{\mu\nu}{B}^{\mu\nu}
\,-\,\frac{1}{2}\Tr {\rho}_{\mu\nu}{\rho}^{\mu\nu}\,
\,-\,\frac{1}{2}\Tr {\rho}_{\mu\nu}{\rho}^{\mu\nu}\nonumber\\
&&+\,
 \frac{{F}^2}{4}\Tr |DS_1|^2+ \frac{{f}^2}{4}\Tr |DS_2|^2+ \frac{{F}^2}{4}\Tr |DS_3|^2\,,
\eeqs
with 
\beqs
S_{1,3}&=&e^{\frac{2i\pi_{1,3}}{F}}\,,\\
S_{2}&=&e^{\frac{2i\pi_{2}}{f}}
\eeqs
and the covariant derivatives
\beqs
DS_1 &=& \partial S_1\,+\,i\left(g{W} S_1 - g_{\rho}S_1{\rho}\right)\,,\\
DS_2 &=& \partial S_2\,+\,i\left(g_{\rho}{\rho} S_2 - g_{\rho}S_2{\rho^{\prime}}\right)\,,\\
DS_3 &=& \partial S_3\,+\,i\left(g_{\rho}{\rho^{\prime}} S_3 - g^{\prime}S_3{B}\right)\,.
\eeqs
In unitary gauge, $S_1=S_2=S_3=1$, and hence all the gauge bosons (aside from the photon) acquire a mass,
via the mass matrix:
\beqs
{\cal M}^2&=&
\frac{1}{4}
\left(\begin{array}{cccc}
g^2 F^2& -g g_{\rho} F^2 & 0&0\cr
-g g_{\rho} F^2 & g_{\rho}^2 (f^2+F^2)& - g_{\rho}^2 f^2&0\cr
0&- g_{\rho}^2 f^2 & g_{\rho}^2 (f^2+F^2)& - g^{\prime} g_{\rho} F^2\cr
0& 0 &  - g^{\prime} g_{\rho}F^2 &  g^{\prime\,2} F^2
\end{array}\right)\,.
\eeqs

In the limit where $f\ll F$, one sees that there is an approximate $SU(2)\times SU(2)$ symmetry.
The spectrum has $M_{\rho}^2\simeq M _{a_1}^2\simeq g_{\rho}^2F^2/4$, $M_Z^2=(g^2+g^{\prime\,2})f^2/4$,
and  $\hat{S}\simeq 2 g^2f^2/(g_{\rho}^2F^2)$.
This explicitly shows  the fact that $\hat{S}$  measures the effect of isospin breaking in the resonances. In particular,
one can suppress $\hat{S}$ by ensuring that the vectorial and axial-vector resonances are quasi-degenerate.
However, notice that for this to work one needs a parametric suppression $f_{\pi} \ll f_{\rho}$, 
which is not what expected in a QCD-like theory.
Some non-trivial dynamics, in which a scale separation exists suppressing $f_{\pi}$ is needed.
In particular, notice that models in which this mechanism is at work cannot be described
at low energy by a 3-site model, because there is no gap allowing to integrate out 
the $a_1$ while keeping the $\rho$ meson. Which is one way of understanding why 
the small-$\hat{S}$ regime is not accessible within the 3-site model, which becomes uncalculable and fine-tuned
in this regime.

Concluding: both perturbative and non-perturbative estimates of the $\hat{S}$ parameter indicate that 
there is a tension between the expectations in a generic model of technicolor and the experimental data.
However, this does not mean that TC is excluded by precision tests.
Fist of all the tension is not dramatic: some models are at the boundary of acceptability.
Also, the NDA estimates are just valid up to unknown numerical factors that are very difficult to compute, even in QCD-like theories.
In walking technicolor, as we will see in next subsection, NDA expectations are expected to be 
violated. Large anomalous dimensions can distort the spectrum of masses and decay constants~\cite{ApS},
and ultimately a more reliable technique is needed in order to perform actual calculations. 
This might be provided by the lattice or, as suggested here, by the 
ideas of gauge/string duality.

\subsection{Fermion masses and FCNC constraints.}

Almost by definition, a technicolor model is an incomplete model of electro-weak symmetry breaking in the SM. 
The mechanism illustrated  in the
previous sections provides a mass for the electro-weak gauge bosons, but not for the SM fermions, which
are exactly massless. This because one cannot write the Yukawa couplings, in the absence of an elementary scalar.
A simple EFT way of solving this problem is to add to the Lagrangian four-fermion interactions coupling
the SM quarks  and leptons $\psi_{SM}$ to the techni-quarks $\psi_{TC}$. When techni-quarks condense, dimensional transmutation
allows to match the theory onto a low-energy EFT in which mass terms are generated for the SM fermions:
\beqs
(\bar{\psi}_{SM}\gamma_{\mu} \psi_{TC})(\bar{\psi}_{TC}\gamma^{\mu} \psi_{SM})
&\rightarrow &\left\langle \bar{\psi}_{TC}\psi_{TC}\right\rangle \bar{\psi}_{SM}\psi_{SM}\,.
\eeqs

There are three, interrelated problems with this idea. First of all, because 
these are higher-order operators, adding them by hand to the fundamental Lagrangian 
would spoil the good UV behavior of the theory that is the original motivation
for technicolor itself. It is hence necessary to extend somehow the new physics sector in such a way
as to dynamically generate these four-fermion interactions at some intermediate scale $\Lambda_{ETC} >  v_W$. 
This is the role of Extended Technicolor (ETC).
The second problem is that in generating the operators needed in order to yield the masses and mixing among 
SM fermions, the ETC interactions are expected in general to produce also dangerous higher-order terms,
coupling for instance four SM fermions, which can easily produce large contributions to FCNC processes,
in excess of what allowed by data on flavor physics.
Avoiding this requires that $\Lambda_{ETC}\gg v_W$, but also a challenging model-building 
exercise, in which the hierarchy of masses is generated dynamically, and some mechanism resembling the GIM mechanism be 
at work. This is the role played by tumbling ETC.
Finally, some of the SM fermions are quite heavy (the top for instance has $m_t\sim v_W$), 
but NDA suggests that the masses obtained from dimension-6 operators should be 
${\cal O}( v_W^3/\Lambda_{ETC}^2) \ll v_W$. Walking dynamics might provide a solution to this problem, by enhancing 
the fermion masses but not the dangerous operators involved in FCNC processes (and in the custodial-symmetry breaking
effects such as $\hat{T}$), as we are going to see.

In this subsection, I start  from recalling the basic mechanisms at work in the SM
and its perturbative extension of relevance for flavor  physics (the GIM and FN mechanisms), 
 then move to strongly-coupled models, and summarize
the concepts of tumbling ETC and the role of walking dynamics. 
Before we embark in this discussion, 
one should be aware of the fact that, to large extent because of the uncertainties intrinsic with strongly-coupled dynamics,
 at present there is no satisfactory, UV-complete model 
encompassing all of the above. It would be interesting to  build such an explicit
UV-complete model of ETC in the context of gauge/string dualities.
It must also be mentioned that some very interesting results have been
obtained with five-dimensional Higgsless theories that
are more closely related to topcolor~\cite{reviews}, rather than to ETC,
in the sense that  the SM fermions propagate in the bulk, and are hence composite objects themselves~\cite{5Dfermions}.

\subsubsection{The GIM mechanism.}

In the MSM, in the simplifying limit in which no right-handed neutrinos are present, and
in which neutrinos are exactly massless, a very rich structure of global family symmetries is present.
At the classical level, in the limit in which all the Yukawa couplings in Eq.~(\ref{Eq:Yukawa}) vanish,
the model has a $U(3)^5$ global symmetry, with each $U(3)$ acting on the three families of matter for each 
fermion species.  The Yukawa couplings break explicitly this symmetry, ultimately generating the 
hierarchy of masses of the SM.
The gauge couplings of  $SU(2)_L\times U(1)_Y$ are universal. The phenomenological distinction between families arises
only because of the masses induced by the Yukawa couplings, and flavor-changing interactions 
are controlled by the mixing matrices that diagonalize the mass matrices.
For concreteness, restrict attention to the quarks only.
The important terms of the Lagrangian after electro-weak symmetry breaking read
\beqs
{\cal L}&=&i\bar{u^{\prime}}_a\,\dslash u^{\prime}_a\,+
i\bar{d^{\prime}}_a\,\dslash d^{\prime}_a\,
-\frac{g}{\sqrt{2}}\left(\bar{u}^{\prime}_L\right)_a\gamma^{\mu}W_{\mu}^{+}
\left(d^{\prime}_L\right)_a\,+\,{\rm h.c.}\nonumber\\
&&-\,M_{ab}\left(\bar{u}^{\prime}_R\right)_{a}\left(u^{\prime}_L\right)_{b}
\,-\,m_{ab}\left(\bar{d}^{\prime}_R\right)_{a}\left(d^{\prime}_L\right)_{b}\,+\,{\rm h.c.}\,,
\eeqs
where $u^{\prime}_a$ and $d^{\prime}_a$ are the up-type and down-type Dirac fermions
describing the quarks in what is called the interaction basis, with family index $a=1,2,3$.

The mass matrices can be diagonalized by the bi-unitary transformations defined as
\beqs
M_d&=&V_R^{\dagger} M V_L\,,\\
m_d&=&U_R^{\dagger} m U_L\,,
\eeqs
where all the rotation matrices are $3\times 3$ unitary matrices 
$
U^{\dagger}_L U_L =1=U^{\dagger}_R U_R=
V^{\dagger}_L V_L=V^{\dagger}_R V_R\,.
$
These transformations define the { mass basis} for the quarks:
\beqs
u^{\prime}_{L,R}&=&V_{L,R}u_{L,R}\,,\\
d^{\prime}_{L,R}&=&U_{L,R}d_{L,R}\,.
\eeqs
In this new basis the Lagrangian density can be rewritten as
\beqs
{\cal L}&=&i\bar{u^{}}_a\,\dslash u^{}_a\,+
i\bar{d^{}}_a\,\dslash d^{}_a\,
-\frac{g}{\sqrt{2}}U_{ab}\left(\bar{u}^{}_L\right)_a\gamma^{\mu}W_{\mu}^{+}
\left(d^{}_L\right)_b\,+\,{\rm h.c.}\nonumber\\
&&-\,(M_d)_{aa}\left(\bar{u}^{}_L\right)_{a}\left(u^{}_R\right)_{a}
\,-\,(m_d)_{aa}\left(\bar{d}^{}_L\right)_{a}\left(d^{}_R\right)_{a}\,+\,{\rm h.c.}\,,
\eeqs
where the matrix $U\equiv V_L^{\dagger}U_L$ is the CKM mixing matrix,
which in general is not diagonal.
All flavor-changing processes involve off-diagonal entries of this matrix, which are known 
to be small~\cite{PDG}. The neutral interactions are diagonal also in the mass basis.

All of this is at the basis of the GIM mechanism for suppressing FCNC transitions.
There are three suppression factors.
 The neutral-current couplings being diagonal, all flavor changing neutral current processes
are induced by weak interactions at the loop level, and are hence suppressed by loop factors in
the weak couplings.  They all involve combinations of the small mixing angles in the CKM matrix,
which can yield significant suppression. Because the origin of the 
mixing is the  Yukawa couplings breaking the family symmetry, additional suppression
comes from factors involving the masses of fermions in the internal lines of loop diagrams.

An example will help illustrating the point.
Consider the semi-leptonic decay of the $K^{+}$ meson, and try to estimate
the ratio
\beqs
x&=&\frac{
\Gamma\left[K^{+}\rightarrow\pi^{+}\,\nu\,\bar{\nu}\right]}
{\Gamma\left[K^{+}\rightarrow\pi^{0}\,e^{+}\,\nu_e\right]}
\,.
\eeqs
Naively, the two final states might look very similar. However, a look at the PDG~\cite{PDG}
reveals that $x \sim  {\cal  O}(10^{-9})$.
A very efficient way to understand this is by integrating out the $W$ boson
and hence building an effective theory describing the interaction such as to include the two
effective operators
\beqs
\label{Eq:4f}
{\cal L} &=& {c_1}\left[\bar{s}_L^{\alpha}\gamma^{\mu} u^{\alpha}_L\right]
\left[\bar{\nu_e}_L\gamma_{\mu} e_L\right]
\,+\,
{c_2}\left[\bar{s}_L^{\alpha}\gamma^{\mu} d^{\alpha}_L\right]
\left[\bar{\nu}_L\gamma_{\mu} \nu_L\right]\,+\,{\rm h.c.}\,,
\eeqs
where $c_1$ and $c_2$ are dimensionful.
It is clear that the kinematics and the matrix elements  of the full calculation are 
very similar for the two processes, and hence $x\simeq 6 |c_2/c_1|^2$,
where a factor of $3$ comes from the three possible neutrino species in the final state,
and a factor of 2 from the matrix elements.
The explanation for the smallness of such ratio must come from the 
weak interactions.

A closer inspection, in the light of the SM, shows that $c_1$ is generated by the tree-level exchange
of a $W$ boson. Hence
\beqs
c_1&\sim&\frac{4G_F}{\sqrt{2}} U_{us}^{\ast}\,,
\eeqs
with $U_{us}  \simeq 0.22$ the Cabibbo angle.
Conversely, $c_2$ is generated by loop diagrams (boxes and penguins) and 
is given by
\beqs
c_2&\sim & \frac{4G_F}{\sqrt{2}} \frac{\alpha}{2\pi\sin^2\theta_W} \sum_i U_{is}^{\ast}U_{id}X\left(\frac{m_i^2}{M_W^2}\right)\,,
\eeqs
where the function $X$ depends on the ratios of the up-type quarks masses in the loop and $M_W$.
 In doing so, contributions that do not depend on the mass of the internal fermions cancel exactly,
 because of  the unitarity of the CKM matrix
 \beqs
 \sum_iU_{is}^{\ast}U_{id}&=&0\,.
 \eeqs
 The non-vanishing terms are weighted by functions 
 of the ratio $m_i^2/M_W^2$ involving the up-type quark masses $m_i$. 
The detailed structure of these functions is calculable and well-known, and can be expanded in powers
and logarithms, such that in the case of degenerate fermion masses the result vanishes identically.
 In the case of this special decay, top and charm contributions are comparable,
 because a suppression of the top contribution due to the CKM angles is compensates by
 the ratio $m_t^2/m_c^2$.
 Hence, one can very roughly estimate $x\sim 10^{-9} - 10^{-8}$, depending on the
 precise value of the function $X$ evaluated on the charm and top contribution, and to short-distance QCD
 effects. This estimate agrees with the data, at this level of accuracy. The function $X$ is known, and 
 this statement can be made more precise.
 
 The important thing is that the huge suppression factor has three origins, ultimately all traceable to 
the GIM mechanism: a (weak-coupling) loop suppression factor $\alpha/\pi$, small CKM elements,
and ratios of masses of up-type quarks. The charm contribution is not severely Cabibbo-suppressed,
but the small charm mass provides another suppression factor, while conversely the top contribution is suppressed by
the CKM angles. All of this comes from the fact that ultimately one might rewrite the physical amplitude in terms
of the up-type Yukawa $(y^{(u)}y^{(u)\dagger})_{12}$ element, which breaks explicitly the aforementioned $U(3)^5$ symmetry.

\subsubsection{Froggat-Nielsen and see-saw mechanisms.}

The fact that in the absence of Yukawa couplings the MSM Lagrangian has a very large 
global family symmetry is at the basis of  the GIM mechanism for suppressing FCNC,
and it is also at the basis of most attempts to  explain the 
hierarchy of masses and smallness of mixing angles, within models in which
the Yukawa couplings are generated dynamically.
The basic idea behind this is the generalization of the Froggat-Nielsen (FN) mechanism~\cite{FN}.
A vast literature exists on the subject, all of which is based on the same principles.
One starts by taking a subgroup of $G_f\subset U(3)^5$, chosen in such a way that 
if $G_f$ were  unbroken then all the small mixing angles and fermion mass ratios would vanish,
because all the Yukawa couplings  would be forbidden. 
The symmetry is broken by the VEVs $\langle \phi_i \rangle \gg v_W$ of some scalar (flavon) field that transforms under $G_f$,
but is a singlet of the SM gauge group. The breaking is communicated to the SM
via the exchange of weakly-coupled, messenger fields with mass $\Lambda_f\gg v_W$.
In doing so, by integrating out the flavons and the messengers one generates the Yukawa couplings of the SM,
and depending on the assignment of SM fermions to representations of $G_f$
the resulting coupling is suppressed by powers of the small ratios
$\epsilon_i\sim \langle \phi_i \rangle/\Lambda_f$, which are typically in the range ${\cal O}(\frac{1}{5}) - {\cal O}(\frac{1}{20})$. 
Appropriate choices of $G_f$, of the representations assigned to the fermions, and of the symmetry-breaking pattern
allow to produce realistic mass matrices. A number of variations on the theme exist in the literature,
in which $G_f$ can be global or local, abelian or non-abelian, continuous or discrete, and combinations thereof,
within or outside grand unified and/or supersymmetric extensions of the MSM.

For all of this to make sense, the most important ingredient is the fact that 
everything must be weakly coupled, and a Higgs scalar be present.
If so, one can integrate out all the flavor physics involved in the FN mechanism below the very large scale 
$\Lambda_f \gg v_W$ of the messengers and flavons.
As a result, at low energy one generates a series of operators to be added to the SM, 
in which the only marginal operators are the Yukawa couplings:
\begin{itemize}
\item the dimension-4 top Yukawa is often assumed to be unaffected by the FN mechanism, and hence unsuppressed,
\item all other dimension-4 Yukawa couplings are suppressed by powers of $\epsilon_i$, leading to the small mixing angles and small mass ratios,
\item dangerous operators inducing new contributions to FCNC emerge at dimension-5, dimension-6 and higher, and are hence suppressed
by powers of $v_W/\Lambda_f$,
\item assuming lepton number be broken in the process, and that no (light) right-handed neutrino singlet be present,
a set of dimension-5 operators generates the mass of the SM neutrinos
via the see-saw mechanism~\cite{seesaw}.
\end{itemize}

More on the last point.
The operators that would produce the mass for the neutrinos read, schematically:
\beqs
{\cal L}_{\nu}&\sim&\frac{1}{\Lambda_f} \left(L H \right) \left(L H\right)\,.
\eeqs
When the Higgs develops a VEV, one obtains the mass
\beqs
m_{\nu}&\sim & \frac{v_W^2}{\Lambda_f}\,.
\eeqs
Taking large values of $\Lambda_f$, this results in a suppression of the neutrino masses,
that goes under the name of see-saw mechanism.
If $m_{\nu}\sim 10^{-2} - 10^{-1}$ eV, one has to require $\Lambda_f \sim 10^{14} - 10^{16}$ GeV,
which is in the range of GUT theories. With such a large scale, all contributions to FCNC are
safely negligible.

\subsubsection{Extended Technicolor, tumbling and walking.}

In a generic extension of the SM, such as ETC, in which all possible higher-order operators generalizing 
those in Eq.~(\ref{Eq:4f})  are produced at some scale $\Lambda_{ETC}$, very strong bounds 
are inferred from data on rare flavor-changing processes. 
For instance, requiring that the ETC contribution to $c_2$ be such as not to spoil the SM
agreement with the data implies that $\Lambda_{ETC}\sim 1/\sqrt{c_2}  > 50$ TeV.
Much more stringent bounds can be obtained analyzing the effect of such operators as
\beqs
\label{Eq:DS2}
\left[\bar{s}_L^{\alpha}\gamma^{\mu} d^{\alpha}_L\right]
\left[\bar{s}_L^{\alpha}\gamma_{\mu} d^{\alpha}_L\right]\,,
\eeqs
and similar operators which affect mixing and CP violation in the $K^0-\bar{K^0}$ system,
from which one learns that $\Lambda_{ETC}\gsim 1000$ TeV.

As anticipated, in Technicolor there is no simple way to write a set of marginally irrelevant  couplings
analogous to the Yukawa's of the MSM. The masses for the SM fermions themselves must arise from 
higher-order operators coupling the SM fermions and the techni-quarks.
This means that the problems of generating the masses, their hierarchies, the mixing angles,
and CP-violation are tangled together with the problem of doing so by suppressing 
FCNC processes.
In other words, the power-counting that made the Froggat-Nielsen and see-saw mechanisms work
in perturbative theories is lost, because there is no simple  way to take the limit in which the parameter
$\Lambda_f$ (or its strongly-coupled equivalent $\Lambda_{ETC}$) to arbitrary large values, while retaining finite fermion
masses and CKM mixing angles.

Let us try to explain all of this in more details.
 Naively, one expects that once some generic mechanism is implement,
it will generate operators of the three generic forms
\beqs
\label{Eq:ETC}
{\cal L}_{ETC}&=&c_{ST}(\bar{\psi}_{SM}\gamma_{\mu} \psi_{TC})(\bar{\psi}_{TC}\gamma^{\mu} \psi_{SM})
+c_{TT}(\bar{\psi}_{TC}\gamma_{\mu} \psi_{TC})(\bar{\psi}_{TC}\gamma^{\mu} \psi_{TC})
+c_{SS}(\bar{\psi}_{SM}\gamma_{\mu} \psi_{SM})(\bar{\psi}_{SM}\gamma^{\mu} \psi_{SM})\,,
\eeqs
possibly with all different admissible  Lorentz and flavor structures.
NDA suggests that in the absence of any other dynamical or symmetry argument
$c_{SS}\sim c_{TT} \sim c_{ST}$.
For the mass of the strange quark, for instance, one finds that
\beqs
m_s&\sim&c_{ST}\left\langle \bar{\psi}_{TC}\psi_{TC}\right\rangle\,\sim\,c_{ST}4\pi v_W^3\,,
\eeqs
so that  the coefficient of the operator in Eq.~(\ref{Eq:DS2}) is expected to be
\beqs
c_{SS}&\sim&c_{ST}\,\sim\,\frac{m_s}{4\pi v_W^3}\,\sim\, (50\,{\rm TeV})^{-2}\,,
\eeqs
and the contribution to $K^0-\bar{K^0}$  mixing results to be somewhere between four and five orders of magnitude too big to be acceptable.
This kind of disaster is what people have in mind  when they say that TC has a problem with FCNC.

Another related problem emerges when considering the generation of mass for the top.
Borrowing the same estimates, including and taking literally the uncertain factor of $4\pi$,
the mass of the top would be obtained with $\Lambda_{ETC} \sim 1/\sqrt{c_{ST}}\sim 1$ TeV.
This  optimistic estimate requires that some strongly-coupled sector responsible for generating
the mass splitting between the bottom and top quarks be present at a very low scale.
Such sector violates custodial symmetry, and hence would generate a contribution to the 
$\hat{T}$ parameter 
$\hat{T}\,\sim\,c_{ST}v_W^2\,\sim\,0.05$,
which is largely in excess of the experimental bounds.
A more sensible estimate requires the $1/\sqrt{c_{ST}}\sim\Lambda_{ETC}\sim 5$ TeV,
in which case $\hat{T}\,\sim\,0.002$, but then the estimate for the top mass
would yield $m_t\,\sim\, 7$ GeV, which is certainly impossible to reconcile with the data.

All of the above highlights  that there is a problem. 
Clearly, a generic model which is reasonably well approximated by NDA counting arguments is 
not compatible with the data.
However, the estimates we did 
rely on the assumption that operators of the same kind,
but involving completely different fields, have coefficients of the same 
value because in applying the NDA rules we assumed that all the operators
in Eq.~(\ref{Eq:ETC}) obey the same counting.
This needs not be true, provided the underlying dynamics yields a modification
of the counting rules. This is based on symmetry considerations within tumbling ETC,
and on dynamical considerations within walking technicolor.
Effectively, tumbling ETC plays the role of the VEVs and messengers in the
FN mechanism in generating different coefficients
for operators of the same kind, but involving different SM fermions.
Walking dynamics changes the rules of dimensional analysis, by changing the power-counting 
of operators that involve techni-quarks (due to the non-perturbative anomalous dimensions), 
in such a way as to make the operators generating the masses
more relevant (lower-dimensional) than the operators producing dangerous FCNC.
The additional suppression factor accounting for the neutrino masses
may arise from the fact that one has to break the lepton number, which requires some specific 
step in the ETC model~\cite{AS}, and ends up making the neutrino masses emerge 
from very high-dimensional operators.

Let us be more precise about these ideas.
In tumbling ETC, one uses a variation of the  idea
that is at the basis of the FN mechanism, namely that a subgroup
$G_f\subset U(3)^5$ be a gauge symmetry of the underlying dynamics, broken sequentially
at several different ETC scales by the dynamical formation of condensates.
Technically, one assumes that there is a ETC gauge theory at some very high scale,
with $G_f\times G_{TC} \subset G_{ETC}$, such that SM fermions and techni-quarks 
transform together in some set of representations  of $G_{ETC}$.
The strongly coupled dynamics then breaks sequentially $G_{ETC}\rightarrow G_1\rightarrow\cdots \rightarrow G_{TC}$,
at scales $\Lambda_1$, $\Lambda_2$, \dots,
in such a way that at lower energy the fermionic field content consists of the techni-quarks and the SM singlets,
with $G_f$ completely broken. The gauge bosons in the coset $G_{ETC}/(G_f\times G_{TC} )$ play the role of the 
messengers in the FN mechanism, and as a result integrating them out yields the four-fermion interactions
of the low-energy theory.
In this way, one can generate the hierarchy of masses in the SM by exploiting the fact that
the $c_{ST}$ coefficients entering the estimate of the mass of different fermions
have a different dynamical origin from different scales $\Lambda_i$, depending on the SM field content
of the operator in consideration.
For example, one can embed the by now familiar example of TC model with $SU(N_T)$ gauge symmetry
into a $SU(N_T+3)$ model, the fundamental representation of which contains the techniquarks and the SM fermions.
At scales $\Lambda_1\sim 1000$ TeV, $\Lambda_2\sim 50$ TeV, $\Lambda_3\sim 5$ TeV~\cite{APS}, 
the breaking takes place as
\beqs
SU(N_T+3)\rightarrow
SU(N_T+2)\rightarrow
SU(N_T+1)\rightarrow
SU(N_T)\,,
\eeqs
so that the $(N_T+3)$-dimensional fundamental representation 
decomposes in one fundamental representation of $SU(N_T)$  and three singlets, identified with the SM fields.

In the estimate of $m_d \ll m_s \ll m_b$ one uses the fact  that the important four-fermion operators be generated 
at scales $\Lambda_1 \gg \Lambda_2 \gg \Lambda_3$ respectively, with larger scales suppressing the mass of the
lighter fermions.
In doing so, one must also explain and generate the small CKM mixing angles, 
in terms of combinations of ratios of such scales $\Lambda_i/\Lambda_j$. 
But then operators such as  Eq.~(\ref{Eq:DS2}),
because they violate the strange number by two units instead of one, 
pick up a suppression factor from the highest scale involved
in generating them, times a  further suppression 
in the form of powers of the ratio entering the mass generation, 
effectively implementing a mild form of  GIM mechanism.
Examples of this construction have been implemented in many ways, and semi-realistic models
exist~\cite{APS}.

A second crucial ingredient in a realistic model is  the fact that the technicolor theory 
 that emerges at the end of the tumbling is assumed to be walking.
This further modifies the counting rules, because strong dynamics is assumed to produce 
anomalous dimensions for operators involving techni-quarks, in particular enhancing the  $c_{ST}$ coefficients.
Assuming the chiral condensate to have anomalous dimension $\gamma \sim {\cal O}(1)$,
and the walking taking place between the electro-weak scale and a scale $\Lambda^{\ast}$
(typically, related to the lowest scale produced by tumbling), the estimate for the mass of the top becomes
\beqs
m_t&\sim&\left(\frac{\Lambda^{\ast}}{v_W}\right)^{\gamma}c_{ST}4\pi v_W^3\,,
\eeqs
which for $\gamma=1$, $\Lambda^{\ast}=1/\sqrt{c_{ST}}\sim 5$ TeV yields $m_t\sim 150$ GeV,
which is acceptably close to the experimental value (this being just an estimate).
Most important, the dangerous $c_{SS}$ coefficients are not affected by walking,
because they involve fermions that do not directly participate in the strong dynamics.
As anticipated, walking changes the counting rules of NDA, making the operators involved in 
the process of mass generation lower-dimensional than their engineering dimension, and hence less irrelevant,
and in this way making ETC more similar to the results of dimensional analysis 
that we summarized in the FN case.

To conclude, it must be noted that the estimate on the top mass is somewhat optimistic.
Nobody really knows for sure of any specific walking theory that reproduces it, particularly because 
the calculation of the anomalous dimension poses a huge challenge.
A specific calculation in a strongly-coupled theory with the right properties would
be needed in order to put all of the above on firm grounds.
The role of walking is certainly that of enhancing the fermion masses, but 
quantitatively it is difficult to draw a firm conclusion, and it might well be that 
the top has to be treated separately.
Other proposals exist that address the specific problem of the top mass.
The basic observation at the basis is that for all practical purposes the mass of the top is 
numerically the same as the scale of the condensate breaking electro-weak symmetry, and it is hence
conceivable that the top plays a special role in the strong dynamics. Models in which
the top quark is a composite object emerging from the strong dynamics itself are generically referred to 
as topcolor models~\cite{reviews}, with some abuse of language. As anticipated earlier on, 
in recent years a variation of this idea has been implemented 
in five-dimensional models, by assuming that the fermions, particularly the top,  be allowed to propagate in the bulk
of the extra-dimension. The interested reader is referred to~\cite{5Dfermions} and references therein.

\subsection{Perturbative unitarity.}

A very important, non-trivial property of the MSM is related to unitarity.
Being a gauge theory, with no anomalies, unphysical degrees of freedom (states with negative-norm or wrong spin-statistics) decouple.
But even more is true: scattering amplitudes can be computed diagrammatically, 
and by looking at their behavior (particularly at large energy),  it is possible to ask under what circumstances the theory 
admits a sensible interpretation, yielding probabilities bounded by $1$.
In the MSM, this procedure yields a relation between the cut-off $\Lambda$, identifying the regime of validity of the 
theory, and its one undetermined parameter, the mass of the Higgs. In particular, this logic yields the classical result that 
an absolute upper bound on the mass of the Higgs $m_h$ exists~\cite{PU}.

In this section, I review the classical derivation of this result, in a
simplified version of the MSM in which I set the gauge coupling $g^{\prime}=0$, so that $M_Z=M_W$,
and compute at the tree-level the (elastic) scattering amplitude between longitudinally polarized
gauge bosons $W_L Z_L\rightarrow W_LZ_L$.
This is a very instructive exercise, which provides deep inside into the functioning of Feynman rules,
and of spontaneous  symmetry breaking in relation to the Goldstone equivalence theorem.
It also teaches us a great lot about the properties of the scalar sector of the spectrum of a strongly-coupled
model of EWSB.

\begin{figure}
\begin{center}
\begin{picture}(360,100)(0,-10)
\Photon(0,0)(30,50){3.5}{5.5}
\Photon(60,0)(30,50){3.5}{5.5}
\Photon(0,100)(30,50){3.5}{5.5}
\Photon(60,100)(30,50){3.5}{5.5}
\GCirc(30,50){4}{0.4}
\Photon(80,0)(110,50){3.5}{5.5}
\Photon(180,0)(150,50){3.5}{5.5}
\Photon(80,100)(110,50){3.5}{5.5}
\Photon(180,100)(150,50){3.5}{5.5}
\Photon(110,50)(150,50){3.5}{4.5}
\GCirc(110,50){4}{0.4}
\GCirc(150,50){4}{0.4}
\Photon(200,0)(230,30){3.5}{4}
\Photon(280,0)(230,70){3.5}{5.5}
\Photon(200,100)(230,70){3.5}{4}
\Photon(280,100)(230,30){3.5}{5.5}
\Photon(230,30)(230,70){3.5}{4.5}
\GCirc(230,30){4}{0.4}
\GCirc(230,70){4}{0.4}
\Photon(300,0)(330,30){3.5}{4}
\Photon(360,0)(330,30){3.5}{4}
\Photon(300,100)(330,70){3.5}{4}
\Photon(360,100)(330,70){3.5}{4}
\DashLine(330,30)(330,70)3
\GCirc(330,30){4}{0.4}
\GCirc(330,70){4}{0.4}
\Text(0,-8)[]{$Z$}
\Text(60,-8)[]{$Z$}
\Text(80,-8)[]{$Z$}
\Text(180,-8)[]{$Z$}
\Text(200,-8)[]{$Z$}
\Text(280,-8)[]{$Z$}
\Text(300,-8)[]{$Z$}
\Text(360,-8)[]{$Z$}
\Text(0,108)[]{$W$}
\Text(60,108)[]{$W$}
\Text(80,108)[]{$W$}
\Text(180,108)[]{$W$}
\Text(200,108)[]{$W$}
\Text(280,108)[]{$W$}
\Text(300,108)[]{$W$}
\Text(360,108)[]{$W$}
\Text(130,62)[]{$W$}
\Text(218,50)[]{$W$}
\Text(338,50)[]{$h$}
\end{picture}
\end{center}
\caption{Diagrams contributing to $W_LZ_L$ scattering in the MSM, yielding the amplitudes
${\cal M}_4$, ${\cal M}_s$, ${\cal M}_u$ and ${\cal M}_h$ (left to right).}
\label{Fig:WL}
\end{figure}
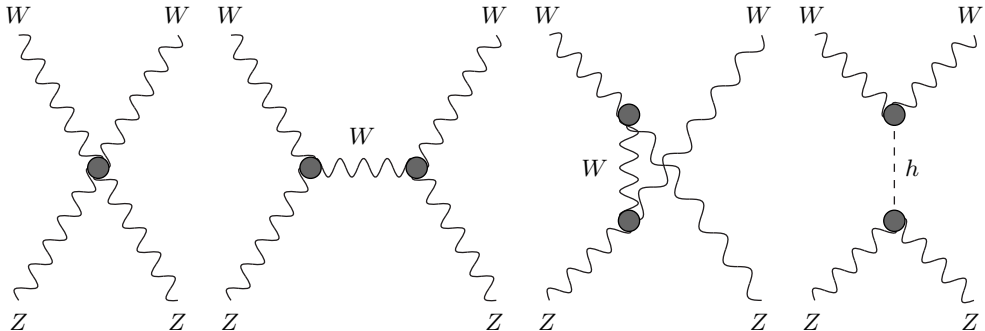

The scattering amplitude of longitudinally polarized gauge bosons
receives four types of contributions at the tree-level in the MSM, depicted in Fig.~\ref{Fig:WL}.
Besides the direct contribution from the 4-gauge boson vertex (${\cal M}_4$)
there are $s$-channel and $u$-channel exchanges of charged gauge bosons
(${\cal M}_s$ and ${\cal M}_u$) and $t$-channel exchanges of neutral scalars  (${\cal M}_h$):
\beqs
-i{\cal M}_4 &=& 
  \frac{g^2}{M_W^4}\left[(c^2-6c-3)E^4+(6c+2)M_W^2E^2\right]\,,\\
  -i{\cal M}_s &=& 
  \frac{1}{M_W^4}\left[c(2E^2+M_W^2)^2(4E^2-4M_W^2)\right]
  \frac{g
  ^2}{4E^2-M_W
  ^2}\,,\\
   -i{\cal M}_u &=& 
  \frac{1}{M_W^4}\left[\frac{}{}(6+10c+2c^2-2c^3)E^6+(-6-26c-18c^2+2c^3)M_W^2E^4\right.\\
&&\left.\frac{}{}  +(2+14c+20c^2)M_W^4E^2+(-2+2c)M_W^6\right]
\frac{g
^2}{2(1+c)(E^2-M_W^2)+M_W
^2}\,,\nonumber\\
   -i{\cal M}_h &=& 
  \frac{1}{M_W^4}\left[M_W^2((c-1)E^2+M_W^2)^2\right]
\frac{g^2
}{2(c-1)(E^2-M_W^2)-m_h^2
}\,.
\eeqs
In the above expressions, written in the center of mass frame, 
$g$ is the  gauge coupling of $SU(2)_L$,
$c=\cos\theta$ is the scattering angle of the outgoing particles measured in respect to the incoming direction, 
and $E$ the center of mass  energy
of the incoming $W$, related to the Mandelstam variable $s=4E^2$.

It is customary to discuss the scattering amplitude in terms of partial wave decomposition.
In particular, for the $J=0$ amplitude:
\beqs
\label{Eq:partialwave}
t_{0}(E)&=&\frac{1}{32\pi}\int_{-1}^{+1}\,\di c\,{\cal M}(E,c)\,.
\eeqs
Unitarity implies that the total cross-section be limited by the geometric cross-section,
and in turns this implies that $|t_{0}(E)|<1/2$. The energy at which this relation stops 
being true gives an estimate of the UV cut-off of the theory $\Lambda$, as a function of the parameters
in the model.

The expression simplifies considerably in the large momentum 
limit $p^2=E^2-M_W^2\rightarrow +\infty$:
\beqs
t_{0}^{SM}&\rightarrow&
\frac{i\,g^2}{64\pi M_W^2}\left(m_h^2-M_W^2+2M_W^2\ln\frac{4p^2}{M_W^2}\right)\,.
\label{Eq:perturbativeunitarity}
\eeqs
Among the terms in parenthesis, the $m_h$ dependence is by far the most important effect.
Neglecting the other two terms gives the anticipated bound $m_h \lsim 1.2$ TeV.
This is a famous result, which puts an upper bound on the MSM Higgs mass.
A more complete and rigorous analysis is done for instance in~\cite{PU}.
In order to understand what this statement actually means,
it is useful to look in details at how it emerges within Eq.~(\ref{Eq:perturbativeunitarity}), when
summing the contributions from the various diagrams.

First, notice how ${\cal M}_s$ cannot contribute to this specific channel, 
because odd in $c$, so that it
cancels in Eq.~(\ref{Eq:partialwave}).
It is useful to consider the behavior at large-$p$ in terms of a series expansion.
Truncating at the $p^2$ term, and hence keeping only terms that asymptotically grow 
as powers of $p$:
\beqs
t_{0}^{{\cal M}_4}&=&\frac{i g^2}{32\pi M_W^4}\left(
-\frac{16}{3}p^4
-\frac{20}{3}p^2M_W^2
\right)\,,\\
t_{0}^{{\cal M}_s}&=&0\,,\\
t_{0}^{{\cal M}_u}&=&\frac{i g^2}{32\pi M_W^4}\left(
+\frac{16}{3}p^4
+\frac{23}{3}p^2M_W^2
\right)\,,\\
t_{0}^{{\cal M}_h}&=&\frac{i g^2}{32\pi M_W^4}\left(
-p^2M_W^2
\right)\,,
\eeqs
which clearly shows that both most problematic terms are cancelled thanks to
gauge symmetry itself.
To be more specific. The ${\cal O}(p^{4})$ terms cancel between diagrams involving the quartic coupling
between gauge bosons and diagrams built with two cubic couplings, which is an immediate 
consequence of the symmetries of the Yang Mills Lagrangian.
More subtle is the cancellation of the ${\cal O}(p^{2})$ term, which involves the Higgs.
Notice that for the cancellation of the coefficient in front of $p^2$ to be exact, 
its contribution form the Higgs cannot depend on $m_h$.
Ultimately, gauge invariance ensures that the coupling $g$ of the Higgs to the gauge bosons be the same as the cubic self-coupling.
Two terms are left not cancelled at the next order: a term logarithmic in $p$, arising from ${\cal M}_u$ (see the analytical structure of the denominator)
and a term proportional to $m_h^2$ from ${\cal M}_h$. The former can be safely neglected,
in the sense that even for $p$ very large it is not growing fast enough to be a source of concern in practical applications.
The latter is the origin of the bound.

Let us open here a brief digression.
This exercise shows one very typical property of Feynman diagrams: the calculation of single individual diagrams has an inherent tendency to
yield non-sensical results. Such are the ${O}(p^4)$ and ${\cal O}(p^2)$ terms, that would indicate, if really present, 
an inherent problem of the theory with unitarity.
However, once all diagrams are summed, such pathology disappears exactly: the diagrammatic procedure, carried out carefully,
contains all the useful information about symmetries and analytical properties of the theory, and is limited only by 
its perturbative nature.
What is even more striking is that the result hence obtained is actually a consequence of a very non-trivial 
property of spontaneously broken gauge theories,  the Goldstone equivalence theorem, which is automatically encoded in the
diagrammatic procedure.

To help making the  statements in the previous paragraph more clear, let us go back and rewrite the Higgs potential as
\beqs
\label{Eq:Higgsb}
{\cal L}_{\cal V} &=& -{\cal V}\,=\,-\,\lambda^{\prime}\left(H^{\dagger}H-\frac{v_W^2}{2}\right)^2\,,
\eeqs
which, provided $\lambda=\lambda^{\prime}$, differs from Eq.~(\ref{Eq:Higgs}) only by an irrelevant additive constant.
The minimum is clearly at
\beqs
H^{\dagger}H&=&\frac{v_W^2}{2}\,,
\eeqs
and the physical mass of the Higgs particle is 
\beqs
m_h^2&=&2\lambda^{\prime}v_W^2\,.
\eeqs
In this way, one important property is highlighted:  the physical Higgs particle 
receives a mass from its coupling to the vacuum, in the same way as the gauge bosons and
fermions of the standard model do. 
Replacing in the Eq.~(\ref{Eq:perturbativeunitarity}) one can rewrite the leading contribution to
the partial-wave amplitude as a bound on $\lambda^{\prime}$
\beqs
\label{Eq:lambdabound}
-i t_{0}^{SM}&\rightarrow&
\frac{\lambda^{\prime}}{8\pi}\,<\,\frac{1}{2}\,,
\eeqs
or $\lambda^{\prime}<4\pi$, which is the familiar bound ensuring that the perturbative expansion makes sense.

This yields a more correct interpretation of the result obtained above: it is perturbation theory that breaks down,
not unitarity, when the Higgs mass is larger than the bound! 
The upper bound on the Higgs mass means that either the Higgs is light, and hence all its couplings are perturbative,
or otherwise the sector responsible for electro-weak symmetry breaking must be strongly coupled at the TeV scale.
This is the basis of the folklore going under the name of {\it no-lose theorem for LHC},
according to which either the LHC will discover a light Higgs boson, for which the experimental sensitivity stretches up to $800-1000$ GeV,
or a new strongly coupled sector, with new bound states must be present at the TeV scale, in which case the experimental sensitivity
might reach somewhat higher masses, good enough for discovery.

A final  parenthetic remark, purely for didactical purposes. Looking at the Eq.~(\ref{Eq:lambdabound}), one might feel puzzled.
We are computing at the tree-level the scattering of gauge bosons, due to a gauge interaction,
and the result for the amplitude is, at the leading-order,  exactly the same as the elastic scattering amplitude of two real scalars in
a theory with no gauge interactions at all, in which the only coupling is the quartic interaction among the scalars.
The reader who is familiar with the electro-weak theory will recognize that this is another indirect manifestation of the
Goldstone equivalence theorem: at large energies, processes involving massive gauge bosons 
are dominated by the contribution coming from the Goldstone bosons that are  eaten up by the Higgs mechanism.
Another well known example is, within the MSM, the decay rate of the Higgs onto $Z$ bosons, 
which at the leading order, and neglecting phase-space corrections, for  $m_h>2M_Z$ reads:
\beqs
\Gamma(h\rightarrow ZZ)&\simeq&\frac{m_h^3}{32\pi v^2}\left(1-4\frac{M_Z^2}{m_h^2}+12\frac{M_Z^4}{m_h^4}\right)\,.
\eeqs 
At large values of $m_h$, the first term dominates, and the decay rate can be rewritten as
\beqs
\Gamma(h\rightarrow ZZ)&\simeq&\frac{\lambda^{\prime}}{16\pi}m_h\,,
\eeqs
which again does not depend at all on the gauge coupling, and can be computed directly from the theory
of the scalars in the $g\rightarrow 0$ limit, from the decay of the Higgs in two Goldstone bosons.

This being a set of lectures on technicolor, it is time to close these digressions
and  go back to seeing what all of the above teaches us about dynamical EWSB.
First of all, this is another manifestation of the non-decoupling of the scalar sector, as anticipated when talking about precision physics.
One might think of building the electro-weak chiral Lagrangian by taking the limit $\lambda^{\prime}\rightarrow +\infty$,
while keeping $v_W$ fixed. The Higgs mass grows to infinity in this (classical) limit. However, the Higgs {\it does not decouple},
because in order to do so one is effectively taking large values of the coupling. 
In this limit, what goes really wrong is that in the last diagram in Fig.~\ref{Fig:WL} one should not, for large-$\lambda$,
use the perturbative propagator for the Higgs, but should replace it with a non-perturbatively computed scalar correlator, formally
written as
\beqs
\frac{1}{p^2-m_h^2}&\rightarrow&\frac{1}{p^2-\Sigma(p)}\,.
\eeqs
The exact form of $\Sigma(p)$ cannot be obtained with simple perturbation theory, and one expects it to have a very non-trivial 
analytical structure of poles and cuts. In a UV-complete TC model, one might also expect that $\Sigma(p)\rightarrow 0$ at high
energies, essentially because all the scales should be generated dynamically in the IR.
And hence, ultimately one expects the ${\cal O}(p^2)$ terms to cancel, and the scattering amplitude to be unitary, but uncalculable.

In Higgsless models, one would not include the diagram yielding ${\cal M}_h$, because, by definition,
there is no Higgs to start with. As a result, the amplitude under study would grow quadratically with $p$,
in the same way as naively obtained from the chiral Lagrangian.
One might be lead to conclude that such models generically violate unitarity already below the TeV scale,
unless completed somehow. 
This statements are misleading, and should be read and used with some caution.
The subtle part is that in the calculation we did here, we have been careful to take $p\rightarrow +\infty$
at fixed $m_h$, before starting to look at large values of $m_h$.
The other way around, in which one first takes $m_h\rightarrow +\infty$, and hence sets ${\cal M}_h=0$,
and then looks at large-$p$, can give sensible answers only provided one can decouple the Higgs.
Which is not the case.
We already anticipated this observation earlier: this scattering amplitude is a particularly bad observable quantity,
that is quadratically sensitive to the mass of the Higgs, and in general to the 
structure of the scalar correlators of the symmetry-breaking sector. 
As such, the amplitude is completely dominated by uncalculable contributions, which cannot be neglected, and hence in general  yields very little useful information about new physics. Contrast this with what happens for oblique precision parameters.

One should also explain what is meant by {\it Higgsless}.
The fact that  EWSB is not induced by an elementary, weakly-coupled scalar sector
does not imply that there are no scalars at all. Whatever its origin, it is always possible to 
excite the vacuum, provided enough energy is at disposal. In doing so, one is probing the structure of the quantum effective potential
expanded around the symmetry-breaking vacuum. 
The fact that we do not know what this potential looks like, does not mean that there are no 
excited, scalar bound states.
The really interesting question has to do with the properties of such excited states: can they be described and detected experimentally as 
particles? Based on the experience of QCD, one might be tempted to give a firm negative answer.
Yet, this is a premature conclusion: answering such a question requires a very detailed knowledge of the strong dynamics itself, 
and dedicated non-perturbative method should be used. 
Unless symmetry arguments become available, which we will discuss in next subsection.

In conclusion. The existence of a perturbative theory describing the sector responsible for electro-weak
symmetry breaking, and that this provides a valid description up to high sales,
 requires the existence of a light scalar degree of freedom with the couplings of the Higgs particle in the MSM,
 in order for the $WW$ scattering amplitudes to be manifestly unitary.
In technicolor, there is no elementary scalar, and hence this statement just amounts to saying that the 
theory must be strongly coupled and non perturbative at the TeV scale, 
which we already new from the start.
One might want to compute the scattering amplitude
of longitudinal gauge bosons in presence of such a strongly coupled sector, starting from the underlying
dynamics, instead of using EFT, but this
is of utmost difficulty. Even in QCD, where data about the analog $\pi\pi$ scattering processes 
are available,  the precise role of the resonances, such as the $\rho$ and the 
scalars, are still subject of studies. 
In more modern versions of technicolor, it is believed that the role of techni-$\rho$ 
be very important in the TeV region, in analogy with QCD, but possible other resonances might contribute.
Corrections to the classical SM results encoded in the coefficients of the chiral Lagrangian,
in particular $\alpha_4$ and $\alpha_5$, might be measurable, or new resonances detectable at the LHC.
In general, this is a very open problem in technicolor, and  many interesting and useful studies
that can be found in the literature~\cite{PU2}. But one has to keep in mind that all of this
is very strongly limited, by the sensitivity of this particular observable to the 
uncalculable structure of the correlation functions involving scalar bound states.

\subsection{About the spectrum of WTC.}

Technicolor being a strongly coupled theory, which confines and produces condensates at low energy,
it is expected to have a rich and complicated spectrum of bound states.
In this section I focus on a few very special such bound states,
the properties of which are related to the fundamental symmetries of the theory: the techni-rho mesons, 
the techni-pions and the techni-dilaton.

One would like to know also what the spectrum of anomalous dimensions is,
at low-energies, where the dynamics is walking.
In this energy region, the theory is approximately conformal and
the dynamics must be governed by a CFT with some set of operators 
with non-trivial anomalous dimensions. These anomalous dimensions should be large,
in such a way as to modify the NDA counting rules, for the reasons explained earlier.
In some classes of supersymmetric theories, supersymmetry itself allows to compute 
these anomalous dimensions, thanks to the relation between R-symmetry and the super-conformal group. 
But in general this is
a very hard task, that requires the use of fully non-perturbative techniques.
In principle, within gauge/gravity dualities the program of classifying the relevant operators and
their dimensions can be carried out, although it still represents a very non-trivial task.

\subsubsection{Techni-rho mesons.}

First of all, the TC sector will have some global symmetry $G$, a subset of which is identified with the SM
gauge group. It is hence expected on general grounds that towers of spin-1 states transforming under this symmetry be present,
analogous to the vectorial $\rho$ meson, axial-vectorial $a_1$ meson and their heavy excitations.
We already saw that these states play a crucial role in the oblique precision observables, particularly in the $\hat{S}$ parameter.
They also play a role in the $WW$ scattering amplitudes. And finally, they are the most accessible new states at the LHC.
This last statement depends crucially on the couplings (and width) of such states, which is the main topic of this subsection.
The discussion will rely almost entirely on the comparison with QCD and with its large-$N$ scaling properties.

Before starting, let us remind the reader about the large-$N$ scaling of the relevant parameters,
which can be summarized by
\beqs
M_{\rho}&\sim&{\cal O}(1)\,,\\
f_{\pi}^2 &\sim&{\cal O}(N)\,,\\
f_{\rho}^2 &\sim&{\cal O}(N)\,,\\
g_{\rho\pi\pi}^2 &\sim&{\cal O}\left(\frac{1}{N}\right)\,,
\eeqs
where the $g_{\rho\pi\pi}$ coupling between technirho and technipion mesons in the HLS language is normalized so that
\beqs
{\cal L}_{\rho\pi\pi}&=&2i g_{\rho\pi\pi} \Tr \rho^{\mu}\left[\pi\,\partial_{\mu}\pi\right]\,.
\eeqs

The masses of the spin-1 states are going to be controlled by the strong-dynamics scale $\Lambda^{\ast}\sim {\cal O}(1)$ TeV,
while $f_{\pi}$ stands for  the electro-weak VEV.
The $g_{\rho\pi\pi}$ coupling 
controls the decay $\rho\rightarrow WW$, which is the analog of the $\pi\pi$ dominant decay of the $\rho$ in QCD,
with branching fraction very close to $1$.
In studying LHC signatures of technicolor, this suggests that a clean signature (at least
in a QCD-like technicolor) is the process $pp\rightarrow \rho \rightarrow WW$.
On general grounds, the number of events expected at fixed center-of-mass energy
is strongly suppressed by the techni-rho mass in the intermediate state. 
At the same time, the facts that the width be quite large,
that the production process might involve vector-fusion or suppressed couplings to the quarks,
and that reconstructing the $WW$ final state is not trivial, 
limits the possibility of distinguishing the signal from the SM background to $M_{\rho}\lsim 1$ TeV,
which is already hard to reconcile with precision electro-weak physics.

A different possibility arises at large-$N$ in a non-QCD-like theory.
Some basic results first, exemplified by QCD.
The techni-rho, in analogy with its QCD analogue, has two main couplings that determine its decays.
There is the cubic coupling to the light $W$ and $Z$ (to the pions), with strength controlled by $g_{\rho\pi\pi}^2\propto 1/N$.
But also, the techni-rho couples to the light SM fermions, with couplings proportional to the
decay constant $f_{\rho}^2\propto N$.
Both yield decays to 2-body final states with effectively massless particles. 
A look at the PDG reveals that in QCD
\beqs
B_{\ell}&\equiv&\frac{\Gamma[\r\rightarrow \ell^{+}\ell^{-}]}{\Gamma[\r\rightarrow \pi^{+}\pi^{-}]}\,\simeq\,4.5\,\times\, 10^{-5}\,,
\eeqs
both for $\ell=e$ and $\ell=\mu$. Which clearly indicates that the $g_{\rho\pi\pi}$ is the dominant coupling to be considered,
as already stated earlier.
Indeed, at the tree-level one can show that
\beqs
\label{Eq:weakdecay}
{\Gamma[\r\rightarrow \ell^{+}\ell^{-}]}&=&\frac{\alpha}{3}\frac{e^2f_{\rho}^2}{M_{\rho}^2}M_{\rho}\,,\\
{\Gamma[\r\rightarrow \pi^{+}\pi^{-}]}&=&\frac{g_{\rho\pi\pi}^2}{48\pi}M_{\rho}\,,
\eeqs
with $e$ the electromagnetic coupling.
Some experimental input: $f_{\rho}\simeq 150$ MeV, $M_{\r}\simeq 770$ MeV, $\Gamma_{\rho}\simeq 150$ MeV,
$\alpha\sim 1/137$.  With these one obtains $g_{\rho\pi\pi}\simeq 5.5$, and hence
\beqs
B_{\ell}&\simeq&\frac{192 \pi^2 \alpha^2 f_{\r}^2}{3 g_{\r\pi\pi}^2M_{\r}^2} \,\simeq\,4.2\,\times\, 10^{-5}\,,
\eeqs
which agrees quite well with the data.

Having established that these approximations are sensible, at least for QCD, take the  result  and rescale it to a large-$N$ QCD-like technicolor theory,
in which the mass of the $\rho$ is very big (such that all the SM-fermion masses can be neglected).
The ratio $B_T$  between partial widths into  SM fermions and into $W$ bosons 
results in
\beqs
B_{T}&=&8 \frac{N_c^2}{3^2} B_{\ell}\,,
\eeqs
where the factor of $8$ comes from summing over all the charges of the SM fermions.
In order that $B_T \sim {\cal O}(1)$, one has to require that $N_c \simeq 12 \pi^2$.
Hence, for the decay into $W$ bosons (pions) to be subdominant in respect to the decay into fermions,
one needs very large values of $N_c$. This is in blatant contradiction with the perturbative
estimates from the $\hat{S}$ parameter. 

However, remember how we clarified that what enters into $\hat{S}$ is $f_{\pi}$,
while the coupling to the currents of interest for the decays in $f_{\rho}$.
In a non-QCD-like large-$N_c$ theory, it is possible that the large value of $N_c$ be compensated
by a ratio of scales making $f_{\pi}\ll f_{\rho}$.
Notice also that for such values of $N_c\simeq 12 \pi^2$ one expects
\beqs
g_{\rho\pi\pi}(N_c)&\sim&\frac{\sqrt{3} g_{\rho\pi\pi}(3)}{\sqrt{N_c}}\,\sim\,1\,,
\eeqs
which means that this would be precisely the regime in which hidden local symmetry calculations
might become reliable.

If $B_T \sim {\cal O}(1)$ or larger, then the phenomenology at the LHC of the techni-rho mesons is going to be drastically different from what
inferred from QCD: they will be relatively narrow resonances, with couplings to the SM currents comparable to those of the $Z$ and $W$ mesons,
and decay predominantly into two fermions. The techni-rhos will look very similar to weakly coupled $Z^{\prime}$ (and $W^{\prime}$)
gauge bosons. In particular, the Drell-Yan $pp\rightarrow \r \rightarrow \ell\ell$ is going to be 
a very favorable experimental signature of such scenarios.
We will come back to this point when discussing the bottom-up approach to holography, but notice how
a classical signature of weakly-coupled extensions of the SM (a new, narrow spin-1 state) might actually emerge as a low-energy 
result of a strongly-coupled theory, at large-$N_c$.

\subsubsection{Technipion, composite Higgses and Little Higgses.}

The second sector of the spectrum that is very important is that of the possible light pseudo-scalars in the theory.
In general, the global symmetry $G$ of the technicolor sector may be much larger  than
the SM gauge group, and the condensates breaking $G$ spontaneously will produce 
a number $n_{\pi}$ of Goldstone bosons. Weakly gauging the $SU(2)_L\times U(1)_Y$ means that $3$ of the Goldstone bosons
disappear from the spectrum, becoming longitudinal components of $W$ and $Z$. Also, the gauging of a subgroup of $G$
in general breaks explicitly the symmetry, and hence most of the $n_{\pi}-3$ will acquire a mass.
The resulting pseudo-Goldstone bosons (PNGB), however, are going to be relatively light, and hence problematic.

As an example,  we can perform explicitly the exercise 
of estimating the masses of the PNGBs 
in the case of the prototypical example of TC model we have already discussed several times,
by using Eq.~(\ref{Eq:CW}).
In the limit in which electro-weak and QCD gauge couplings are turned off
there is a $SU(8)_L\times SU(8)_R$ global symmetry, acting on the vectorial techni-quarks,
which is  broken spontaneously to $SU(8)$.
Which means there are a total of $n_{\pi}=63$ PNGBs. A classification of their properties can be found in~\cite{AW}.
An estimate of their masses can be obtained from the non-linear sigma-model description
of the 63 pions $\pi^i$:
\beqs
{\cal L}_8&=&\frac{f^2}{4}\Tr |D \Sigma_8 |^2\,,
\eeqs
where 
\beqs
\Sigma_8&=&\exp \frac{2 i \pi}{f}\,,
\eeqs
the pions describe the $SU(8)\times SU(8)/SU(8)$ coset,
and the covariant derivative is
\beqs
D \Sigma &=&\partial \Sigma \,+\,i\left(g^{\prime} B + g W + g_s G\right) \Sigma \,-\, i\Sigma \left( g^{\prime} B + g_s G\right)\,,
\eeqs
with $W$ the matrix of the $SU(2)_L$ gauge bosons, $B$ the matrix of the $U(1)_Y$, and $G$ the gluons.
With these normalizations, one deduces masses for the electro-weak gauge bosons $M_W^2=g^2 f^2$
and $M_Z^2=(g^2+g^{\prime\,2})f^2$, implying that $f^2=v_W^2/4$.

By using Eq.~(\ref{Eq:CW}), one sees that the QCD interactions give quadratically divergent masses to
the majority of the PNGBs~\footnote{
The weakly-coupled effective potential induced by the explicit symmetry-breaking couplings is
 also important for the study of vacuum alignment~\cite{Peskin}.}. 
To be more precise, if $\Lambda$ is the cut-off in Eq.~(\ref{Eq:CW}), one finds
\beqs
m_{24}^2&=&\frac{g_s^2\Lambda^2}{4\pi^2}\,,\\
m_8^2&=&\frac{3g_s^2\Lambda^2}{8\pi^2}\,,\\
m_{24}^{\prime\,2}&=&\frac{9g_s^2\Lambda^2}{16\pi^2}\,,
\eeqs
where the index $24$ and $8$ indicates the multiplicity, and where corrections due to electro-weak couplings have been neglected
because the QCD coupling $g_s$ is bigger. 
These masses are all estimated to be ${\cal O} (200)$ GeV (up to very large uncertainties signaled by the quadratic divergence).
But notice that all the corresponding states transform  as triplets or octets of QCD $SU(3)_c$,
and hence carry strong interactions.

There are hence $63-24-8-24=7$ massless Goldstone bosons at this stage of the calculation,
which do not receive quadratically divergent contributions to their mass from any of the 
standard-model gauge interactions. 
Two of them obtain a logarithmically divergent mass from Eq.~(\ref{Eq:CW}), 
\beqs
m_{2}^{2}&=&\frac{3g^2g^{\prime\,2}f^2}{16\pi^2}\,c\,,
\eeqs
where $c\sim\log \Lambda/f$ is some ${\cal O}(1)$ number.
These two masses are much lighter, $m_{2}\sim {\cal O} (10)$ GeV, and the corresponding states electrically charged and 
colorless.

Finally, three of the remaining five Goldstone bosons are the longitudinal components of the
$W$ and $Z$ bosons. The last two are exact Goldstone bosons, which are electrically neutral 
and which remain massless
as a result of the fact that a global $U(1)\times U(1)$ symmetry is not broken by the gauge interactions.
They are sometimes referred to as techni-axions.
Their existence is a very major source of concern for technicolor.
However, another source of symmetry breaking might be ETC. Among the 4-fermion interactions 
generated by ETC, some might break the global symmetry $G$, effectively giving mass to 
the two undesirable light Goldstone bosons. 
Walking might also play a role in enhancing dynamically all such masses, because some of the operators produced by ETC, which
break  explicitly the large global symmetry of the model, will involve techni-quarks, and will hence develop large anomalous dimensions,
making them more relevant than NDA would indicate.

Notice one curious fact. The mechanisms giving rise to the very existence, and to the small masses, 
of the technipions
also suggests an intriguing alternative possibility.
A similar mechanism has been exploited in order to construct variations of strongly coupled
theories in which one set of such PNGBs constitutes the Higgs field $H$ of the MSM itself.
The symmetry properties of the Goldstone bosons in this case (partially) protect the 
Higgs potential from dangerous divergences.
In these proposals, the strong-coupling scale has nothing to do with electro-weak symmetry breaking,
which is triggered by the VEV of the Higgs, emerging at a lower scale.
These variations go under the name of composite Higgs~\cite{compositeHiggs}, and of little Higgs~\cite{LH} models.
The difference between the former and the latter being that in little Higgs models the same
mechanism of collective breaking that anomalously  suppressed the masses $m^2_2$ of the example
given above (the fact that the quadratically-divergent contribution to the effective potential vanishes at 1-loop)
is fully exploited in order to enhance the scale separation between the cut-off $\Lambda$ and the electro-weak scale.

Coming back to our topic, namely technicolor. 
This is one particularly simple example,  rich enough to yield generic, rather than general, indications of what to expect. 
The spectrum and interactions of the
techni-pions depend on the global symmetries of the specific realization of technicolor one is interested in,
and one should  analyze the details of the symmetry structure on a model-by-model basis,
but all of the above clearly shows that  techni-pions are a generic source of concern for technicolor, possibly the biggest.
On the other hand, the existence of  techni-pions 
carrying QCD interactions
is possibly testable, because of their relatively light mass.
The fact that light scalars be present which carry electro-weak interactions,
possibly very light,  is also interesting phenomenologically.
And the fact that the number of such states can easily be very large 
is also interesting in LHC perspective, provided one identifies models
in which all of the techni-pions could have escaped detection up to now.

\subsubsection{Techni-dilaton.}

Finally, a very open problem, which is not a general property of dynamical electro-weak symmetry breaking
but rather specifically related to the nature of walking technicolor. 
A light scalar particle might be present as a consequence of the spontaneous breaking of dilatation symmetry.
Its phenomenological properties are so striking to deserve a systematic discussion.

To understand the implications of this possibility, it is useful to look back at the Lagrangian of the MSM,
in which we write the potential in the form of Eq.~(\ref{Eq:Higgsb}) rather than Eq.~(\ref{Eq:Higgs}).
At the classical level, in the limit in which $\lambda^{\prime}$ vanishes, this Lagrangian is exactly scale-invariant,
containing only marginal operators. The minimum of the Higgs potential  breaks scale invariance spontaneously,
at the scale $v_W$. The coupling $\lambda^{\prime}$, and all the quantum correction  giving rise to anomalous dimensions,
represent the  explicit breaking of scale invariance.
As a result, there is a spontaneously broken approximate global symmetry, and hence in the spectrum there must be
a dilaton, the PNGB associated with dilatations, whose mass is controlled by the symmetry breaking terms.
Indeed, such dilaton is the Higgs particle $h$, and the expression for its mass
\beqs
m_h^2&=&2 \lambda^{\prime} v_W^2\,,
\eeqs
is similar to the relation yielding the mass of the pion in QCD
\beqs
m_{\pi}^2&\propto& m_q f_{\pi}\,.
\eeqs
In the latter, $m_q$ is the quark mass, that breaks explicitly the chiral symmetry,
while $f_{\pi}$ breaks it spontaneously. For the Higgs/dilaton,  the $\lambda^{\prime}$
coupling plays the same role as $m_q$ for the pions.
Also, notice that  the pion is sensibly described as a PNGB only provided $m_q$ be small.
For instance the mass of the $B$ mesons, which have quantum numbers that make them the analog
of the pions when the flavor symmetry of QCD is extended to $SU(5)\times SU(5)$,
scales as $m_B^2\sim m_b^2$. The breaking of chiral symmetry is in this case  
so large that it does not make sense to speak about $SU(5)\times SU(5)/SU(5)$ Goldstone bosons.
In the same sense, only provided $\lambda^{\prime}$ be very small, is the Higgs a dilaton.

The striking consequence of thinking about the Higgs as a dilaton, is that one automatically
can infer that the couplings of the Higgs to all the SM fields are given by
\beqs
{\cal L}&=&2\frac{h}{v_W}M_W^2W_{\mu}W^{\mu}\,+\,\frac{h}{v_W}M_Z^2Z_{\mu}Z^{\mu}\,-\,
\frac{h}{v_W}m_{\psi}\bar{\psi}\psi\,,
\eeqs
which again is what expected: each appearance of the Higgs $h$ is accompanied by a $1/v_W$
factor taking into account the scale of spontaneous symmetry breaking (as in the NDA rules), and the coupling is then proportional to 
the explicit breaking encoded in the masses of the bosons and fermions in the associated dilatation current.

Provided the Higgs be perturbative (i.~e. light), most of the main calculations needed to 
discuss its experimental searches are done at tree-level, while  higher-order operators generated by
new physics are going to give subleading corrections.
For instance, the decay rate into two SM fermions $\psi$ is
\beqs
\Gamma[h\rightarrow \psi\psi]&\simeq&\frac{{\cal N} m_\psi^2 m_h}{8\pi v_W^2}\,,
\eeqs
where ${\cal N}=3$ for quarks and ${\cal N}=1$ for leptons.
Which means that whatever the theory responsible for electro-weak symmetry breaking,
if it has a light dilaton in the spectrum, not only its quantum numbers, but also most of its coupling,
and all of its phenomenology, are going to be identical to the higgs $h$ of the MSM.
Only at higher energies, and/or measuring very precisely the subleading corrections, 
one might be able to tell the difference.
Which implies that knowing if a walking technicolor has or not a light dilaton in the spectrum is
of very direct and immediate relevance for the LHC program.
Notice also that such a state would then enter directly into the arguments 
about  perturbative unitarity deduced from the $WW$ scattering, and of course
enter logarithmically in the oblique parameters, precisely in the same way as the Higgs
field $h$ does.
This is another example of the fact that this non-QCD-like technicolor models might 
be more difficult to distinguish from weakly-coupled extensions of the SM than naively expected
(see a related comment in the subsection devoted to the techni-rho mesons).

The reason why one might think that a walking technicolor theory
might have a light dilaton  in the spectrum is not immediately apparent.
For instance, in QCD it is known that such a state does not exist.
One might associate it with the $\sigma$. But then again, the $\sigma$ is not light,
it is certainly not weakly-coupled, and it has a possibly complicated internal structure,
consisting of an admixture of quark and pion bound states, implying that its couplings are very non-trivial.
Also, it is somewhat difficult to argue that there is a sense in which at
the $\Lambda_{QCD}$ scale there is dilatation symmetry, 
and hence nothing to spontaneously break in order to produce a dilaton.

By contrast, the basic idea of walking technicolor is that the theory be {approximately} scale
invariant in the IR. Barring the problem that a sensible definition of the word {\it  approximate}
would be needed, this is clearly suggesting that one cannot exclude the presence of such a light dilaton,
provided the dynamics be very different from QCD-like.
Making this statement quantitative is very hard. 
If the underlying dynamics is some form of gauge theory, the explicit breaking of conformal symmetry 
is presumably related to the coefficient of the $\beta$-function of the gauge coupling being parametrically smaller
than the coupling itself.
But also  this is a confusing and ambiguous statement: at what scale, and in what scheme, should one 
calculate this coefficient, and hence the resulting mass for the dilaton?
One might say that if the gauge coupling runs as illustrated in the third panel in Fig.~\ref{Fig:running},
then it is reasonable to believe that conformal symmetry be broken spontaneously
at the lower end $\Lambda_{IR}$ of the plateau region, and that at that scale the $\beta$ function is indeed small.
But then one might ask what is the effect of the lowest scale in the problem, at which the coupling diverges,
the theory confines, and the $\beta$-function is not small.

An actual calculation is needed in order to answer this question.
Examples exist in the literature, done using the most diverse approaches
in order to model the strongly-coupled sector and gain in calculability,
and date back to the early days of walking technicolor~\cite{dilaton}.
Unfortunately, they disagree with each other.
This is perhaps the biggest open problem that gauge/gravity dualities might help to solve~\cite{ENP,dilaton2,dilaton5D}, and
I will spend more time on it later in these lectures.

\subsection{Summary:  questions on walking TC.}
\label{Sec:questions}

Walking technicolor provides a very natural solution to the 
big hierarchy problem, and is a way of softening the 
little hierarchy problem of more generic technicolor models.
On the basis of what we said so far in this lectures,
we can produce a list of open problems that we want to address
using holography and gauge/gravity correspondence:
\begin{itemize}
\item[i)] what is the field content of a model yielding walking dynamics, in particular what are the conditions of the number of colors,
type of representation, number of flavor,
\item[ii)] what are the values of the dynamically generated scales, in particular the value of the chiral condensate,
and how are they related to each other and to the quantities in i),
\item[iii)] what are the anomalous dimensions, and in particular how do  they depend parametrically on the choice of the theory and
on the scale at which they are evaluated,
\item[iv)] what are the oblique precision parameters, how do they depend on the details in point i), on the scales at point ii) and iii)
and  on the specific way in which the technicolor sector is coupled to the SM,
\item[v)] what are the coefficients of the chiral Lagrangian, including those that correct the cubic 
and quartic interactions among gauge bosons,
\item[vi)] what does a fully realistic ETC theory look like, what is its field content, what are the ETC scales, what is the 
pattern of symmetry breaking, and how can one effectively describe the dynamics of tumbling,
\item[vii)] what are the masses of the fermions, as a function of the points vi) and iii),
\item[viii)] what are the contribution to FCNC from operators generated by ETC, and how do they depend on the 
properties of the theory in point vi),
\item[ix)] what is the spectrum of the spin-1 mesons, and what are their couplings and decay constants,
 how do they depend on i) and ii), and what is the precise relation to the oblique parameters
in point iv),
\item[x)] what are the global symmetries of the TC theory at the electro-weak scale, and what is the spectrum of
PNGBs as a function of the operators  generated by ETC and of the anomalous dimensions,
\item[xi)] is there a light dilaton, and what are its mass and couplings as a function of i) and ii),
\item[xii)] what is the $WW$ scattering amplitude going to look like at large energies, what are the 
intermediate states that are relevant from points ix) x) and xi), and what is its analytical structure.
\end{itemize}

This is an incomplete set of questions. All of them are very difficult, and all of them ultimately need
to be answered in order
to ask the most fundamental questions:
\begin{center}
{\it
Are there calculable models of dynamical electro-weak symmetry breaking that 
are compatible with all experimental data? If so, how do we discover them and
characterize them at the LHC?}
\end{center}

The main aim of these lecture notes is to highlight the
fact that gauge/gravity dualities represent  one very powerful tool for such a
program of investigation of strong dynamics to be carried out, and the next sections will be devoted to
this topic. The reader should be aware of the fact that other approaches exist. 
In very recent years a lot of  resources have been focused 
on trying to answer these questions (in particular  those at point i)
with the non-perturbative instruments of lattice field theory.
The mere number of papers that appeared recently on the topic
gives a measure of how interesting, and difficult, these studies are. 
An incomplete list includes~\cite{lattice}.

\newpage
\section{Holographic technicolor: bottom-up approach.}

As we have seen, hidden local symmetry is useful, but affected by two related problems: 
the calculability is limited both by the proliferation of independent parameters when including 
more states, and by the fact that some couplings take large values, making the perturbative expansion questionable.
One basic idea behind the bottom-up holographic approach is to keep the successes of hidden local symmetry 
(particularly, its simplicity) and overcome these main difficulties by replacing the chain of 
links and sites with an extra-dimension~\cite{AdSTC,AdSTC2}. A background metric inspired by the AdS/CFT correspondence is assumed,
which is justified by the assumption of walking,
and the effective couplings are kept small by the assumption that the model hence obtained be dual to a large-$N_c$ theory.
The gravity formalism can be used because a large value for  the large 't Hooft coupling $\lambda = g^2 N_c$ is assumed 
for the dual field theory. 

This idea has been used also in order to model low-energy QCD~\cite{AdSQCD}. The three assumptions about $N_c$, $\lambda$
and the AdS background are hardly justified in this case. Yet, many interesting and useful results have been obtained,
and one can argue that they are in acceptable agreement with actual QCD data, at least in a  variety of interesting examples.

I will not review all the literature on the subject, which is vast, but focus on one very specific example~\cite{AdSTC}.
The main result that I want to convince the reader of is that a sensible description of all the physics of
electro-weak gauge bosons can be obtained in which only very few parameters with very specific meaning are needed,
and that it is possible to choose such parameters in a way that renders $\hat{S}$ compatible with the bounds.
Once this is done, this set up allows to make actual predictions for LHC signatures.
The example discussed has the only special feature of being very minimal. In the end, it contains only one more parameter than the MSM,
but describes an infinite number of new resonances. Infinite numbers of non-minimal variations of this
set-up can be formulated, and many have been explored in the literature.

\subsection{Holography and AdS/TC: a simple model.}

\subsubsection{The Model.}

A summary of  the minimal requirements for a viable model of dynamical EWSB
in four dimensions allows to identify the properties of the five-dimensional dual description. 
Many of these assumptions can in principle be relaxed, but they constitute 
a simple and natural place to start.
There must be a new strong sector possessing the global symmetry
$SU(2)_L\times U(1)_Y$ of the standard model.
The new interaction must confine, and a symmetry breaking condensate
must form. The (weak) gauging of the global symmetry of the strong sector
gives the massive SM gauge bosons and the photon.
The strong sector has to be close to conformal over the energy range 
between the electro-weak  scale and a much larger ETC scale.
Large anomalous dimensions are present
so  that the chiral condensate
has scaling dimension very different from $d=3$ (for concreteness, here $d=2$). 
All the SM fermions are elementary, and do not carry
quantum numbers of the new strong interactions, hence ensuring 
universality of the electro-weak gauge coupling.

The (quasi-conformal) energy window just above the electro-weak scale
is described by a slice of $AdS_5$, 
i.e. by a five-dimensional space-time containing a warped gravity 
background given by the metric:
\beqs
\di s^2 &=& \left(\frac{L}{z}\right)^{2}\left( \eta_{\mu\nu}\di x^{\mu}\di x^{\nu}\,-\,\di z^2\right)\,,
\eeqs
where $x^{\mu}$ are four-dimensional coordinates, $\eta_{\mu\nu}$ the Minkoski 
metric with signature $(+,-,-,-)$, and $z$ is the (warped) extra-dimension.
The dimensionful parameter $L$ is the $AdS_5$ curvature, and sets the 
overall scale of the model.
Conformal symmetry is broken by  the boundaries 
\beqs
L_0\,<\,z\,<\,L_1\,,
\eeqs
with $L_0>L$, where $L_0$ and $L_1$ correspond to the UV and IR cut-offs of the
conformal theory, i.~e. to the ETC scale and to the confinement scale respectively.

The field content in the bulk of the five-dimensional  model 
consists of a  complex scalar $\Phi$ transforming 
as a $(2,1/2)$ of the gauged  $SU(2)_L\times U(1)_Y$.
The generator of $SU(2)_L$ are $T^a=\tau^a/2$ with $\tau^a$ the 
Pauli matrices.

The bulk action for $\Phi$ and the gauge bosons $W = W^a T^a$ of $SU(2)_L$  
and $B$ of $U(1)_Y$ is 
\beqs
{\cal S}_{5} &=& \int\di^4 x \int_{L_0}^{L_1}\di z\,\sqrt{G}\left[\frac{}{}
\left(G^{MN}(D_M\Phi)^{\dagger} D_N\Phi -M^2|\Phi|^2\right)\,
\nonumber\right.\\ &&\left. 
\left(-\frac{1}{2}\Tr\left(W_{MN}W_{RS}\right)-\frac{1}{4}B_{MN}B_{RS}\right)G^{MR}G^{NS}\right]\,,
\eeqs
and the boundary terms are 
\beqs
{\cal S}_{4} &=& \int\di^4 x \int_{L_0}^{L_1}\di z \,\sqrt{G}\left\{ \frac{}{}\delta(z-L_0)\,D
\left[-\frac{1}{2}\Tr\left[W_{\mu\nu}W_{\rho\sigma}\right]-\frac{1}{4}B_{\mu\nu}B_{\rho\sigma}\right]
G^{\mu\rho}G^{\nu\sigma}\nonumber\right.\\
&& -\delta(z-L_0)\,2\lambda_0 \left(|\Phi|^2-\frac{\mbox{v}_0^2}{2}\right)^2
\left. -\delta(z-L_1)\,2\lambda_1 \left(|\Phi|^2-\frac{\mbox{v}_1^2}{2}\right)^2\right\}\,,
\eeqs
where the covariant derivative is given by
\beqs
D_M\Phi &=& \partial_M \Phi + i (g W_M \Phi +\frac{1}{2} g^{\prime}B_M  \Phi )\,,\nonumber
\eeqs
and where the Yang-Mills action is written in terms of the antisymmetric 
field-strength tensors $W_{\mu\nu}$ and $B_{\mu\nu}$.
In the action, $M^2$ is a bulk mass term for the scalar, 
and $g$ and $g^{\prime}$ are
the (dimensionful) gauge couplings in five-dimensions.
The coefficient $D$ will be fixed later.

Without loss of generality, the VEV of the $\Phi$ field can  be written as
\beqs
\langle \Phi \rangle &=& \frac{\mbox{v}(z)}{\sqrt{2}}\left(\begin{array}{c}
0\cr 1\end{array}\right)\,.
\eeqs
The localized potentials enforce a non-vanishing ${\rm v}(z)$ that induces 
electro-weak symmetry breaking.  For $M^2=-4/L^2$, in the $\lambda_i\rightarrow +\infty$ limit,
 (in which the transverse degrees of freedom become infinitely massive and decouple from the spectrum) the bulk equation of the motion 
\beqs
\partial_z\left(\frac{L^3}{z^3}\partial_z{\rm v}\right)-\frac{L^5}{z^5}M^2{\rm v}=0\,,
\eeqs
admits the solution
\beqs
\mbox{v}(z)&=&\frac{\mbox{v}_1}{L_1^2}z^2\,=\,\frac{\mbox{v}_0}{L_0^2}z^2\,,
\eeqs
by appropriately choosing  $\mbox{v}_{0}/\mbox{v}_1=L_0^2/L_1^2$, 
so as to describe a chiral condensate of dimension $d=2$.

The localized kinetic terms for the gauge bosons
are required by holographic renormalization~\cite{HR} 
in order to remove a logarithmic divergence, and retain finite SM gauge couplings
in the $L_0\rightarrow 0$ limit, renormalizing the otherwise divergent
 kinetic terms of the SM gauge bosons.
This procedure ensures that  the SM gauge couplings
be independent of the strength of the bulk coupling, 
and that all physical quantities be independent of any 
UV-sensitive details, such as the precise value of $L_0$, or the 
structure of the UV boundary.

\subsubsection{Electro-weak Phenomenology.}

Focusing on the spin-1 sector,
and defining 
\beqs
V^{M}&\equiv & \frac{g^{\prime}W^{3\,M}+gB^{M}}{\sqrt{g^2+g^{\prime\,2}}}\,,\\
A^{M}&\equiv & \frac{g W^{3\,M}-g^{\prime}B^{M}}{\sqrt{g^2+g^{\prime\,2}}}\,,
\eeqs
one obtains the electro-weak equivalent of the vector and axial-vector sectors of the chiral Lagrangian.
In particular the massless mode of $V^{\mu}$ is the photon, and the lightest
mode of $A^{\mu}$ is the $Z$ boson.

After Fourier transformation in the four-dimensional Minkowski coordinates one can write
\beqs
A^{\mu}(q,z)&\equiv&A^{\mu}(q)v_Z(z,q)\,,
\eeqs
and analogous for $W_{1,2}$ and $V$, 
where $q=\sqrt{q^2}$ is the four-dimensional momentum.
In the limit in which one neglects 
the five-dimensional gauge coupling, and hence cubic and quartic self-interactions, the
bulk equations are:
\beqs
\partial_z\frac{L}{z}\partial_z v_i-\mu^4_{i} L z  v_i&=&-q^{2}\frac{L}{z}v_i\,,
\eeqs
where $i=V,Z,W$, with $\mu_V=0$,  $\mu^4_W=1/4g^2\mbox{v}_0^2/L^2$ and 
$\mu^4_Z=1/4(g^2+g^{\prime 2})\mbox{v}_0^2/L^2$.
This approximation and the approximation of restricting all the analysis
at the tree-level are valid provided the effective couplings $g/\sqrt{L},g^{\prime}/\sqrt{L}\lsim O(1)$,
which corresponds to  the large-$N$ limit.

The bulk equations can be solved exactly.  
 Choosing Neumann boundary conditions in the IR (which are compatible with the gauge symmetry)
leaves a non-vanishing UV-localized action, that can be interpreted in terms of yielding the
vacuum polarizations for the gauge bosons in the form
\beqs
{\cal L}&=&-\frac{1}{2}A_{\mu}^i\,\pi_{i,j} P^{\mu\nu}A^{j}_{\nu}\,,
\eeqs
where $P^{\mu\nu}\equiv\eta^{\mu\nu}-q^{\mu}q^{\nu}/q^2$,
and where $i=B,W^a$.
The matrix of the vacuum polarizations $\pi_{i,j}(q^2)$
of the SM gauge bosons results in
\begin{widetext}
\beqs
\frac{\pi_{+}}{{\cal N}_W^2}&=&Dq^2+\frac{\partial_zv_{W}}{v_{W}}(q^2,L_0)\,,
\\
\frac{\pi_{BB}}{{\cal N}_B^2}&=&
Dq^2 +\frac{g^2}{g^2+g^{\prime\,2}}\frac{\partial_zv_{V}}{v_V}(q^2,L_0)
+\frac{g^{\prime\,2}}{g^2+g^{\prime\,2}}\frac{\partial_zv_{Z}}{v_Z}(q^2,L_0)\,,
\\
\frac{\pi_{WB}}{{\cal N}_W{\cal N}_B}&=&
\frac{g g^{\prime}}{g^2+g^{\prime\,2}}\left(\frac{\partial_zv_{V}}{v_V}(q^2,L_0)
-\frac{\partial_zv_{Z}}{v_Z}(q^2,L_0)\right)\,,\\
\frac{\pi_{WW}}{{\cal N}_W^2}&=&
Dq^2 +
\frac{g^{\prime\,2}}{g^2+g^{\prime\,2}}\frac{\partial_zv_{V}}{v_V}(q^2,L_0)
+\frac{g^{2}}{g^2+g^{\prime\,2}}\frac{\partial_zv_{Z}}{v_Z}(q^2,L_0)\,.
\eeqs
\end{widetext}
The precision electro-weak parameters have been defined earlier,
in terms of the vacuum polarizations,
with the convention of choosing the normalizations ${\cal N}_i$
of the fields so that  $\pi_{BB}^{\prime}(0)=\pi_{+}^{\prime}(0)=1$.

By taking for simplicity $L_0\rightarrow L$, and by  expanding  for small  $L_0\rightarrow 0$,
one has~\cite{AdSTC}:
\begin{widetext}
\beqs
\frac{\partial_zv_{V}}{v_V}(q^2,L_0)&=&q^2L_0\left(\frac{\pi}{2}
\frac{Y_0(qL_1)}{J_0(qL_1)}-\left(\gamma_E+\ln \frac{qL_0}{2} \right)\right)\,,\\
\frac{\partial_zv_{Z}}{v_Z}(q^2,L_0)&=&
L_0\left\{
\mu_Z^2\,-\,q^2\left[
\gamma_E+\ln(\mu_Z L_0)+\frac{1}{2}\psi\left(-\frac{q^2}{4\mu_Z^2}\right)-\frac{c_2}{2c_1}\Gamma\left(-\frac{q^2}{4\mu_Z^2}\right)
\right]\right\}\,,
\eeqs
where, having  imposed Neumann boundary conditions in the IR, 
\beqs
c_1&=&2L\left(-1+\frac{q^2}{4\mu_Z^2},\mu_Z^2L_1^2\right)+L\left(\frac{q^2}{4\mu_Z^2},-1,\mu_Z^2L_1^2\right)\,,\\
c_2&=&-U\left(-\frac{q^2}{4\mu_Z^2},0,\mu_Z^2L_1^2\right)+\frac{q^2}{2\mu_Z^2}
U\left(1-\frac{q^2}{4\mu_Z^2},1,\mu_Z^2L_1^2\right)\,,
\eeqs
\end{widetext}
and where $\partial_zv_{W}/v_{W}=
\partial_zv_{Z}/v_{Z}(\mu_Z\rightarrow \mu_W)$. 

Bounds on oblique precision parameters require that the  
symmetry-breaking parameter ($\mu_Z$ here) be somewhat small compared to the 
confinement scale ($L_1$). One can see that this must be the case directly 
by noticing that
\beqs
\frac{\partial_zv_{Z}}{v_Z}(0,L_0)&=&-L_0\mu_Z^2\tanh\frac{\mu_Z^2 L_1^2}{2}\,,
\eeqs
that $\mu_Z^2\propto g$, and that hence one must require $\mu_Z^2L_1^2 \ll 1$, otherwise the mass
of the gauge bosons would be proportional to $g$, instead of $g^2$!
Assuming hence that  $\mu_Z^2L_1^2 \ll 1$, we can expand also in $\mu_Z^2L_1^2$, and as a result
\beqs
\frac{\partial_zv_{v}}{v_v}(q^2,L_0)&\simeq&L_0\left(q^2 \log\frac{L_1}{L_0}\,+{\cal O}(q^4)\right)\,,\\
\frac{\partial_zv_{Z}}{v_Z}(q^2,L_0)&\simeq&L_0\left(-\frac{\mu_Z^4L_1^2}{2}-\frac{3}{16}\mu_Z^4L_1^4 q^2+q^2\log\frac{L_1}{L_0}\,+{\cal O}(q^4)\right)\,.
\eeqs

The  localized counter-term 
\beqs
D&=&L_0\left(\ln\frac{L_0}{L_1}+\frac{1}{\varepsilon^2}\right)\,
\eeqs
cancels the logarithmic divergences, and choosing a universal normalization 
 ${\cal N}^2=\varepsilon^2/L_0$  all the dependence on $L_0$  disappears (at leading order in $L_0$), 
the limit $L_0\rightarrow 0$ can be taken,
and the model is renormalized, with finite (dimensionless) SM couplings
$g_4^{(\prime)\,2}= \varepsilon^{2}g^{(\prime)\,2}/L$.
The resulting vacuum polarizations (truncating at the order needed to extract $\hat{S}$) are
\begin{widetext}
\beqs
{\pi_{+}}&=&q^2+\varepsilon^2g^2\kappa\left(-\frac{1}{2L_1^2}-\frac{3}{16} q^2\right)\,,
\\
{\pi_{BB}}&=&
q^2 
+\varepsilon^2g^{\prime\,2}\kappa\left(-\frac{1}{2L_1^2}-\frac{3}{16} q^2\right)\,,
\\
{\pi_{WB}}&=&
-\varepsilon^2gg^{\prime}\kappa\left(-\frac{1}{2L_1^2}-\frac{3}{16} q^2\right)\,,\\
{\pi_{WW}}&=&
q^2 
+
\varepsilon^2g^2\kappa\left(-\frac{1}{2L_1^2}-\frac{3}{16} q^2\right)\,,
\eeqs
\end{widetext}
where for notational simplicity the symmetry-breaking parameter $\kappa=\mu^4_WL_1^4/g^2$ has been introduced,
and only terms at the leading order in $\kappa$ have been retained.
Notice that the normalization of the $\pi^{\prime}$ terms are not identically $1$, there are small corrections
proportional to $\kappa$. One should  redefine (in a $SU(2)\times U(1)$ invariant way) the normalization of the gauge bosons,
in such a way to eliminate this. But because $\kappa$  is very small anyhow, this process just makes the algebra heavier, without changing
the final results, and hence I am not going to do so.

From these expressions, one reads that, at the leading order in $\kappa$,
\beqs
M_W^2&\simeq&\frac{1}{2}\varepsilon^2g^2\kappa\frac{1}{L_1^2}\,,
\eeqs
and hence that 
\beqs
\hat{S}&=&\frac{3}{16}\varepsilon^2g^2\kappa\,\simeq\,\frac{3}{8}M_W^2L_1^2\,.
\eeqs
Notice that because $M_{\rho}\propto L_1^{-1}$, the expression for $\hat{S}$ agrees with usual expectations
yielding $\hat{S}\propto M_W^2/M_{\rho}^2$.
In order to compute $\hat{T}$ one has to carefully  include sub-leading corrections (in $\kappa$)
in the symmetry-breaking parameters. In particular, the fact derived here that $\pi_+=\pi_{WW}$ is not an exact result, 
but holds only at the leading order in the expansion in $\kappa$. 
The approximations made are precise enough only for $\hat{S}$.
A complete treatment of all precision parameters can be found elsewhere~\cite{AdSTC},
and goes beyond our present scope.

We can take as indicative of the experimentally allowed range
(at the $3\sigma$ level):
$
\hat{S}_{exp}= (-0.9\pm 3.9) \times 10^{-3}$  from~\cite{Barbieri}. These bounds  are extrapolated to
the case of a Higgs boson with mass of $800$ GeV. Form the approximate expression for $\hat{S}$ comes the limit on the
confinement scale $L_1$ of the model:
\beqs
L_1^2&\lsim&
\,\frac{1}{(900\, {\rm GeV})^2}\,.
\eeqs

\subsubsection{LHC phenomenology}

In order to talk about phenomenology, one has to talk about the spectrum of spin-1 states first.
Because the precision parameters can  agree with the data only provided $\kappa \ll 1$,
while the SM gauge couplings are $g_4^2=\varepsilon^2g^2/L\sim{\cal O}(1)$, this means that
the splitting between vector and axial-vector states is going to be very suppressed.
Hence, it suffices to consider the spectrum of the excited states of the 
photon, by reading it from 
\beqs
\pi_{V}&=&{\cal N}^2 \left(D q^2 +\frac{\partial_zv_V}{v_V}(q^2,L_0)\right)\\
&\simeq&q^2\left[1+
\varepsilon \left(\frac{\pi}{2}
\frac{Y_0(qL_1)}{J_0(qL_1)}-\gamma_E-\ln \frac{qL_1}{2} \right)\right]\,,
\eeqs
and this will give a good approximation for the spectrum also of the other states (the excitations of the $Z$ and $W$ mesons).

One might be tempted to think that the spectrum of heavy states be given by the zeros of the Bessel $J_0(qL_1)$.
This is a good approximation for the zeros of $\pi_{V}(q^2)$ when $\varepsilon \ll 1$,
otherwise it is not. Unfortunately, one cannot make the assumption $\varepsilon \ll 1$.
For instance, one sees that  (at large-$q$)
$\pi_V \sim \varepsilon q^2\ln q$
which, if matched brutally to the OPE of large-$N_c$ QCD-like theory suggests that
$\varepsilon \sim N_c/(12\pi^2)$. Most importantly, the large-$N_c$ regime is entered when $\varepsilon \sim {\cal O}(1)$
or larger, as seen by the fact that $g_4=\varepsilon g/\sqrt{L}\sim \varepsilon g_{\r}$. Indeed, because $g_4\sim {\cal O}(1)$,
and because we are using the tree-level (supergravity) approximation, at most one can have $g_{\r}\sim {\cal O}(1)\propto1/\sqrt{N_c}$.

Finding the actual spectrum requires some numerical work.
The result of which is that $M_{\r}L_1 \sim 2.4 - 5.5$, where the upper part of the interval is 
obtained for large values of $\varepsilon$.
Hence the bound on $M_{\r}$ turns out to be quite stringent, and depending on how 
small choices of $\varepsilon$ we allow for, at best $M_{\rho} > 2.2$ TeV, possibly another factor of 2 higher for large values of $\varepsilon$.
This is a problem: these masses are hard to discover at the LHC.  In particular, the second set of resonances will be 
roughly  $M_{\r^{\prime}}\simeq M_{\r}+\pi/L_1$, which for $L_1^{-1}\sim 1$ TeV means a $3$ TeV gap. 
This almost completely excludes the possibility that LHC may
observe several copies, with heavier masses, of this resonance, which would be the unmistakable signature of a strongly-coupled model
(equivalently, these would be the Kaluza-Klein excitation of a compact  extra-dimensions).

There are hence two important open problems: under what circumstances can one see the techni-rho mesons
at the LHC? If the techni-rho is discovered, how can one show that it is related to a strongly-coupled sector?
There are many studies of this type of LHC searches, which depend very strongly on the precise details
of the model (in particular, enhancement of the production cross-section is possible if the top quark is
playing a role in the new strong dynamics). It goes far beyond the scope of these lectures to give a full account 
on the topic.  I want here just to signal two interesting facts that might happen if 
the model is actually well-described by the large-$N_c$ expectations.

\begin{figure*}[ht]
\begin{center}
\includegraphics[width=0.6\linewidth]{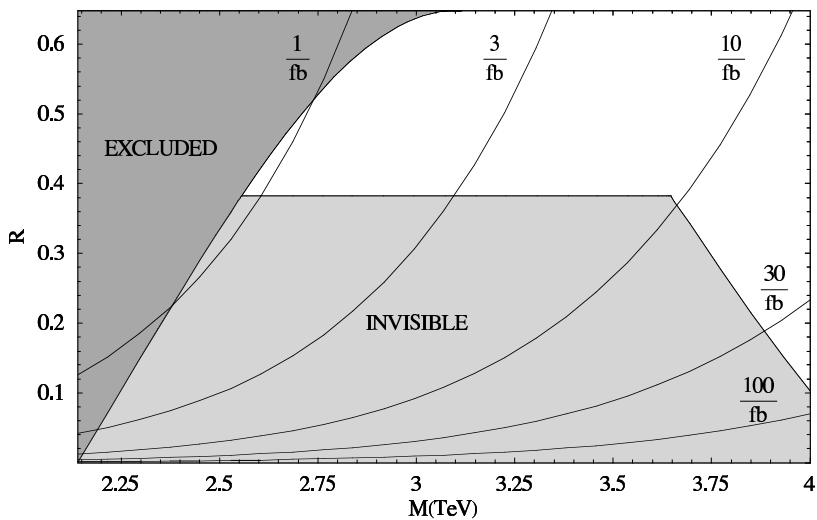}
\caption{LHC exclusion/discovery reach from $pp\rightarrow \mu^{+}\mu^{-}\,+\,X$
as a function of $M_{\r}=M\simeq M_{\gamma^{\prime}}\simeq M_{W^{\prime}}
\simeq M_{Z^{\prime}}$ and of $R$ defined in the text. The curves are 
obtained by requiring $10$ events, for integrated luminosity of $1,3,10,30,100$
 fb$^{-1}$. The darkest region is excluded by indirect limits from $\hat{S}<3\times 10^{-3}$.
The light-grey shaded region is allowed by precision data only for
$\varepsilon < 0.5$, so that the dominant decay mode of
the techni-rho is into longitudinally polarized SM gauge bosons (and hence
the signal in SM fermions strongly attenuated, or {\it invisible}) and the
large-$N$ approximations used in the analysis performed here do not hold.
\label{Fig:exclusions10}
}
\end{center}
\end{figure*}

Using the result highlighted earlier on $\varepsilon = g_4 \sqrt{L}/g \simeq N_c/(12 \pi^2)$, 
the comments we made in discussing the phenomenology of techni-rho mesons
show that for values of $\varepsilon \gsim {\cal O} (1)$, the most important coupling
of the techni-rho is to the SM fermions, not to the pions. This can be checked numerically to hold
by explicitly evaluating the partial width $\Gamma[\rho\rightarrow WW]$ from the five-dimensional set-up~\cite{AdSTC}.
We also stressed that only  for $\varepsilon \gsim {\cal O}(1)$ the calculations performed up to now are trustable.
Hence, in the regime of validity of what we did in the previous subsection,
 the dominant decay mode of the techni-rho
is to SM fermions, not to $WW$. At least as long as we want to trust the calculation of the precision parameter $\hat{S}$.

Notice that all of the above agrees with the large-$N$ result that the resonances should become narrow
in the limit in which $N_c$ is very large, and if the electro-weak couplings are neglected the techni-rho 
becomes effectively stable. In order to study the LHC phenomenology one can use the result
on the vacuum polarizations and look at the LHC process $pp\rightarrow \r \rightarrow f\bar{f}$.
In respect to the decay into $WW$ final states, this process is much cleaner at the LHC, because 
if one focuses attention of the final states involving electrons and muons, this can be reconstructed very precisely,
and there is no missing energy.

The very simplest signature of the model  is 
hence the Drell-Yan production of the heavy partners of the electro-weak gauge bosons.
This should be detectable as a narrow resonance  in the few-TeV region of the invariant mass distribution of 
final states with two fast leptons (muons or electrons).
At the partonic level, the dominant tree-level contribution can be very easily computed
 by simply replacing the complete propagators (inverse of the $\pi(q^2)$ computed earlier) in place of the 
propagators of the SM gauge bosons.
With one caveat: one needs to modify the propagators in order to include the width.
Provided one is in the regime in which the decay to SM fermions dominates such width, and the width itself is small,
one can pragmatically implement the width by replacing in the Feynman diagrams
\beqs
\label{Eq:propagators}
P_{\gamma}(q^2)=\frac{1}{q^2+i\epsilon}&\rightarrow&P_{\gamma}(q^2)=\frac{1}{\pi_V(q^2) + i \frac{8\alpha}{3} q^2 +i\epsilon}\,,
\eeqs
and analog for the $Z$ and $W$.
Notice that this very simple procedure is acceptably accurate only because we assumed that the SM-fermion final state dominates,
and that the fermions be localize at the UV boundary. This result is what follows from taking the imaginary part of 
diagrams in which the five-dimensional photon receives corrections from 1-loop diagrams localized on the four-dimensional boundary.
The procedure for including the width due to strongly-coupled decays is ways more complicated, because not only all the gauge bosons 
propagate in the bulk of the five-dimensions, so that the effective cubic couplings depend on complicated integrals convoluting the 
bulk wave functions, but also heavier resonance  can decay into combinations of lower modes. 

As long as the weak-coupling calculations are acceptably accurate, the partonic cross-section of interest is
\beqs
\hat{\sigma}_{q\bar{q}}(\hat{s})\equiv \sigma(q\bar{q}\rightarrow \ell^+\ell^-)&=&
\frac{\hat{s}}{48\pi}e^2\sum_{A,B}|G_{AB}(\hat{s})|^2\,,
\eeqs
where $A,B=L,R$, where $\hat{s}=q^2$ is the partonic momentum, where
\beqs
G_{AB}(\hat{s})&=&Q^{(q)}P_{\gamma}(\hat{s})+\frac{4}{\sin^22\theta_W}g_A^{(q)}g_B^{(\ell)}P_{Z}(\hat{s})\,,\\
g_L^{(f)}&=&T^{3\,(f)}-Q^{(f)}\sin^2\theta_W\,,\\
g_R^{(f)}&=&-Q^{(f)}\sin^2\theta_W\,,
\eeqs
and where $T^{3\,(f)}$ and $Q^{(f)}$ are the isospin and electric charge of the fermion $f$. 

Finally, the differential cross-section for the LHC process as a function of $s$ and $\hat{s}$  is
\beqs
\frac{\di \sigma(pp\rightarrow \ell^{+} \ell^{-} +X)}{\di \hat{s}}(s,\hat{s})&=&\frac{1}{3}\sum_q\hat{\sigma}_{q\bar{q}}(\hat{s})\
\frac{1}{s}\int_{\hat{s}/s}^1\frac{\di \eta}{\eta}\left[\phi_q(\eta)\phi_{\bar{q}}\left(\frac{\hat{s}}{\eta s}\right)
+\phi_{\bar{q}}(\eta)\phi_{{q}}\left(\frac{\hat{s}}{\eta s}\right)\right]\,,
\eeqs
where the functions $\phi_a(\eta)$ are the parton distribution functions yielding the probability of finding a parton of species $a$ 
with momentum fraction $\eta$ within the proton, where  $\sqrt{s}=14$ TeV is the center of mass energy of the $pp$ collision,
 and where the factor of $1/3$ takes into account the average over
color in the partonic initial state.

One can then do the exercise of using the results at the previous section to select the input parameters such as $L_1$,
 and of computing the number of events expected
at the LHC as a function of the machine parameters (luminosity and energy) and of the relevant parameters in the model,
namely the mass of the techni-rho $M_{\rho}$ and its coupling to the 
SM fermions relative to the coupling of the SM gauge bosons $R$ (which generalizes the factor $e^2f_{\r}^2/M_{\r}^2$ in Eq.(\ref{Eq:weakdecay})),
which is a complicated function of $\varepsilon$.

The results are plotted in Fig.~\ref{Fig:exclusions10}.
A few words of comment. The dark shaded region is excluded by the bound on $\hat{S}$. Notice that 
the actual exclusion is a function not only of the mass of the techni-rho mesons, but also of the coupling $R$,
because the latter depends on $\varepsilon$, and we have seen that there is a non-trivial functional dependence between $M_{\r}$, $L_1$ and $\varepsilon$.
The region at small $R$ is the region in which neglecting the contribution of $\r\rightarrow WW$ to the width and decay rates 
of the techni-rho is not justified. In this region, this  signal is actually very suppressed in comparison to the final state with two gauge-bosons.

Provided $\varepsilon \sim {\cal O}(1)$, the LHC phenomenology is very similar to what one would have in
a weakly-coupled extension of the SM with an extra $SU(2)\times U(1)$.
The four heavy gauge bosons would be almost degenerate in mass. So much so that it would be not
possible to distinguish in the invariant-mass distribution of di-lepton final states
the peaks due to the two neutral states. In order to see that the peak is due to more than
one resonance, much more sophisticated data analysis would be needed, for example looking at the forward-backward asymmetry~\cite{PR}.

\subsection{Final Remarks.}

The results obtained for this simple model depend on the specific set-up,
and many variations exists of the basic ideas.
Before commenting on some very interesting possibilities, it is important to stress that
the way in which the bounds from electro-weak precision is accommodated relies
on the fact that $f_{\rho}$ and $f_{\pi}$ have two different dynamical origins.
To be more specific, the former is controlled by the IR boundary $L_1$, the latter
by the value of the condensate ${\rm v}_0$, which in this model are treated as independent.
Both scale as $f_{\pi}^2\propto f_{\rho}^2 \propto N$, but the smallness of the 
$\hat{S}$ parameter is obtained by assuming the ratio of the two  be a somewhat small numerical constant.
For this reason, one is allowed to take the large-N limit, and hence treat the spin-1 states as weakly coupled, 
while at the same time making their masses large so that  precision physics bounds do not exclude the model.

There is a number of possible problems arising with this approach.
First, in a fully dynamical model, $f_{\pi}$ and $f_{\rho}$ will be related, 
and there is no known  simple example in which their ratio be parametrically  small.
Second (to be taken  with all the caveats we discussed earlier on) 
if the mass of the techni-rho mesons is in the few TeV range,
 it is not possible to show explicitly that
the elastic scattering amplitude can be treated perturbatively 
and stays below the unitarity bounds all the way up to this large mass scale.
As such, at least some part of the theory must be strongly-coupled, which raises questions
about the tree-level calculations been done. But remember how we stressed that
the $WW$ scattering amplitude is a particularly bad observable, much more sensitive 
to the details of the strongly-coupled part of the symmetry-breaking sector than 
the oblique parameters.

Finally, if one takes literally the relation  $\varepsilon \sim N_c/(12\pi^2)$,
values of $\varepsilon \sim {\cal O}(1)$ require very large values of $N_c$.
This means that a large number of fermions in the dual sector is coupled to the electro-weak gauge bosons,
making the $SU(2)$ gauge coupling blow-up at relatively small scales, not far above the  scale set by $L_1$.
This problem is indirectly related to the perturbative-unitarity bound, and also to the fact that if one looks at the ${\cal O}(q^4)$
precision parameter $W$ and $Y$ one finds that they grow with $\varepsilon$.
Yet, this model is simple, calculations are doable, the results can be reconciled with precision
electro-weak tests, and the model is testable a the LHC, which is ultimately the most important thing.
If such a signature shows in the data, there will be plenty of time to go back to this model and
refine its study, addressing the objections raised here.

Infinite number of variations of this model are possible. Here is a brief and non-exhaustive list of 
things that can be done (and have been done).
\begin{itemize}
\item Custodial symmetry can be implemented, by gauging a complete $SU(2)\times SU(2)$ in the bulk,
and then arranging for its breaking via some mechanism localized in the UV, in such a way as to recover 
a spectrum in which only four gauge bosons are light, but in which $\hat{T}$ is small.
\item The bulk profile of the condensate can be changed. The extreme case is the higgless extra-dimension set-up,
in which the field $\Phi$ is exactly localized on the IR boundary, effectively modifying the IR boundary conditions~\cite{Higgsless}.
\item More condensates can be present. In particular one may envision the case in which 
the bulk scalars affect also the dynamics of the vectorial five-dimensional fields, not only the axial-vectorial fields.
Also, phenomenologically more realistic models of confinement can be obtained, by introducing modifications of the
background AdS metric in the deep IR~\cite{AdSTC2,softwall}.
\item A larger symmetry can be gauged in the bulk, hence allowing the low energy spectrum to contain
also pseudo-Goldstone bosons. An extreme variation of this model is dual to composite Higgs models~\cite{compositeHiggs}.
\item The fermions can be allowed to propagate in the bulk. In particular the third-family, such as the top.
In this way it is possible to enhance the large mass of the top~\cite{5Dfermions}.
\end{itemize}

This huge freedom shows one of the limitations of these models: because they are not fully dynamical,
the description they provide has a limited predictive power.
However, for many practical purposes they are very effective: they allow to relate in sensible and predictive ways
the results from precision studies of the electro-weak gauge bosons to the spectrum, 
couplings and LHC phenomenology of the 
new particles they predict (new vector states, new axial-vector states, new pseudo-Goldstone bosons, \dots).
These models are hence very useful, in the dawn of the LHC era, because they give us a plethora of new 
signatures to look for, within calculable frameworks in which the comparison to the data can be made quantitative,
cross-relations between observables exist, and favorable regions of parameter space can be identified and tested experimentally.

\newpage
\section{Holographic technicolor: top-down approach.}

\subsection{Why a full string-theory  model?}

We have seen that, because of the strong coupling,
 four-dimensional techniques are  limited  
 when applied to technicolor models. We have also seen that five-dimensional models provide a substantial 
 improvement over hidden local symmetry, incorporating some of its ideas and features,
 but reducing drastically the number of parameters needed to describe 
 the spectrum of composite states, and hence allowing to compute the precision parameters and relate them to measurable 
 quantities, such as masses and decay constants of spin-1 resonances.
 With the LHC approaching, this framework provides a wonderful opportunity 
 to study the phenomenology of the lightest resonances of a strongly-coupled model,
 relating their masses and couplings to the precision measurements performed in the last two decades.
 However, the bottom-up approach  has itself a limited power: it is based on assuming that the answer to many of the questions
 highlighted in subsection~\ref{Sec:questions} is known a priori, and then uses 
 these answers in order to build the five-dimensional action.
 
 To be more specific, let us look back at each of the questions and 
 see what can be said from the bottom-up approach.
 \begin{itemize}
\item[i)] The  bottom-up approach describes only bound states, and their low-energy properties, 
 hence it has only indirect information about the fundamental theory (its gauge group and field content are not known).
 \item[ii)] The dynamical scales are put in by hand, choosing the position of the boundaries, the form of the metric, 
 the presence of condensates.
 \item[iii)] The anomalous dimensions are chosen by hand, they are completely free parameters, bounded only by very general considerations.
 \item[iv)] The precision parameter can be computed. However, they still depend on how the light SM fermions are coupled
 to the strong sector, and a certain amount of freedom is left.
 \item[v)] Computing the coefficients of the chiral Lagrangian is possible, as a function of what 
 said in ii). The only limitation being that one is using large-$N$ expansions.
 \item[vi,vii,viii)] This approach has little  to say about ETC,  the fermion mass hierarchies are controlled by free parameters,
 not by the underlying dynamics. 
 \item[ix)] One can derive relations between the spectrum and couplings of spin-1 states,
 as a function of the choices in points ii) and iii). 
 \item[x)] The global symmetries are chosen by hand, and their breaking pattern too, there is no real sense in which 
 symmetry-breaking arises dynamically within this framework.
 \item[xi)] Determining under what conditions a light dilaton is part of the spectrum
 requires a dynamical study of the underlying theory, with fully back-reacted background geometry.
 \item[xii)] Perturbative unitarity is in part connected with sum rules on the couplings of spin-1 fields,
 yet the dominant contributions from the non-decoupled scalar sector is  missing and uncalculable.
 \end{itemize}

Ultimately, the bottom-up approach yields a very major improvement over hidden local symmetry,
and is a very useful and practical way of constructing models that are testable at the LHC.
But it cannot answer to those most fundamental questions regarding dynamical scales, spectrum and symmetries 
that are intrinsically dependent on the strong dynamics.
For this reason, it is important to try and study the extra-dimension dual of the
full theory, in a context in which the background and all the condensates can be computed 
from a fundamental action.
This is the topic of this last section.

\subsubsection{From ${\cal N}=4$ towards walking technicolor.}

The most celebrated example of gauge/gravity duality 
is the conjectured correspondence between four-dimensional, superconformal 
${\cal N}=4$ Yang-Mills theory with $SU(N_c)$ gauge group and Type IIB string theory on a background with
AdS$_5\times S^5$ geometry~\cite{AdSCFT}.
The $SO(4,2)$ isometry group of the AdS space is related to the four-dimensional conformal group,
while the $SO(6)$ symmetry of the internal $S^5$ space is related to the $SU(4)_R$ symmetry of ${\cal N}=4$.
The value of $N_c$ is related to the flux of the RR $F_5$ form of type-IIB.
One important reason why this relation is very useful is that
taking the large-$N_c$ limit, at fixed (but large!) 't Hooft coupling $\lambda=g^2 N_c$ 
one enters the regime in which the dual description reduces to a weakly-coupled 
supergravity theory in 10 dimensions.
Hence, very non-trivial calculations involving strong dynamics in the gauge theory can be rephrased
in terms of a more accessible weakly-coupled gravitational system.

For phenomenological purposes,  the AdS$_5\times S^5$ background has ways too much symmetry: realistic models
of nature are neither conformal, nor do they have so much supersymmetry.
In the last ten years, the same idea of looking for weakly-coupled duals to strongly coupled four-dimensional theories
has been pushed in the direction of looking for sensible supergravity backgrounds  which 
 depart from the AdS geometry, and in which a lesser degree of supersymmetry 
is encoded in a less symmetric structure of the internal five-dimensional manifold.
Very interesting models exist, proposing for instance backgrounds with ${\cal N}=1$ supersymmetry (such as~\cite{MN,KS}),
or no supersymmetry at all~\cite{WSS,PS}, in which the geometry is far away from AdS, and describes the dual of 
a theory in which all the couplings run. 
There also exist 10-dimensional models where the geometry is well approximated by an AdS space in the far UV and by a different AdS geometry 
deep in the  IR (at large and small values of the radial coordinate, respectively). These models provide a description
of a four-dimensional theory  which flows from a UV fixed-point to an IR fixed point~\cite{flow}.

Many very interesting aspects of strongly-coupled field theories can be studied 
using the techniques of gauge/gravity dualities. The physics near fixed points (in particular,
the calculation of anomalous dimensions) can be  studied quantitatively. 
Correlation-functions can be computed. Certain RG flows can be studied in details.
New and effective way of describing confinement and the formation of 
symmetry-breaking condensates exist. A huge amount of formal work has been done in order to 
put all of these ideas on firm grounds, by producing rigorous prescriptions for
performing the calculation of physical quantities of interest in the dual field theory
starting from the supergravity (or superstring) framework (for instance, holographic renormalization is  
a precise prescription for computing correlation functions of local operators~\cite{HR}, while
the use of probe-strings provides a tool for the calculation of the Wilson loops, that I will discuss in some detail later).

The reader will immediately realize that many of these observations
are (in subtle ways) related to the
 very large body of very hard field theory questions
that emerged in the context of dynamical electro-weak symmetry breaking
in the first two sections of these lecture notes.
Hence, it is very natural to try and reformulate some of those questions 
in terms of  dual theories, and look for the answers in this (weakly-coupled) context.

In a perfect world, one might even hope to rewrite the whole
theory of WTC/ETC in terms of its dual 10-dimensional description.
This putative theory should be asymptotically AdS, hence ensuring the
UV-completeness of the symmetry-breaking sector of the SM,
which would approach a fixed point (not necessarily weakly coupled) in the UV.
Relevant deformations would cause the theory to flow away from the UV fixed point,
and undergo a set of symmetry-breaking transitions (tumbling), 
effectively reducing the gauge group of the dual ETC theory down to TC,
and in the process producing all the four-fermion operators needed to generate the 
SM fermion masses~\footnote{Something vaguely reminiscent of this scenario exists 
in the supergravity context~\cite{KS,cascade}, although the superficial similarities between tumbling ETC and the cascade
are to very large extent overshadowed by the important differences.}.
At the end of this tumbling process, one should find that the supergravity background 
has a geometry which is approximately AdS, providing a dual description of walking,
and could compute all the anomalous dimensions, and the running of the 
aforementioned higher-order operators, hopefully showing that enhancement factors are generated
along the lines of what we discussed in the second section.
This geometry should also have a global symmetry containing the SM $SU(2)_L\times U(1)_Y$
gauge group.
After walking for a while, the dual supergravity geometry should 
substantially deviate from AdS, to signify the fact that the IR fixed point 
the theory was approaching is only approximate.
At low energies a set of condensates should form, signaled by the breaking of the internal global symmetry,
which corresponds to EWSB in the dual theory.
Finally, there should be an end of space in the radial direction
in the IR, and the properties of the geometry near this end of space should be 
healthy enough to allow for a sensible interpretation of it in terms of confinement.
The weak gauging of the $SU(2)_L\times U(1)_Y$ subgroup would produce
the massive gauge bosons $W$ and $Z$, and the presence of higher-order operators enhanced by walking 
produce the SM fermion masses, but not unacceptably large new sources of FCNC transitions.
In principle, given such a complete theory one would be able to compute all possible observables of relevance
for phenomenology: spectra, couplings, scattering amplitudes, production and decay rates of all the possible 
composite objects in the theory, and their contribution to precision electro-weak parameters and FCNC transitions.

Of course, the real world is far from perfect: such a construction does not exist (yet),
and much of what I wrote in the previous paragraph is  wishful thinking,
with limited scientific support.
But encouraging steps in this direction have been taken~\cite{NPP,ENP,NPR,Mintakevich:2009wz}.
In this final section, I spend some time working out in some details 
a few examples of these steps.
The discussion is far from  systematic and pedagogical. The reader is assumed to be familiar with
the basic ideas of gauge/gravity dualities (if you are not, a good place to start
is~\cite{reviewAdSTC}), since  part of the discussion will be rather technical,
and limited  to one very specific system, 
that exemplifies what can be done with these powerful 
techniques.

Before beginning along these lines, a clarifying comment, anticipating some
of the ideas to come.
The model(s) I am about to discuss are not, strictly speaking, technicolor models,
because they do not implement electro-weak symmetry breaking.
Yet, they are used here as illustrations of what the dynamics of walking technicolor might 
yield, because these dynamical systems are interpreted in terms of a dual four-dimensional theory
that exhibits some of the features that are assumed to be at the basis of walking technicolor.
The notion of walking itself is a field-theory property that not necessarily requires 
coupling to the electro-weak sector. Confinement, and the formation of symmetry-breaking condensates,
are well-defined concept even outside the context of electro-weak interactions.
And many of the questions we are interested in, when talking about strongly-coupled EWSB,
can be formulated in field-theory terms that do not necessarily require EWSB itself.
For these reasons, the exercises we are going to do are going to teach us some very important lesson 
about walking technicolor, even if none of the  models illustrated is the actual dual of a technicolor model.

\subsection{Wrapped-$D5$ system.}

The set-up we focus on is based 
on the geometry produced by stacking on top of each other $N_c$ 
$D5$-branes that wrap an
$S^2$ inside a CY3-fold and then taking
the strongly-coupled limit of the gauge theory on this stack,
in the (type-IIB) supergravity approximation. 
We start by recalling the basic definitions 
that yield the general class of backgrounds 
obtained from the $D5$ system, 
which includes~\cite{MN} as a very special case.
Truncating type-IIB supergravity to include only gravity, dilaton $\Phi$ and  RR 3-form $F_3$ the 10-dimensional action is
\beq
S_{IIB}=\frac{1}{G_{10}}\int d^{10}x \sqrt{-g}\Big[
R-\frac{1}{2}(\partial\Phi)^2 -\frac{e^{\phi}}{12}F_3^2  \Big],
\eeq

Defining the $SU(2)$ left-invariant one-forms as 
\bea\lab{su2}
\tilde{\w}_1\,=\, \cos\psi d\tilde\theta\,+\,\sin\psi\sin\tilde\theta
d\tilde\phi\,\,,\,
\tilde{\w}_2\,=\,-\sin\psi d\tilde\theta\,+\,\cos\psi\sin\tilde\theta
d\tilde\phi\,\,,\,
\tilde{\w}_3\,=\,d\psi\,+\,\cos\tilde\theta d\tilde\phi\,\,,
\eea
and using an ansatz
which assumes the functions appearing in the background depend 
only the radial coordinate $\r$ (but not on the Minkowski coordinates $x^{\mu}$ 
nor the 5 angles $\theta,\tilde{\theta},
\phi,\tilde{\phi},\psi$ of the compact internal manifold)
allows to write the background (in string frame) as
\bea
ds^2 &=& \alpha' g_s e^{ \Phi(\rho)} \Big[\frac{dx_{1,3}^2}{\alpha' g_s} +
e^{2k(\rho)}d\rho^2
+ e^{2 h(\rho)}
(d\theta^2 + \sin^2\theta d\phi^2) +\nonumber\\
&+&\frac{e^{2 {g}(\rho)}}{4}
\left((\tilde{\omega}_1+a(\rho)d\theta)^2
+ (\tilde{\omega}_2-a(\rho)\sin\theta d\phi)^2\right)
 + \frac{e^{2 k(\rho)}}{4}
(\tilde{\omega}_3 + \cos\theta d\phi)^2\Big], \nonumber\\
F_{3} &=&\frac{N_c}{4}\Bigg[-(\tilde{\omega}_1+b(\rho) d\theta)\wedge
(\tilde{\omega}_2-b(\rho) \sin\theta d\phi)\wedge
(\tilde{\omega}_3 + \cos\theta d\phi)+\nonumber\\
& & b'd\rho \wedge (-d\theta \wedge \tilde{\omega}_1  +
\sin\theta d\phi
\wedge
\tilde{\omega}_2) + (1-b(\rho)^2) \sin\theta d\theta\wedge d\phi \wedge
\tilde{\omega}_3\Bigg].
\label{nonabmetric424}
\eea
The full background is then determined by solving the equations of motion for the 
functions $(a,b,\Phi,g,h,k)$.

The system of BPS equations derived using this ansatz 
can be rearranged in a convenient form, by rewriting the
functions of the background in terms of a set of functions $P(\rho),
Q(\rho),Y(\rho), \tau(\rho), \sigma(\rho)$ as~\cite{HNP}
\beq
4 e^{2h}=\frac{P^2-Q^2}{P\cosh\tau -Q}, \;\; e^{2{g}}= P\cosh\tau -Q,\;\;
e^{2k}= 4 Y,\;\; a=\frac{P\sinh\tau}{P\cosh\tau -Q},\;\; N_c b= \sigma.
\label{functions}
\eeq
Using these new variables, one can manipulate the BPS equations to obtain a
single decoupled second order equation for $P(\rho)$, while all other functions are
obtained from $P(\rho)$ as follows:
\bea
& & Q(\rho)=(Q_0+ N_c)\cosh\tau + N_c (2\rho \cosh\tau -1),\nonumber\\
& & \sinh\tau(\rho)=\frac{1}{\sinh(2\rho-2\hat{\rho_0})},\quad \cosh\t(\r)=\coth(2\r-2\hat{\r_0}),\nonumber\\
& & Y(\rho)=\frac{P'}{8},\nonumber\\
& & e^{4\Phi}=\frac{e^{4\Phi_o} \cosh(2\hat{\rho_0})^2}{(P^2-Q^2) Y
\sinh^2\tau},\nonumber\\
& & \sigma=\tanh\tau (Q+N_c)= \frac{(2N_c\rho + Q_o + N_c)}{\sinh(2\rho
-2\hat{\rho_0})}.
\label{BPSeqs}
\eea
The second order equation mentioned above reads
\beq
P'' + P'\Big(\frac{P'+Q'}{P-Q} +\frac{P'-Q'}{P+Q} - 4 
\coth(2\rho-2\hat{\rho}_0)
\Big)=0.
\label{Eq:master}
\eeq
In the following I will fix the integration constant $Q_0=-N_c$, so that no singularity appears in the function $Q(\r)$. 
When convenient, I will  also choose $\hat{\r}_o=0$, together with $\alpha^{\prime}g_s=1$, in order to simplify 
the notation. 

\subsubsection{Gauge coupling}

As we saw, one clear signal of a walking theory is the existence of an energy range
over which the $\beta$-function is anomalously small, and hence the gauge coupling 
does not run. One needs hence to relate the four dimensional gauge coupling
of the dual theory to quantities that are well defined in the 10-dimensional geometry.
I will keep the factors of
$\alpha'$ and $g_s$ explicit, but set $\hat{\rho}_o=0$ in this subsection.

The six-dimensional theory on
the D5 branes has a 't Hooft coupling given by the dimensionful $\lambda_6=g_s \alpha' N_c$,
and the supergravity limit is taken by keeping this fixed~\cite{Itzhaki:1998dd}.
The  branes wrap a small two-cycle $\Sigma_2$,
so that at
low energies an effectively  four-dimensional theory emerges. A natural way of
defining its (dimensionless) gauge coupling  is given by combining
\beqs
g_{4}^2&\equiv& \frac{g_{6}^2}{Vol \Sigma_2}.
\eeqs
Following~\cite{VLM}, that consider a  five brane (in probe approximation)
extended along the Minkowski directions and the two-cycle defined by
\beq
\Sigma_2=[\theta=\tilde{\theta},\;\;
\phi=2\pi-\tilde{\phi},\;\;\psi=\pi],
\label{2cycle}
\eeq
 one arrives to
\beq
\frac{8\pi^2}{g_4^2 }= 2 [e^{2h}+\frac{e^{2g}}{4}(a-1)^2]= P e^{-\tau}\,=\,\frac{P}{\coth \r}.
\label{coupling}
\eeq
With this result, together with a particular radius-energy relation, it was shown in
\cite{VLM} that one particular solution $\hat{P}$ of Eq.~(\ref{Eq:master})
leads to reproducing  the NSVZ beta function. I will sketch in part this result later on.

Before proceeding, four final comments are due.
A technical one first.
The five-branes on the
submanifold $R^{1,3}\times \Sigma_2$ preserve supersymmetry only if the probe brane is at
infinite radial distance from the end of the space (i.~e. when $\rho\to \infty$)~\cite{Nunez:2003cf}.
This result is valid for the particular solution considered there and does
not necessarily extend to the other  backgrounds, because  a five brane in the
configuration described above does not preserve the same spinors as the
background itself.

More about the physics.
One has to keep in mind that the expansion yielding the supergravity approximation
involves the six-dimensional coupling $g_6$, and not
the four-dimensional coupling defined here.
When $g_4$ becomes small (and it does in the far UV at $\rho\rightarrow +\infty$, as we will see),
this does not necessarily mean that the supergravity approximation is breaking down,
nor that perturbation theory in the dual gauge theory is useful. At large-$\rho$,
 in a putative calculation done in terms of four-dimensional degrees of freedom, one should include
a large number of excited modes living on the  $\Sigma_2$,
and this will compensate for the small coupling of each individual one, 
ensuring that the four-dimensional dual system is indeed strongly coupled.

There will also be divergences at finite values of $\r$. 
In particular at $\r=\hat{\r}_o$. This is in general problematic, 
because the singularity might mean that
  unphysical results 
appear in various amplitudes, and ultimately
the supergravity approximation breaks down,
spoiling predictivity.
However, I will concentrate on backgrounds where the only singularity 
is at $\r\rightarrow \hat{\r}_0$, and for which the Ricci curvature $R$ be finite everywhere.
While per se this does not ensure that the background defines a fully  sensible theory,
a pragmatic
 way of testing if this is the case is to compute actual physical quantities,
and to check how they behave. In particular, we will see that the IR-divergence is associated 
with confinement, via the study of Wilson loops and the related quark-antiquark potential.

A final, more general comment.
Given a walking theory, its coupling must be approximately constant over some energy range.
However, away from an exact fixed point, the very definition of gauge coupling is
ambiguous, and affected by scheme-dependence.
One might worry that the presence of a plateau in the gauge coupling defined here,
especially in considerations of all the caveats that go into its derivation,
might be just a scheme-dependent artifact.
These arguments apply to  any possible definition of gauge coupling in any 
theory away from the fixed points, and ultimately derive from the fact that in a generic field theory
the gauge coupling per se is not a very well defined observable quantity.
The fact that there are solutions in which the gauge coupling flattens at finite values
has to be understood as a first indication of the fact that something very peculiar is happening 
to the theory. In particular, it indicates the existence of two possibly distinct dynamical scales at which the behavior of the theory
changes drastically.
In order to understand how physical this is, one has to use the background to compute actual physical quantities, 
(such as scattering amplitudes or masses)
and ask if they show an interesting change in behavior related to the scales indicated by the gauge coupling.

\subsubsection{Regular Maldacena-Nunez. Asymptotic behaviors.}

Solving the equation for $P$ is not easy. As apparent, Eq.~(\ref{Eq:master}) is very non-linear.
And potentially plagued by possible nasty divergences arising from
the denominators.
One needs to use  approximate and/or numerical 
methods in order to learn about the properties of its solutions. 
A couple of very general results are useful in doing so.

First, it is known that there exist two different classes of 
UV-asymptotic solutions for $P$ valid at large-$\r$~\cite{HNP}:
\beqs
\label{Eq:classI}
P&\sim&2N_c \r \,+\,{\cal O}(e^{-4\r})\,\,\,\,{\rm (class \,I)}\,,\\
P&\sim&{\cal O}(e^{4/3\r})\,+\,{\cal O}(e^{-4/3\r})\,+\,{\cal O}(e^{-8/3\r})\,\,\,\,{\rm (class\, II)}\,.
\label{Eq:classII}
\eeqs
Notice that both diverge with $\r$, implying that the gauge coupling $g_4$ is becoming small,
which we already commented about in the previous subsection.

Second, the simplest and better understood solution of 
Eq.~(\ref{Eq:master}), which is the only non-trivial exact solution known in closed form,
is~\cite{MN}
\beqs
\label{Eq:MN}
\hat{P}&=&2 N_c \r\,.
\eeqs
It belongs to class I, and we already mentioned it earlier on.
For this solution, the gauge coupling is
\beqs
\hat{\lambda}&=&\frac{\hat{g}_4^2 N_c}{8\pi^2}\,=\,\frac{N_c\coth\r}{\hat{P}}\,=\,\frac{\coth\r}{2\r}\,.
\eeqs

Identifying 
\beqs
\r&=&\frac{3}{2}\ln \mu\,,
\eeqs
with $\mu$ the renormalization scale in units of a reference scale, and expanding at large $\r$
yields
\beqs
\hat{\lambda}&\simeq&\frac{1}{3\ln\mu}\,,
\eeqs
which agrees with Eq.~(\ref{Eq:qcd}) at asymptotically large $\mu$, provided $b_0=3N_c$, and $N_f=0$
as in Super-Yang-Mills (SYM) theory with $SU(N_c)$ gauge symmetry. 
This is one simple way to see that  this background is  related (in a non-trivial way) to
the dual to SYM (more precise statement about the dual theory
can be found in~\cite{Andrews:2006aw}).

\subsubsection{Walking solutions in class II.}

\begin{figure}[h]
\begin{center}
\begin{picture}(300,200)
\put(0,0){\includegraphics[height=7cm]{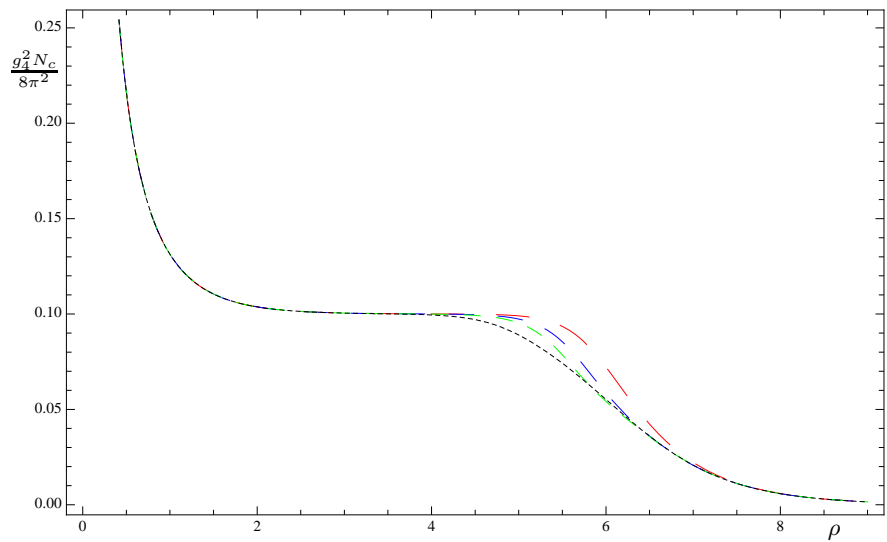}}
\put(-10,173){$\frac{g^2_4N_c}{8\pi^2}$}
\put(300,0){$\rho$}
\end{picture} 
\caption{The 't Hooft coupling $g^2_4N_c/(8\pi^2)$
as a function of $\rho$  for $N_c=10$,  $c=100$, $\a=0.0005$ for (iii). The red (long dashes) curve
is the $\co(c)$ approximation in the expansion of Eq.~(\ref{solution}), the blue (medium dashes) line is the $\co(1/c)$ approximation,
the green (short dashes) line is the $\co(1/c^3)$ approximation, and the black (dotted) line is the numerical solution~\cite{NPP}.}
\label{Fig:plotcoupling}
\end{center}
\end{figure}

\begin{figure}[h]
\centering
\begin{picture}(350,220)
\put(0,120){\includegraphics[height=3.5cm]{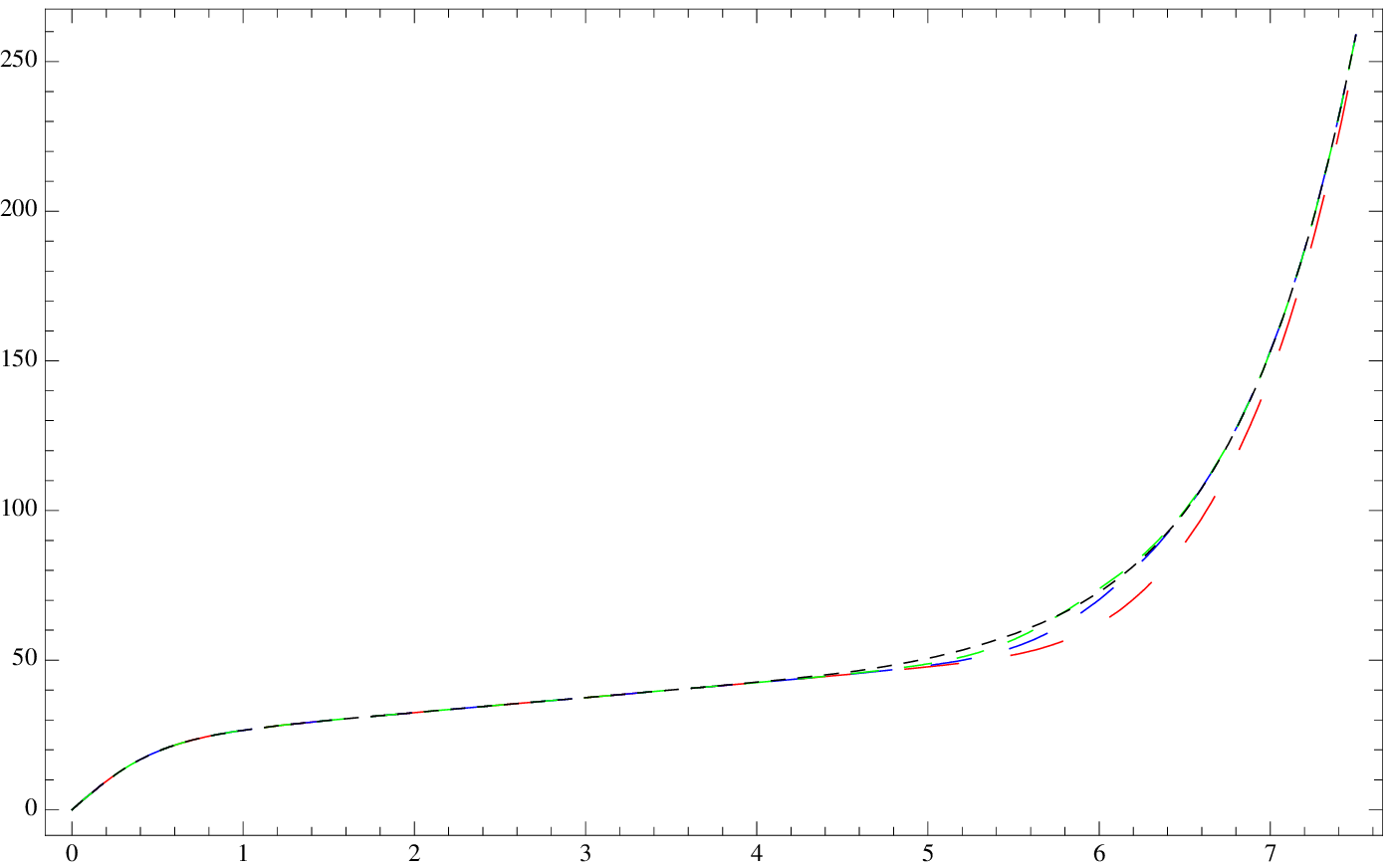}}
\put(190,120){\includegraphics[height=3.5cm]{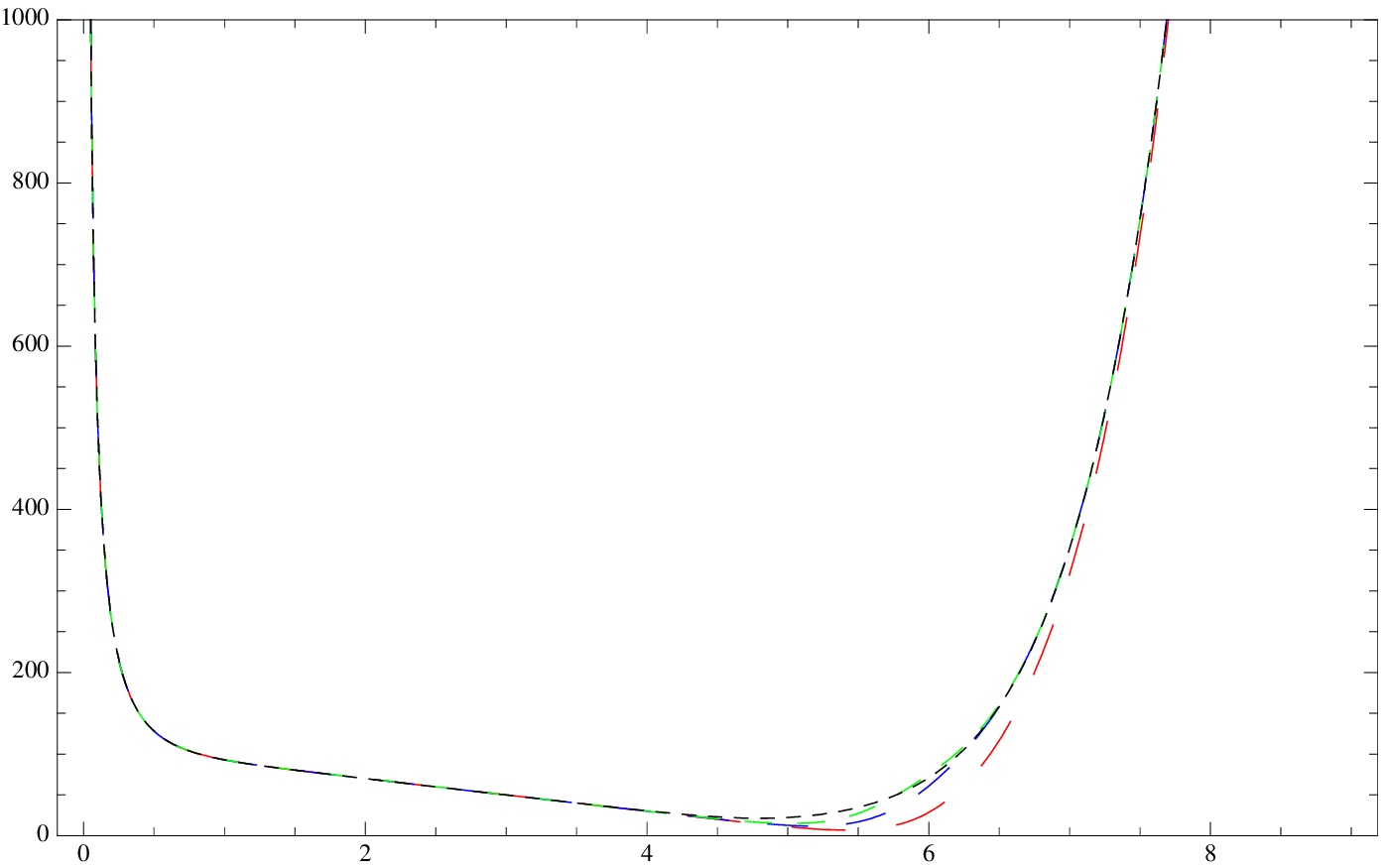}}
\put(0,0){\includegraphics[height=3.5cm]{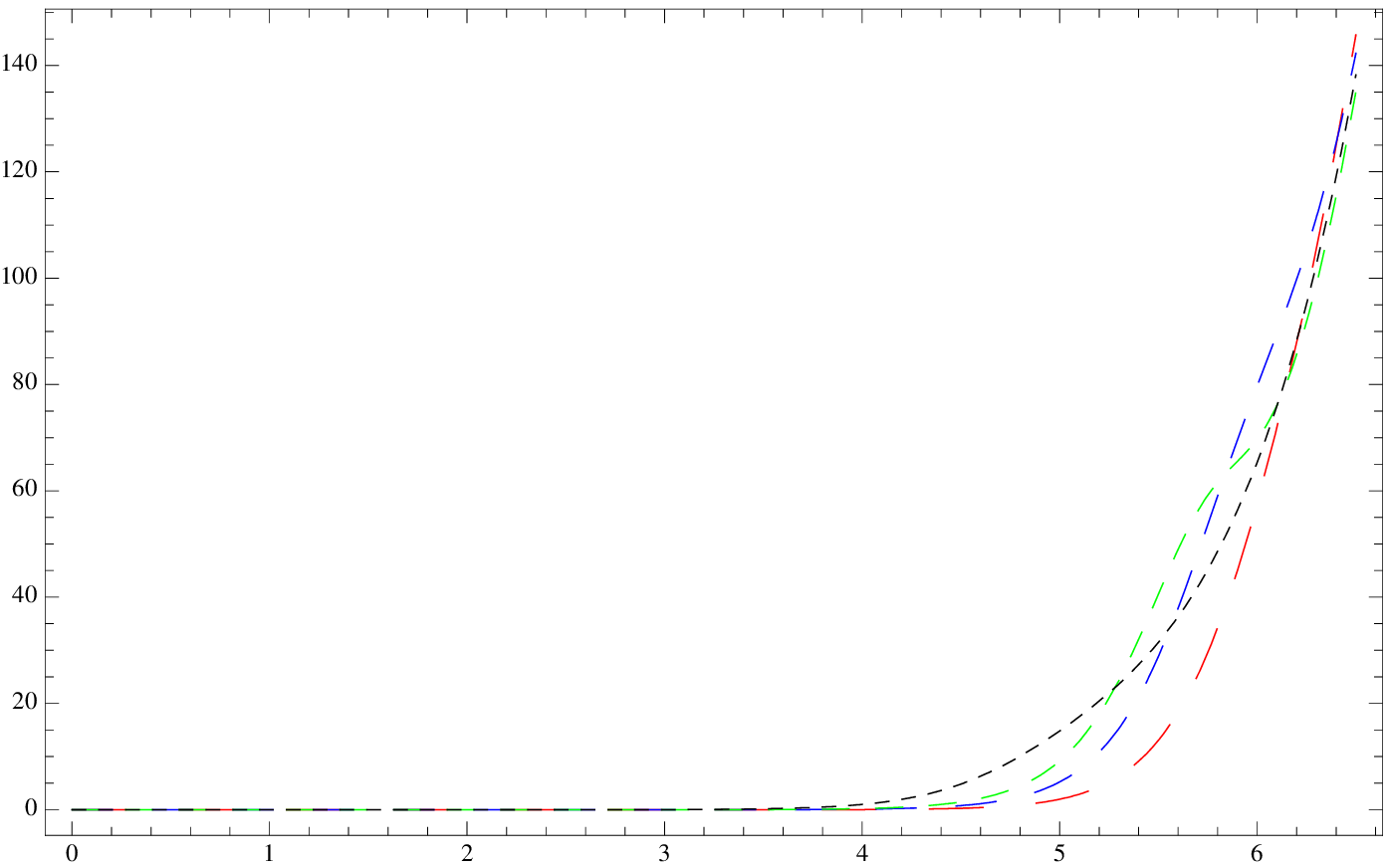}}
\put(190,0){\includegraphics[height=3.5cm]{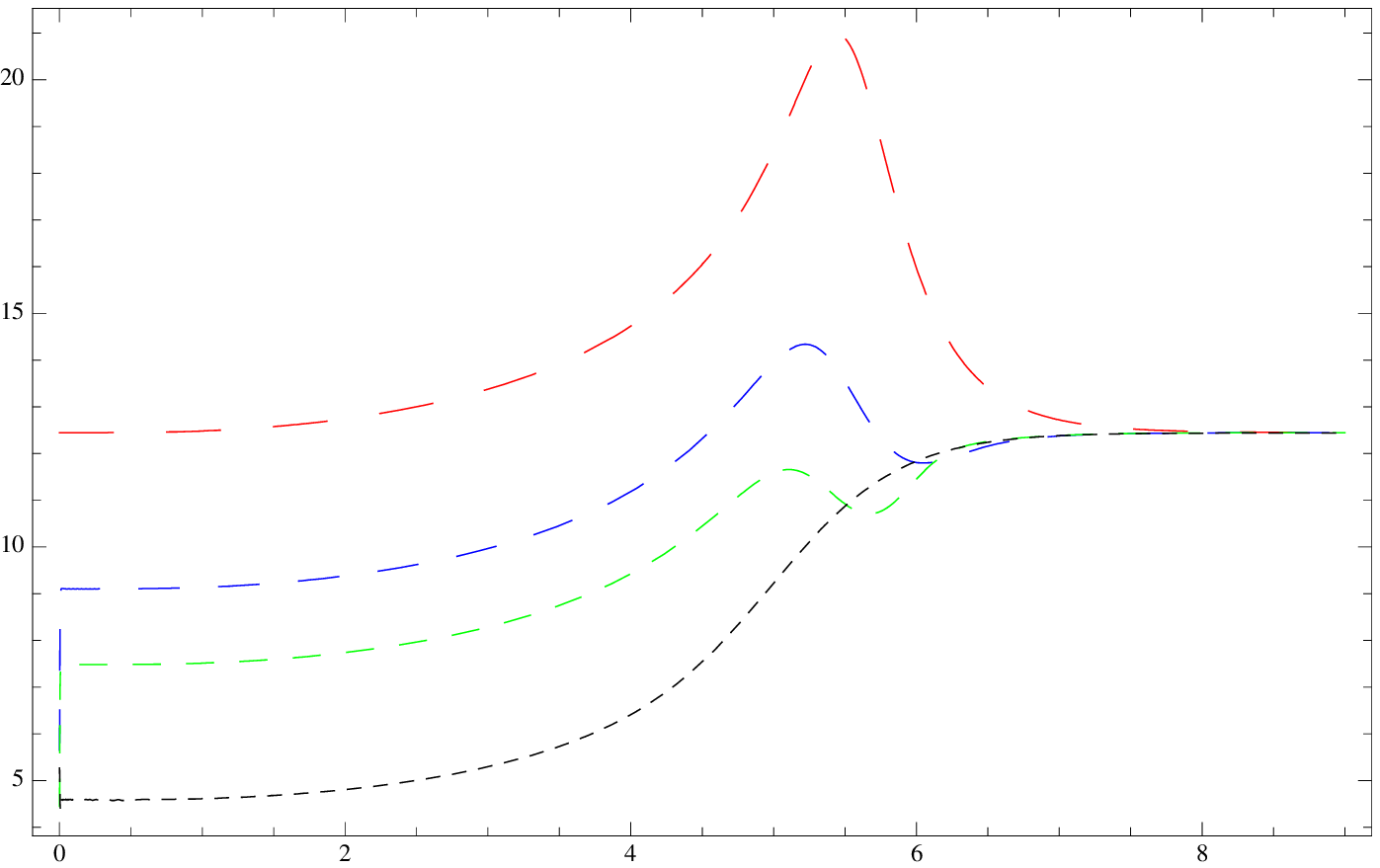}}
\put(-12,205){$e^{2h}$}
\put(180,205){$e^{2g}$}
\put(-12,80){$e^{2k}$}
\put(170,80){$e^{\Phi-\Phi_0}$}
\put(330,-6){$\rho$}
\put(140,-6){$\rho$}
\put(330,114){$\rho$}
\put(140,114){$\rho$}
\end{picture} 
\caption{The background functions $h$, $g$, $k$ and $\Phi$ as a function of $\r$, for the same parameters as in Fig.~\ref{Fig:plotcoupling}, and with the same color-coding.  It is clear that the expansion in Eq.~(\ref{solution}) 
 converges sufficiently fast for $g$ and $h$, but some caution has to be used with $\Phi$ and $k$~\cite{NPP}.}
\label{Fig:background}
\centering
\end{figure}

The first class of solutions to Eq.~(\ref{Eq:master}) which yield the running gauge coupling of a walking theory
was found in~\cite{NPP}, and
belongs to class II. This is a 2-parameter family of solutions, the parameters being the integration constants of the second-order 
differential equation. It has the nice property that it can be obtained 
starting from an approximate solution, and that all corrections can be in principle computed in a formal power-series, 
in which the expansion parameter is related to one of the two integration constants.
Without loss of generality I will set $\hat{\r}_o=0$ in the following. One must require that 
the solution $P(\r)$ exists and be finite for every $\r>\hat{\r}_o$, which  imposes some restriction
on the integration constants.

Let us see how to construct such solutions.
The basic observation  is that Eq.~(\ref{Eq:master}) would be exactly solvable if one  could set $Q=0$.
It is hence sensible to look for solutions such that $P\gg Q$ for any $\r$.
Following \cite{HNP} we can write $P$ in a formal expansion in inverse powers of $c$ as~\cite{NPP}
\beqs\label{solution}
P&=&\sum_{n=0}^\infty c^{1-n} P_{1-n}.
\eeqs
where
\be
R(\r)\equiv\left(\cos^3\a+\sin^3\a(\sinh(4\r)-4\r)\right)^{1/3},
\ee
and where $c$ and $\a$ are the two integration constants.
Taking $c$ large compared to $N_c$ yields $P\gg Q$, and  one
obtains the approximate solution
\beqs\label{approx_sol}
P&\simeq& P_1 =  c R(\r).
\eeqs

For this to be a well defined solution one needs to ensure that $P(\r)> Q(\r)$ for all $\r\geq 0$,
since any $\r$ where $P=Q$ is a singular point of Eq.~(\ref{Eq:master}).
At asymptotically  small values of $\r$, 
$Q(\r)=\co(\r^2)$, while $P(\r)\approx c\cos\a+\co(\r^3)$ and so $P>Q$ is ensured by requiring $\cos\a>0$.
For asymptotically  large $\r$, $Q(\r)\sim 2 N_c \r$, while $P(\r)\sim 2^{-1/3} c \sin\a e^{4\r/3}$.
So that requiring that $\sin\a >0$ is again sufficient to ensure that $P>Q$ for large $\r$. 
Because $P$ and $P^{\prime}$ are monotonically increasing functions of $\r$,
$P\geq c\cos\a$ for all $\r$. At some special value of $\r=\r_*$,  $\sinh(4\r_*)-4\r_*\approx \cot^3\a$ and $P$ starts deviating
from the constant value $c\cos\a$, being approximated by the UV-asymptotic exponential dependence on $\r$. 
In order to allow for a large region of walking behavior we  need to take
$\r_{\ast} \gg 1$, in which case
\be
\r_*\approx \frac14\left(\log 2+3\log\cot\a\right)\,\gg\,1.
\ee
In order to ensure that $P>Q$ everywhere, it is hence sufficient to require that $P(\r_*)> Q(\r_*)$, which
puts an upper bound on $\cot\a$, namely
\be\label{approx}
1\,\ll\,\cot\a \lesssim \exp\left(\frac{2^{4/3}}{3}\frac{c}{N_c}\right).
\ee
 In this approximation $P(\r)$ remains almost constant for $0\leq \r\lesssim \r_*$.
The fact that $P$ is almost constant up to very large
scales $\r_*$ produces an intermediate energy
region over which the four-dimensional gauge coupling is
almost constant, as can be seen from Fig.~{\ref{Fig:plotcoupling}.
The larger $\cot\a$, the wider this region  is, with the only limitation provided by
the upper bound on the value of $\cot\a$, which depends on the ratio $c/N_c$.

 Fig.~\ref{Fig:background} shows  the other functions appearing in the background.
Notice the divergence of $e^{2g}$ in the IR.
Interesting is the behavior of the dilaton:  it is practically constant above and below $\r_{\ast}$, and 
at $\rho_{\ast}$ the IR and UV behaviors are smoothly connected.

This is the first very important result obtained with these methods. We found a background 
such that the running of the dual gauge coupling
has many of the properties of a putative walking theory. 
There is a region, delimited by two scales $\r_{IR}$ and $\r_{\ast}$, where 
the running almost flattens at a finite value.
Below $\r_{IR}$ the running reappears, and finally there is a scale $\hat{\r}_o$ at which the space ends,
deep in the IR.
This is a very encouraging starting point. For all the reasons discussed earlier, 
in order to understand how physical all of this is, one needs to use this background 
in order to compute some physical quantity.

\subsubsection{Walking solutions in class I.}

 \begin{figure}[htpb]
\includegraphics[width=12cm]{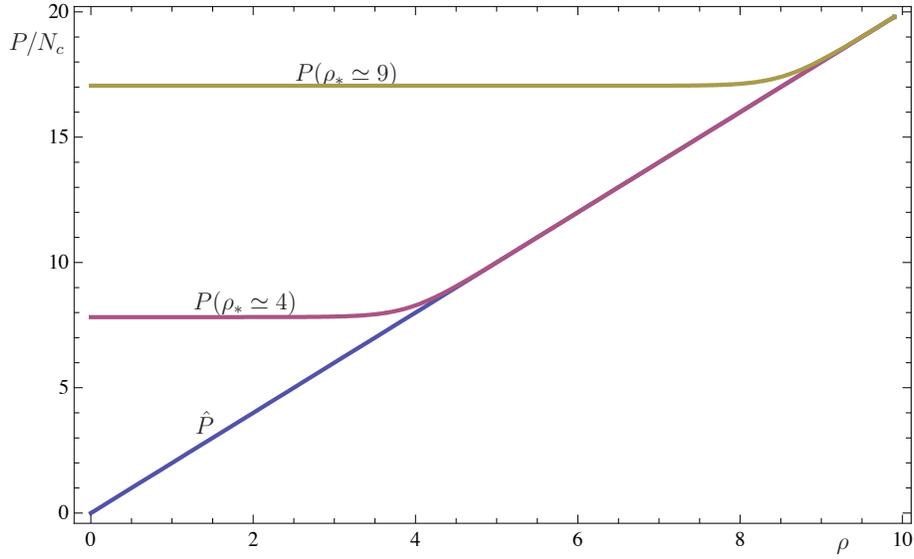}
\caption{The numerical solutions for $P(\r)/N_c$ used in the analysis as an example, for  $N_c=100$. 
The three solutions correspond to the $\hat{P}$ case
with $\r_{\ast}=0$, and to two new numerical solutions with, respectively, $\r_{\ast}\simeq 4$ and
$\r_{\ast}\simeq 9$~\cite{NPR}.}
\label{Fig:numericalP}
\end{figure}

\begin{figure}[htpb]
\includegraphics[width=7cm]{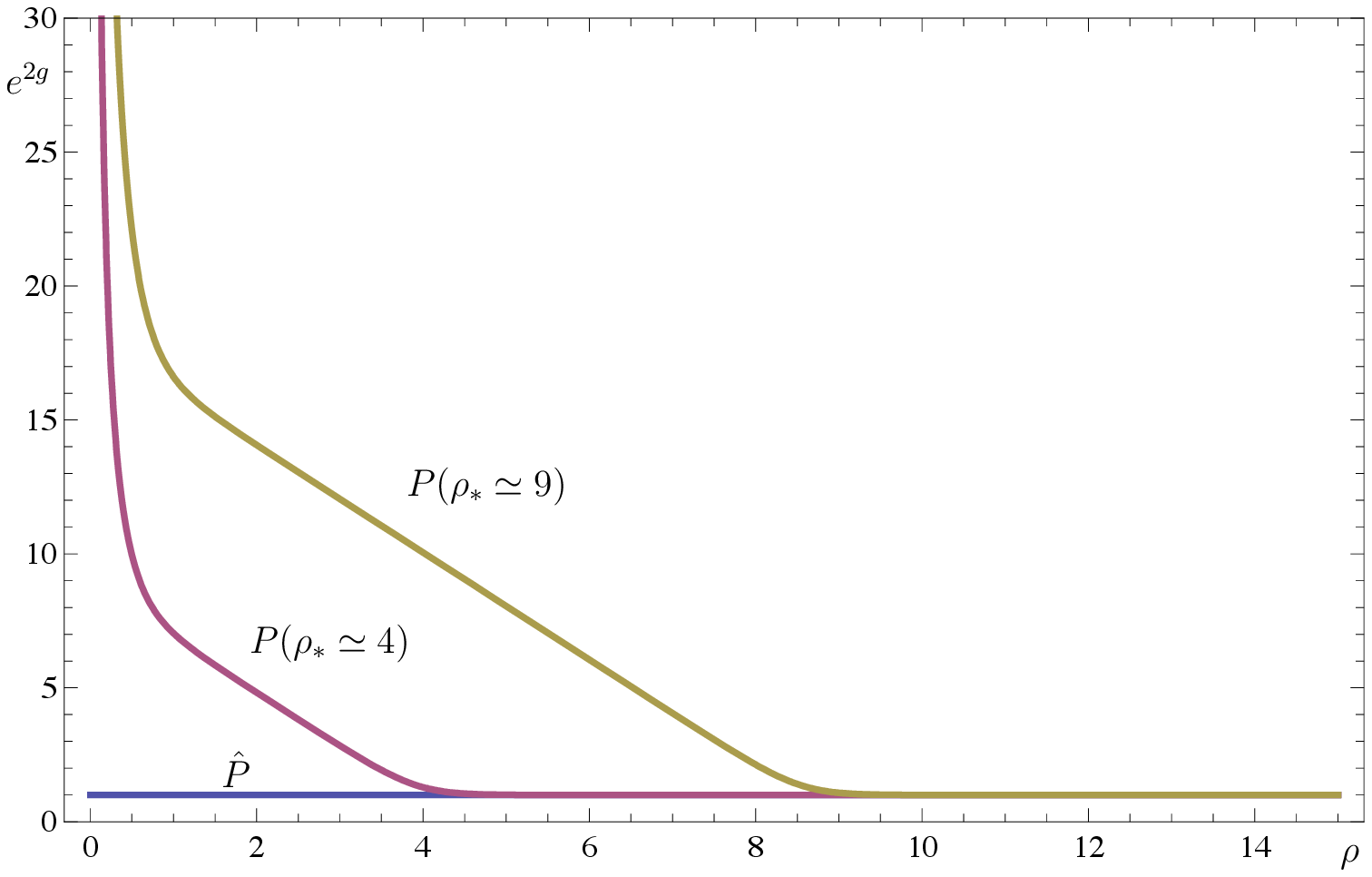}\includegraphics[width=7cm]{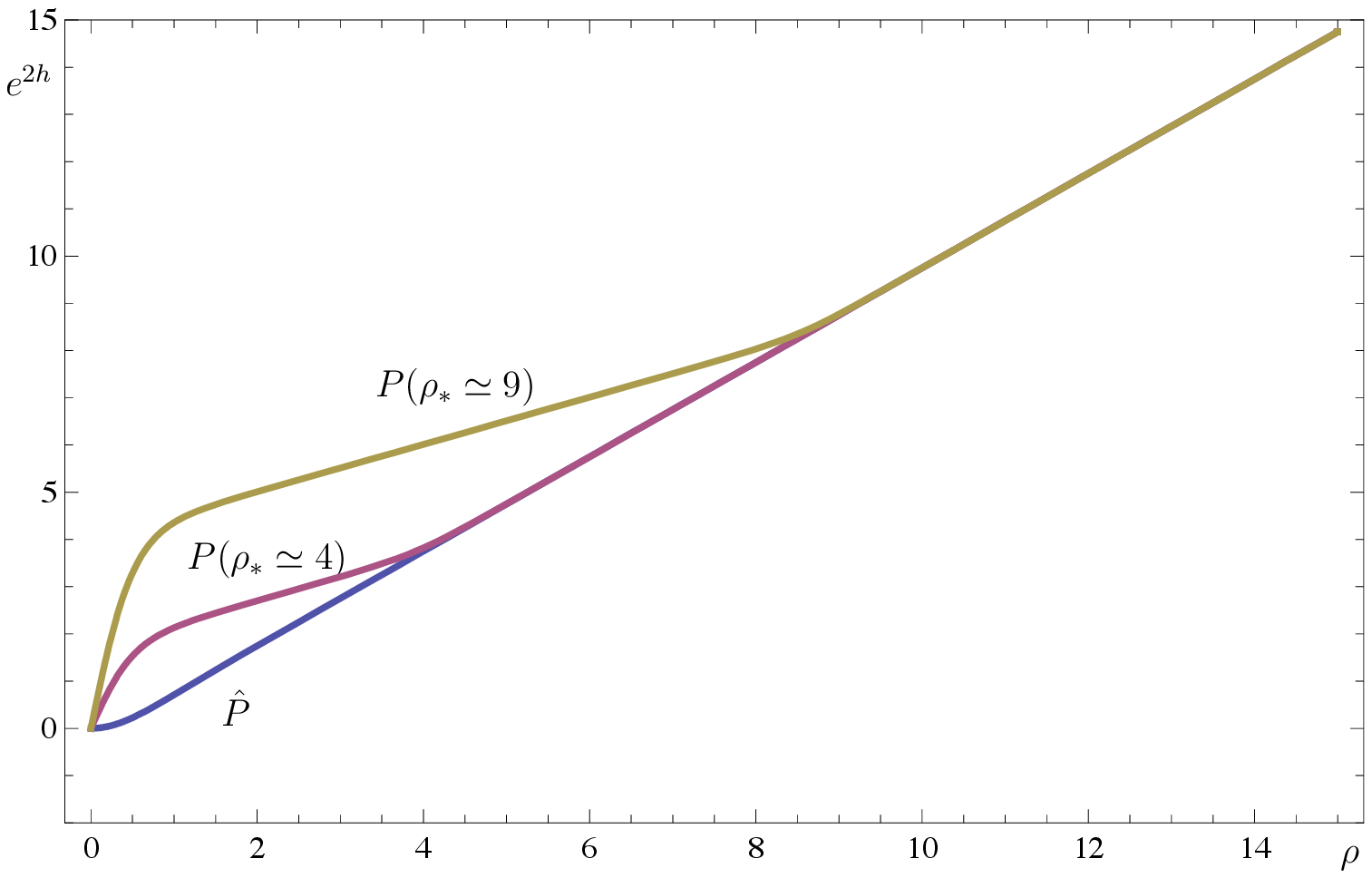}
\includegraphics[width=7.1cm]{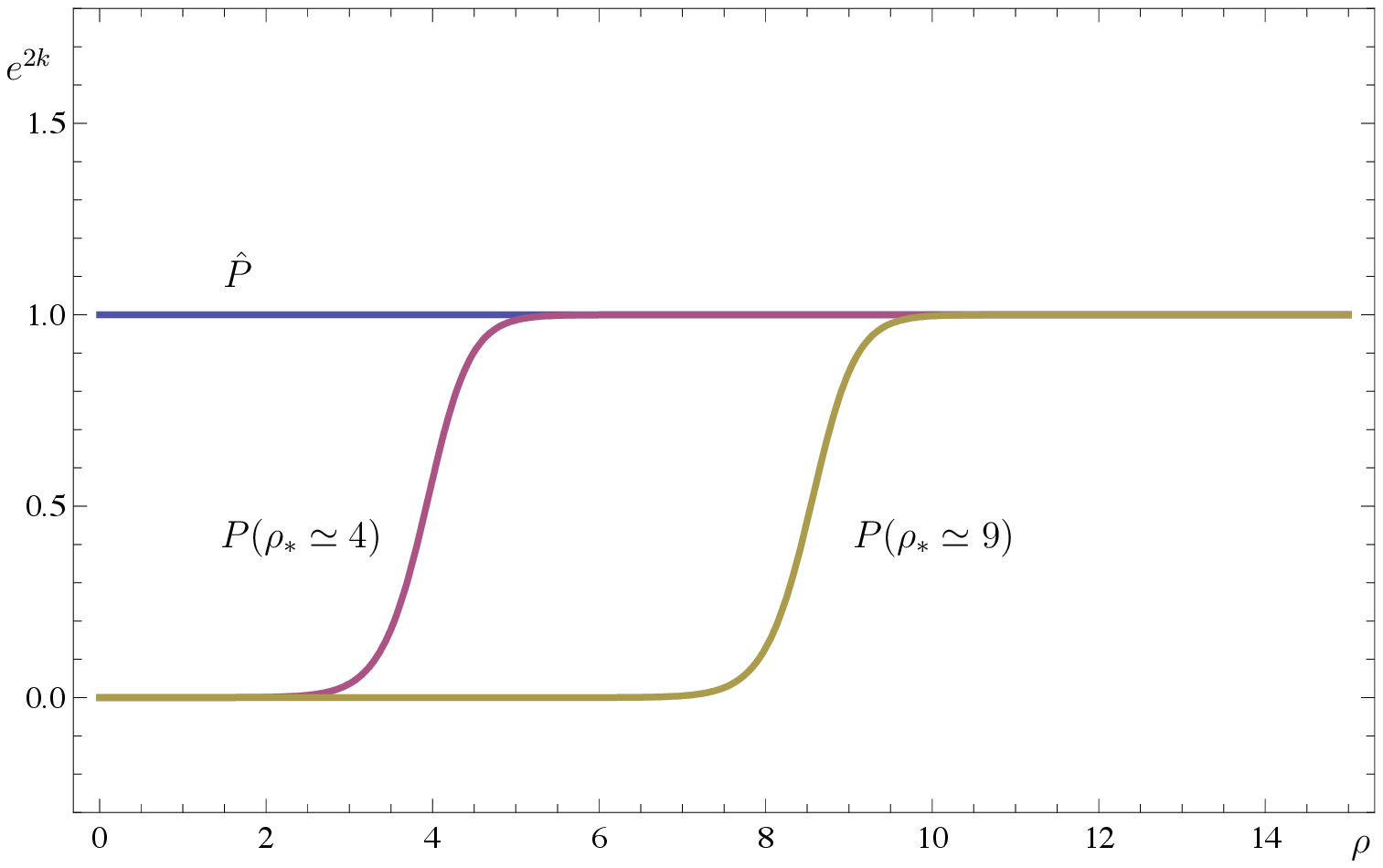}\includegraphics[width=7cm]{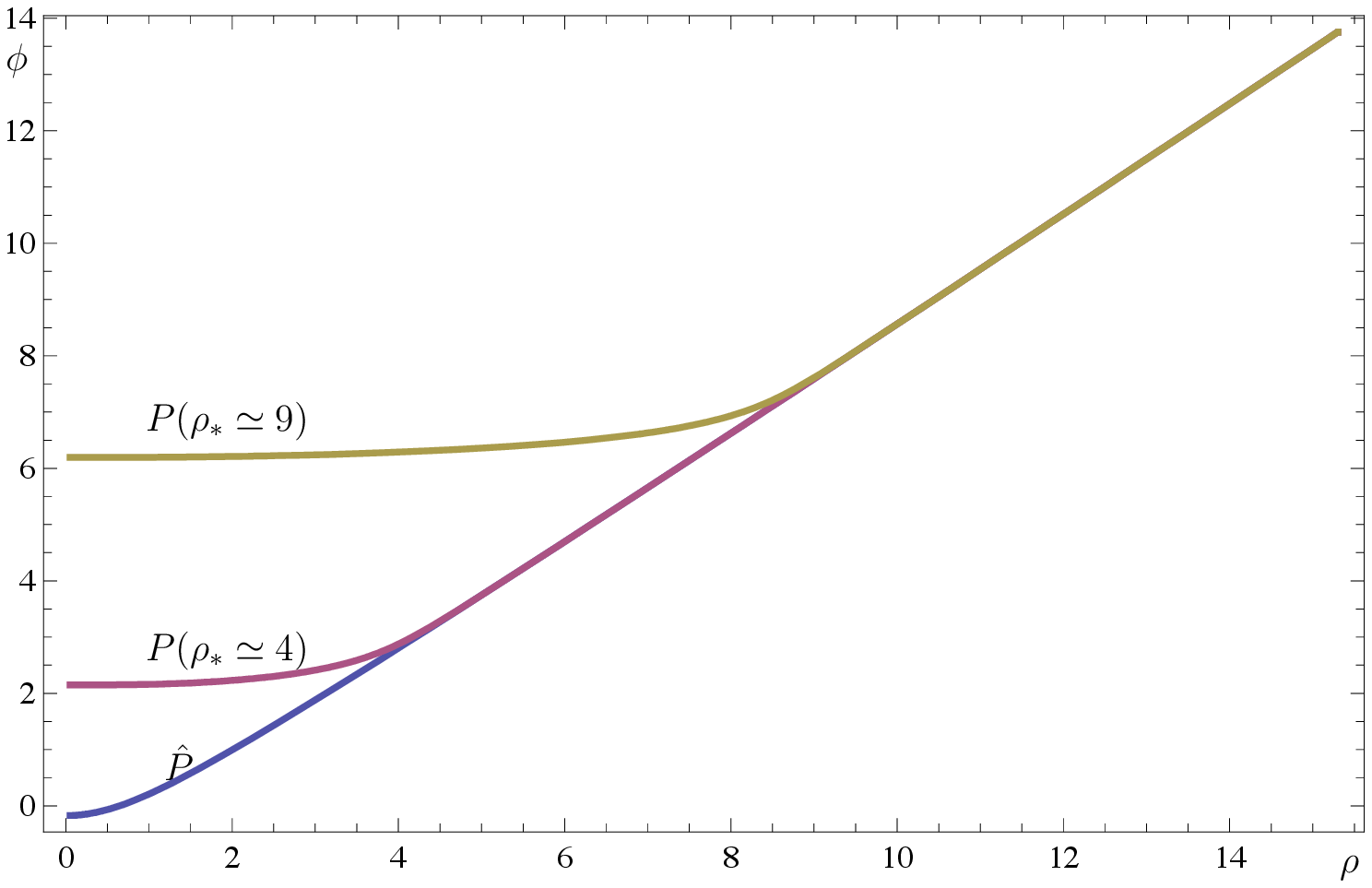}
\caption{The functions $(e^{2g},e^{2h},e^{2k},\Phi)$ appearing  in the metric
for the same solutions as in Fig.~\ref{Fig:numericalP}, computed rescaling 
$P\rightarrow P/N_c$, and $Q\rightarrow Q/N_c$~\cite{NPR}.}
\label{Fig:backgroundI}
\end{figure}

The walking behavior described in the previous subsection is appearing in the IR, for $\r<\r_{\ast}$. 
It is hence important to understand if it
is in any way related to the choice of UV-asymptotic behavior of the theory.
Also, for several reasons it is more sensible to perform  calculations of physical quantities in backgrounds
of class I, and it is hence important to know that walking solutions exist also in this class.
In principle, one might try and construct numerically such solutions
by observing that the walking solutions, for small-$\r$, can be written as
\beqs
\label{Eq:Pexpan}
P(\r)=P_0+k_3 P_0\r^3+\frac{4 k_3 P_0\r^5}{5}
     -k_3^2P_0\r^6
+\frac{16 \left(2 P_0^2 k_3-5 k_3 N_c^2\right)\r^7}{105
   P_0}\,
+\cdots\,,
\eeqs
with $P_0$ and $k_3$ two integration constants.
Notice how this expansion does not contain a term linear in $\r$. 
The  strategy  based on setting up the boundary conditions in the IR in such a way as to reproduce this
asymptotic behavior, and then solving numerically towards the UV, is not very efficient, because as we will see 
the class of walking solutions with UV behavior of class I depends only on one parameter.
Hence, one would need to know the precise relation between the integration constants in the IR-expansion, 
such that asymptotically in the UV one obtains a class-I solution. This is not possible with present knowledge.

In order to build numerically the solution, we start by expanding 
Eq.~(\ref{Eq:master}), by assuming that the solution can be written as
\beqs
P(\r)&=&\hat{P}(\r)\,+\,\varepsilon f(\r)\,+\,{\cal O}(\varepsilon^2)\,,
\label{Eq:linear}
\eeqs
with $\varepsilon \ll 1$, and expanding at large-$\rho$ by assuming that $\r\gg 0$.
Asymptotically this means that the linearized equations
admit solutions that, dropping power-law corrections for notational simplicity,  behave as
\beqs
f(\r)&\simeq&c_1 e^{-4 \r} \,+c_2 e^{2\r}\,,
\eeqs
which implies that consistency of the perturbative expansion from Eq.~(\ref{Eq:linear})
enforces the choice $c_2=0$.
Indeed, there are no asymptotic (in the UV) solutions
that behave as $e^{2\r}$.

We have now obtained an important result: at least for large
values of $\r$, there exists a 1-parameter class of solutions that approach asymptotically the 
$\hat{P}$ solution.
We cannot prove that such solutions are well behaved all the way to $\r\rightarrow 0$.
However, one  can use this result  in setting up the boundary conditions 
(at large-$\r$) and numerically solve Eq.~(\ref{Eq:master}) towards the IR~\cite{ENP,NPR}.
By inspection, these solutions turn out to be precisely the ones we 
are looking for. They start deviating significantly from $\hat{P}$ 
below some $\r_{\ast}>0$, below which $P$ is approximately constant.
We plot in  Fig.~\ref{Fig:numericalP} two such solutions,
with $\r_{\ast}\simeq 4$ and  $\r_{\ast}\simeq 9$, together with 
the $\hat{P}$ solution for the same value of $N_c$.
We also plot in Fig.~\ref{Fig:backgroundI} the functions appearing in
metric $(e^{2g},e^{2h},e^{2k},\phi)$ for the same solutions.
Notice the behavior of $e^{2g}$ for $\rho\rightarrow 0$. 
Also notice that the dilaton $\Phi$ is finite at $\r\to 0$.

\subsection{Relation to other systems.}

This subsection contains a formal digression that  will become useful later on.
The Maldacena-Nunez system, and in general the wrapped $D5$ system,
can be thought of within a more  general class of solutions to  type-IIB supergravity
that is often referred to as Papadopoulos-Tseytlin ansatz~\cite{PTa}.
This ansatz contains as special cases
a variety of important models, in particular  the  Klebanov-Witten (conifold, conformal)~\cite{KW}, the Klebanov-Tseytlin (singular conifold)~\cite{KT} and the Klebanov-Strassler (resolved conifold)~\cite{KS} ones.
Also, it is possible to show that the 10-dimensional equations yielding the background can be 
re-derived from a five-dimensional action, in which a sigma-model theory of a set of scalars
is coupled to gravity.
This section will also introduce a certain amount of notation, which may help the reader in relating the conventions
used by several different authors.

Following the notation in~\cite{BHM}, we describe the system using an effective five-dimensional 
action that reads, up to an overall normalization,
\beqs
{\cal S}&=&\int\di^5y\sqrt{-g}\left[\frac{1}{4}R\,-\,\frac{1}{2}G_{ab}g^{MN}\partial_M\Phi^a\partial_N\Phi^b\,-\,V(\phi)\right]\,,
\eeqs
where $\Phi^a=(\tilde{g},x,p,a,b,\Phi,h_1,h_2,\chi,{\cal K})$ and $y^M=(x^{\mu},z)$.
One uses the two constraints
\beqs
{\cal K}&=&M+2N(h_1 + b h_2)\,,\\
\partial_M\chi&=&\frac{(e^{2\tilde{g}}+2a^2+e^{-2}a^4-e^{-2\tilde{g}})\partial_M h_1+2a(1-e^{-2\tilde{g}}+a^2e^{-2\tilde{g}})\partial_Mh_2}{e^{2\tilde{g}}+(1-a^2)^2e^{-2\tilde{g}}+2a^2}\,,
\eeqs
where ${\cal K}$ is the normalization of the $F_5$ form in 10 dimensions, 
and $\chi$, $h_1$ and $h_2$ appear in the NS $B_2$ antisymmetric tensor of Type-IIB. 
$N$ is the normalization of the $F_3$ form, and essentially counts how many $D5$ branes are present,
while $M$ would count the number of $D3$ branes if $N=0$.

The constraints allow to remove $\chi$ and ${\cal K}$ from the sigma-model, which is hence defined by
\beqs
G_{ab}\partial_M\Phi^a\partial_N\Phi^b
&=&
\frac{1}{2}\partial_M \tilde{g} \partial_N \tilde{g} 
\,+\,\partial_M x \partial_N x
\,+\,6\partial_M p \partial_N p
\,+\,\frac{1}{4}\partial_M \Phi \partial_N \Phi \\
&&+\,\frac{1}{2}e^{-2g}\partial_M a \partial_N a
+\frac{1}{2}N^2e^{\Phi-2x}\partial_M b \partial_N b \nonumber\\
\nonumber &&
+\frac{e^{-\Phi-2x}}{e^{2\tilde{g}}+2a^2+e^{-2\tilde{g}}(1-a^2)^2}
\left[\frac{}{}(1+2e^{-2\tilde{g}}a^2)\partial_M h_1 \partial_N h_1\right.\\ &&\left.
+\frac{1}{2}(e^{2\tilde{g}}+2a^2+e^{-2\tilde{g}}(1+a^2)^2)\partial_M h_2 \partial_N h_2
+2a(e^{-2\tilde{g}}(a^2+1)+1)\partial_M h_1 \partial_N h_2\right]\,.\nonumber
\eeqs
The potential is
\beqs
V&=&-\frac{1}{2}e^{2p-2x}(e^{\tilde{g}}+(1+a^2)e^{-g})\\
&&\,+\,\frac{1}{8}e^{-4p-4x}(e^{2\tilde{g}}+(a^2-1)^2e^{-2\tilde{g}}+2a^2)\nonumber\\
&&\,+\,\frac{1}{4}a^2e^{-2\tilde{g}+8p}\nonumber\\
&&\,+\,\frac{1}{8}N^2e^{\Phi-2x+8p}\left[e^{2\tilde{g}}+e^{-2\tilde{g}}(a^2-2a b +1)^2 +2 (a-b)^2\right]\nonumber\\
&&\,+\,\frac{1}{4}e^{-\Phi-2x+8p}h_2^2\nonumber\\
&&\,+\,\frac{1}{8}e^{8p-4x}(M+2N(h_1+b h_2))^2\,.\nonumber
\eeqs
The five-dimensional metric is written as
\beqs
\label{Eq:5Dmetric}
\di y^2 
&=& e^{2A} \eta_{\mu\nu}\di x^{\mu}\di x^{\nu}\,+\,  \di z^2\,,
\eeqs
where we use the convention in which the metric is mostly
$+$, and the {\it warp factor} $A$ is to be determined by the Einstein equations.

The basic idea is that any solution to the equations of motion derived from this 5-dimensional system
can be lifted to a full solution of  10-dimensional type-IIB supergravity.
The 10-dimensional metric is (in Einstein frame)
\beqs
\label{Eq:10dmetric}
\di s^2_E&=&e^{2p-x}\di y^2\,+(e^{x+\tilde{g}}+a^2e^{x-\tilde{g}})(e_1^2+e_2^2)
+e^{x-\tilde{g}}(e_3^2+e_4^2-2a(e_1e_3+e_2e_4))+e^{-6p-x}e_5^2\,,
\eeqs
where the metric on the internal manifold is written in terms of
\beqs
e_1&=&-\sin\theta\di \phi\,,\\
e_2&=&\di \theta\,,\\
e_3&=&\cos\psi\sin\tilde{\theta}\di \tilde{\phi}\,-\,\sin\psi\di\tilde{\theta}\,,\\
e_4&=&\sin\psi\sin\tilde{\theta}\di \tilde{\phi}\,+\,\cos\psi\di\tilde{\theta}\,,\\
e_5&=&\di \psi+\cos\tilde{\theta}\di \tilde{\phi}+\cos{\theta}\di {\phi}\,.
\eeqs
All the other functions in the type-IIB background ($F_3$, $F_5$, $H_3$, $C$)
can be found explicitly in~\cite{BHM} and references therein.

In looking for solutions to the background, we assume that all the functions
have a non-trivial dependence only on the radial direction $z$.
Some very interesting backgrounds can be described by this formalism.
When $N=0$, ${\cal K} =M$ is the normalization of  $F_5$. 
In this case, there is no $F_3$, and hence no $b$.
Further, it is not difficult to show that in this case $H_3=0$ satisfies the equations of motion,
and hence $h_1=0=h_2=\chi$. This can be seen directly by minimizing the potential $V$
by brute force.
The dilaton $\Phi$  disappears from the sigma-model metric and from the potential $V$,
hence its equation of motion is solved by any constant, in particular $\Phi=0$.
Only $(\tilde{g},x,p,a)$ have to be solved for.
There exists in this case a stable minimum for the potential with
\beqs
a&=&0\,,\,
\tilde{g}\,=\,0\,,\,
x\,=\,-\frac{1}{2}\ln\frac{4}{3M}\,,\,
p\,=\,\frac{1}{6}\ln\frac{2}{M}\,.
\eeqs
At the fixed points, the Einstein equations imply that
\beqs
(\partial_z A)^{2}&=&-\frac{1}{3} V_0\,,
\eeqs
where $V_0$ is  the potential evaluated at the minimum. Replacing in the background one finds
\beqs
\di s^2_E&=&\frac{2^{4/3}}{\sqrt{3}M^{5/6}}\di y^2\,+3\sqrt{3M}\left(\frac{1}{6}(e_1^2+e_2^2
+e_3^2+e_4^2)+\frac{1}{9}e_5^2\right)\,,\\
\di y^2&=&e^{\frac{2^{5/3}}{3 M^{2/3}} z} \eta_{\mu\nu}\di x^{\mu}\di x^{\nu}\,+\,\di z^2\,,
\eeqs
and 
\beqs
e_1^2+e_2^2
+e_3^2+e_4^2&=&\sin^2\theta \di \phi^2 +\di \theta^2+\sin^2\tilde{\theta} \di \tilde{\phi}^2 +\di \tilde{\theta}^2
\eeqs
is the metric on $S^2\times S^2$.
The background has  
$AdS_5\times T^{1,1}$ geometry (the six extra-dimensions describe the conifold). 
This is the Klebanov-Witten fixed point~\cite{KW}, which describes
a ${\cal N}=1$ supersymmetric, conformal theory with $SU(2)\times SU(2)\times U(1)$
global symmetry.
Choosing for instance $M=2/\sqrt{27}$, so that the AdS curvature is $L=1$:
\beqs
\di s^2_E&=&\sqrt{2}3^{3/4}\left(e^{{2 z}} \eta_{\mu\nu}\di x^{\mu}\di x^{\nu}\,+\,\di z^2
+\frac{1}{6}(e_1^2+e_2^2
+e_3^2+e_4^2)+\frac{1}{9}e_5^2\right)\,.
\eeqs

One good reason for introducing all of this formalism is that it allows for a very simple exercise to be performed: the calculation 
of  the (quantum) dimensions of the operators of the dual field theory which correspond to the supergravity scalars.
By simply expanding the potential around the minimum (and carefully normalizing the kinetic terms),
working at unit curvature with $N=0$ and $M=2/\sqrt{27}$,
one finds that $a$ has mass $m^2=-3$,  $\tilde{g}$ has $m^2=-4$, $h_2$ and $b$ mix, the eigenvalues being $m^2=-3$ and $m^2=21$, 
$x$ and $p$ mix, the eigenvalues of the masses being $m^2=12$ and $m^2=32$, while $h_1$ and $\Phi$ are massless.
Using the celebrated relation $m^2=(\Delta-4)\Delta$ yields the scaling dimensions $\Delta$~\cite{BCPZ}:
\beqs
m^2=-4&\rightarrow&\Delta=2\,,\\
m^2=-3&\rightarrow&\Delta=1,3\,,\\
m^2=0&\rightarrow&\Delta=0,4\,,\\
m^2=12&\rightarrow&\Delta=-2,6\,,\\
m^2=21&\rightarrow&\Delta=-3,7\,,\\
m^2=32&\rightarrow&\Delta=-4,8\,.
\eeqs
Notice the presence of a set of non-trivial
higher-dimensional operators (of dimension 6, 7 and 8).
(The reader is assumed to be familiar with the interpretation of $m^2$ in terms of the
dimension of the dual operator and its coupling~\cite{KlW}).
A more complete discussion of this model can be found elsewhere, the main lesson we learn from here
is that computing the (non-perturbative) dimensions of many important operators in the dual theory is relatively easy
(compared for instance to the formidable task that doing it in four dimensions represents).
Notice also that this exercise clearly shows that this is not simply the dual description of ${\cal N}=1$
super-Yang-Mills: this set-up yields a very rich structure, with many operators being present,
all of which in general have very important non-perturbative implications.

An important comment about internal symmetries.
Looking back at Eq.~(\ref{Eq:10dmetric}) one sees that for $a=0=\tilde{g}$ the internal space gains the structure
of $T^{1,1}$, and its $SU(2)\times SU(2)\times U(1)$ symmetry.
In particular, the functions $a$ and $b$ (the latter appears in the $F_3$ and $B_2$ fields), when non-vanishing,
induce symmetry-breaking. 
For instance,  in the wrapped $D5$ background defined earlier on,
the equations yield $a$ and $b$ that vanish in the UV, but are non-vanishing below $\r_{IR}$, signaling that they encode the fact
that spontaneous symmetry breaking is taking place at the scale corresponding to $\r_{IR}$.
The exercise we just did means that the non-vanishing of $a$ and $b$ is related to a dimension-3 condensate
in the dual theory, usually identified with the gaugino condensate~\footnote{
Important subtleties related to the anomalies should be taken into account in discussing the internal symmetries,
but for our present purposes they do not change the substance of these results.}. 

Another important class of solutions can be found with the restriction
\beqs
a&=&\tanh y\,,\\
e^{-g}&=&\cosh y\,.
\eeqs
With both $M\neq 0 \neq N$,
one is left with a system of equations that admits the Klebanov-Strasser background as regular solution.
This system is known to provide a description of the duality cascade~\cite{cascade}.

Finally, the Maldacena-Nunez system of wrapped $D5$ that is the focus of this section
 is obtained by setting $M=0$, and $N=N_c/4$,
in which case one can consistently set $h_1=h_2=\chi={\cal K}=0$, reducing to six the number of scalar functions
controlling the background.
One important thing to stress here is that all of these backgrounds are related to each other,
belonging to the same general class~\cite{MM}.

\subsection{Glueball spectrum.}

By making use of the five-dimensional formalism introduced in the previous subsection, 
we can compute the spectrum of scalar excitations (glueballs) for the walking solutions
with type-I asymptotic behavior in the UV.
Following~\cite{BHM}, we  relate the variables
$[A,\tilde{g},p,x,\Phi,a,b]$ to the functions describing the background in Eq.~(\ref{nonabmetric424}) as
\bea
& & \frac{\Phi}{4} = A + p - \frac{x}{2}, \;\;
g = -A - \frac{\tilde{g}}{2} - p + x + \log 2, \nonumber\\
& & h = -A + \frac{\tilde{g}}{2} - p + x, \;\;\; k = -A - 4 p + \log 2\,,
\eea
and relate the radial coordinates according to
$
2 \di \r e^{-4p} = \di\r e^{A+k}= \di z
$\,.
The resulting effective 5-dimensional non-linear sigma model with fields 
$\Phi^a=[\tilde{g},p,x,\Phi,a,b]$ coupled to gravity can be studied from: 
\beq
{\cal L}_{5d}^{eff}=\frac{4(\alpha^{\prime}g_s N_c)^2(4\pi)^3}{G_{10}}
\sqrt{-g}\left[\frac{R}{4}-\frac{1}{2}G_{ab}\partial \Phi^a \partial \Phi^b 
-V(\vec{\Phi})\right] \,,
\label{5deffnonab}
\eeq
where we now made explicit all the constants,
and the five-dimensional metric  has the form in Eq.~(\ref{Eq:5Dmetric}).

The sigma-model metric (replacing $N=N_c/4$) is
\bea
\label{sigmamodelmetric}
 4G_{\phi\phi}= 2 G_{gg}= 
G_{xx}=\frac{G_{pp}}{6}=1\,,\\
\nonumber G_{aa}=\frac{e^{-2\tilde{g}}}{2},\;\; G_{bb}= 
\frac{N_c^2 e^{\Phi-2x}}{32
},
\label{kahler}
\eea
and the potential $V$ is given by (in agreement with~\cite{BHM})
\begin{widetext}
\SP{
	V = \frac{e^{-2 (\tilde{g}+2 (p+x))}}{128} \Bigg[& 16 \left(a^4+2
   \left(\left(e^{\tilde{g}}-e^{6 p+2 x}\right)^2-1\right) a^2+e^{4 \tilde{g}}-4
   e^{\tilde{g}+6 p+2 x} \left(1+e^{2 \tilde{g}}\right)+1\right)+ \\& e^{12 p+2 x+\Phi
   } \left(2 e^{2 \tilde{g}} (a-b)^2+e^{4 \tilde{g}}+\left(a^2-2 b
   a+1\right)^2\right) N_c^2 \Bigg].
}
\end{widetext}

From here on, we work in units where $\alpha^{\prime}g_s N_c=1$.
Following~\cite{BHM}, after changing coordinates 
$dz = e^{A+k} d\r$,  writing the fluctuations as $\mathfrak{a}(x,\r)=e^{i K x}\mathfrak{a}(\r)$, 
replacing $\Box$ by $-K^2$ (=$M^2$) and using the equations
for the 5d gravity fluctuations,  
 we obtain the system of linearized equations for the scalar perturbations~\cite{Elander:2009bm}
\SP{
\label{eq:eomfluc}
	\Big[ D_z^2 + 4 A' D_z - e^{-2A} K^2 \Big] \mathfrak{a}^a - \Big[ V^a_{|c} - \mathcal{R}^a_{\ bcd} \Phi'^b \Phi'^d +
	 \frac{4 (\Phi'^a V_c + V^a \Phi'_c )}{3 A'} + \frac{16 V \Phi'^a \Phi'_c}{9 A'^2} \Big] \mathfrak{a}^c = 0&.
}

One important comment. The six $\mathfrak{a}^c$ fields solving 
Eq.~(\ref{eq:eomfluc}) are not simply the fluctuations of the scalars $\Phi^a$ in the sigma-model Lagrangian. 
In fluctuating the geometry, the scalars mix non-trivially with the fluctuations of the metric. The formalism
of~\cite{BHM}, from which  Eq.~(\ref{eq:eomfluc}) is derived, 
has the advantage of explicitly making use of all the equations of motion (including the 
Einstein equations) in such a way that the system of six fluctuations 
written here is already written in the physical basis.
In Eq.~(\ref{eq:eomfluc}), primed quantities mean derivatives in respect to $z$,
$V$ is the potential, $V_a$ its derivative in respect to the field $\Phi^a$,
in $V^b$ the index is raised with the inverse sigma-model metric $G^{ab}$,
$V^a_{|c}$ is the (sigma-model) covariant derivative of $V^a$, 
${\cal R}^a_{\ bcd}$ is the Riemann tensor of the sigma-model metric
and $D_z$ is the  background covariant derivative defined in~\cite{Elander:2009bm}.

The values of $K^2 = -M^2$ for which the whole system is solved while at the same time satisfying 
appropriate boundary conditions
  give us the glueball spectrum of the dual field theory. 
In order to find them, we  employ a numerical method described in \cite{BHM}, that 
in effect evolves solutions from both the IR and the UV, and then determines whether they  match  smoothly at a midpoint.

In the UV (for $\r\rightarrow +\infty$), Eq.~(\ref{eq:eomfluc})
can be diagonalized by a change to a  basis  
in which the fluctuations behave as
$\psi^i = e^{C_i \rho} \sum_n a_{i,n} \rho^{b_{i,n}}$,
where the exponents $b_{i,n}$ in general are non-integer, while
\SP{
    C_{1,2,6} = -1 \pm \sqrt{9 - M^2}, \\
    C_{3,4,5} = -1 \pm \sqrt{1 - M^2}.
}
Notice that this behavior implies the presence of cuts in the two-point functions
for $M^2>1$ and $M^2>9$.
The study of the discrete spectrum requires to chose UV boundary conditions such as to
select the exponentially suppressed UV-behavior.
The IR boundary conditions are more subtle, and the reader who is interested 
in the details can find them in~\cite{ENP}.

 \begin{figure}[htpb]
\includegraphics[width=9cm]{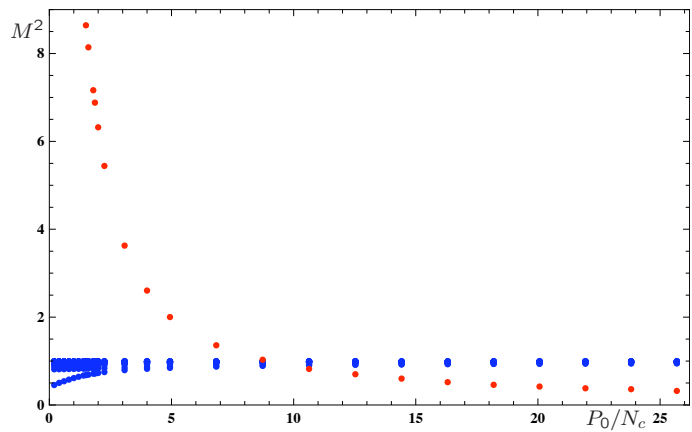}
\caption{The masses $M^2$ plotted against   $P_0/N_c \simeq 2\r_{\ast}$~\cite{ENP}.
\label{Fig:walking}}
\end{figure}

 The numerical results from the analysis in~\cite{ENP} are plotted in Fig.~\ref{Fig:walking}. 
 The spectrum for $\hat{P}$ ($P_0\rightarrow 0$) consists of a series of poles approaching $M^2=1$ (in agreement with~\cite{BHM}).
For large values of $P_0$ (equivalently, of $\r_{\ast}$), all the masses in the series 
approach the branch points, so that  the discrete spectrum effectively disappears into the
continuum. With one notable exception: one isolated state
becomes lighter, and as $\r_{\ast}$ is increased, its mass is pushed 
below both the continuum thresholds. For $\r_{\ast}\rightarrow \infty$, this state becomes massless.

The appearance of a single light spin-0 state in the spectrum 
is typical of systems having some approximate, spontaneously broken symmetry.
If this symmetry  is some form of scale invariance,
the light scalar might be  a light dilaton, 
and  hence its couplings should be dictated by 
this property.
With present  information, we are not able to reconstruct 
its composition in terms of the 
original degrees of freedom in the sigma-model.  Notice in particular 
that the metric is
not asymptotically AdS  in the UV, hence the rigorous procedure for holographic renormalization is not known,
nor is it  known how to characterize this state as normalizable 
or non-normalizable. 
Also, the background is not approximately AdS in the walking region, which would be a clear indication
of conformal invariance.

The appearance of a very light scalar state, whose mass is parametrically
lighter than the dynamical scale controlling the mass of all other composite states,
would have profound implication
in models of dynamical electro-weak symmetry breaking,
affecting precision physics (remember that in order to avoid divergences we extracted indicative bounds on $\hat{S}$
from the $m_h\sim 800$ GeV case),
and  gauge boson scattering amplitudes.
It would be the obvious candidate for a LHC signature. It is remarkable that,
if its couplings are dictated by scale invariance, it would be hard to distinguish it from the 
Higgs of the MSM.

\subsection{Wilson loops.}

The Wilson loop~\cite{Wilson:1974sk} along the curve ${\cal C}$  is defined as
\beq
W({\cal C})\equiv \frac{1}{N_c} Tr P e^{i\oint_{\cal C}A_\mu dx^\mu}.
\label{wilson1}
\eeq
It is one of the most useful gauge-invariant objects, when trying to extract non-perturbative information
about a gauge theory.
It is computed in the dual string 
theory by calculating the action of a string bounded by  ${\cal C}$ at 
the four-dimensional  boundary of the space:
\beq
\left\langle \frac{}{}W({\cal C})\right\rangle= \int_{\partial F({\cal C})} {\cal D} F e^{- S_{NG}[F]}
\label{definitionwilson}\,,
\eeq
where $F$ denotes all the fields of the string theory and $\partial F$ their 
boundary values. A good approximation to this path integral is by steepest 
descent.  The Wilson loop is then related to the area of the minimal 
surface  bounded by 
the curve ${\cal C}$, 
spanned by classical string  configurations
(with Nambu-Goto action $S_{NG}$) that 
explore the bulk of the space.

All of  this was first proposed ten years ago in 
\cite{Maldacena:1998im}.
In the meantime, this proposal motivated lots of developments, see 
\cite{wilson} for beautiful  papers along this line.
See also~\cite{Sonnenschein:1999if}
for a review. 
In this subsection we are going to compute the Wilson loop using this prescription, for the walking backgrounds of Type I.
 In doing so, we will be able to show that the scales $\r_0$, $\r_{IR}$ and $\r_{\ast}$
 correspond to physical scales, at which the functional relation between quark-antiquark potential 
 and quark-antiquark separation changes.

\subsubsection{General treatment}

We start by summarizing the general treatment,
which  uses the ideas of~\cite{Maldacena:1998im} (see also~\cite{Kinar:1998vq}).

We study the action for a string in a background of the generic form
\beq
ds^2=-g_{tt} dt^2+ g_{xx} d\vec{x}^2+ g_{\r\r}d\r^2+ g_{ij}d\theta^i 
d\theta^j .
\label{back}
\eeq
We assume that the functions $(g_{tt}, g_{xx}, g_{\r\r})$ depend only on the 
radial coordinate $\r$. 
By contrast, $g_{ij}$ for the internal (typically compact) space can 
also depend on other coordinates. 
But we will choose a configuration for a probe string that is 
not excited on the $\theta^i$ directions, and  hence  we will 
ignore  the  internal space.

The configuration we  choose is,
\beq
t=\tau,\;\;\;\; x=x(\sigma),\;\;\;\; \r=\r(\sigma).
\label{ansa}
\eeq
and compute the Nambu-Goto action
\beq
S= \frac{1}{2\pi \alpha'}\int_{[0, T]} d\tau \int_{[0,2\pi]} d\sigma 
\sqrt{-\det G_{\alpha\beta}}. 
\label{ng}
\eeq
The induced metric on the 2-d world-volume is 
$G_{\alpha\beta}=g_{\mu\nu}\partial_\alpha X^\mu \partial_\beta X^\nu$, 
where
\beq
G_{\tau\tau}=-g_{tt},\;\;\; G_{\sigma\sigma}=g_{xx}(\frac{dx}{d\sigma})^2+ 
g_{\r\r} 
(\frac{d\r}{d\sigma})^2 \,.
\label{zxzx}
\eeq
Defining for convenience
$ f(\r)^2\equiv g_{tt}g_{xx},\; g(\r)^2= g_{tt}g_{\r\r}$, 
the Nambu-Goto action  is
\beq
S= \frac{T}{2\pi \alpha'} \int_{0}^{2\pi} d\sigma
\sqrt{f^2 x'(\sigma)^2 + g^2 \r'(\sigma)^2}
\,\equiv\,
 \frac{T}{2\pi \alpha'} \int_{0}^{2\pi} d\sigma L\,.
\label{ng2}
\eeq
Notice that we consider the situation in which the string does not 
couple to the NS $B_2$-field. 

We first compute the Euler-Lagrange equations  from Eq.~(\ref{ng}) and  
then we specify them  for the ansatz in Eq.~(\ref{ansa}). We also assume a 
background metric independent of time (we consider the system at the equilibrium). 
Defining 
\beqs
V_{eff}(\r)&\equiv&\frac{f(\r)}{C g(\r)}\sqrt{f^2(\r)- C^2}\,,
\label{nnzz}
\eeqs
where $C$ is an integration constant,
we rewrite the equations of motion in terms of only one equation
\beq
\label{drdxfinal}
\frac{d\r}{d\sigma}=\pm\frac{dx}{d\sigma} V_{eff}(\r)\leftrightarrow \frac{d\r}{d x}= 
\pm V_{eff}(\r)\,.
\eeq

The kind of solution we are interested in can be depicted as follows: a string that hangs from infinite radial position at $x=0$ and drops down towards smaller $\r$ as $x$ increases. Once it arrives at the smallest $\r$ compatible with the equations, dubbed $\r_0$, it starts growing in the radial direction up to infinite $\r$ where $x=L_{QQ}$, see Fig.~\ref{Fig:stringset}.
\begin{figure}[h]
\begin{center}
\begin{picture}(160,225)
\put(0,0){\includegraphics[height=7cm]{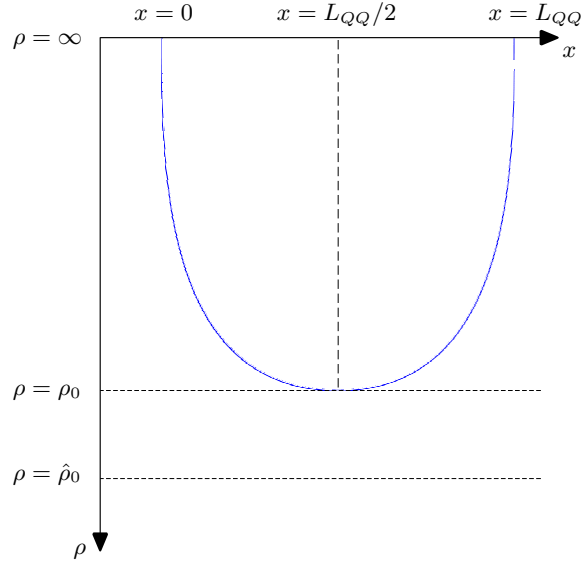}}
\put(-30,193){$\rho=\infty$}
\put(-30,60){$\rho=\r_0$}
\put(-30,28){$\rho=\hat{\r}_0$}
\put(-7,0){$\rho$}
\put(16,202){$x=0$}
\put(150,202){$x=L_{QQ}$}
\put(70,202){$x=L_{QQ}/2$}
\put(178,188){$x$}
\end{picture} 
\caption{Setting of the string.}
\label{Fig:stringset}
\end{center}
\end{figure}
This means that in the two distinct regions $x<L_{QQ}/2$ and $x>L_{QQ}/2$ the equations of motion will differ only in a sign
\begin{eqnarray}
\label{Eq:eqofmot}
x<\frac{L_{QQ}}{2}\,, && \frac{d\r}{d x}= -V_{eff}(\r)\nonumber\\
x>\frac{L_{QQ}}{2}\,, && \frac{d\r}{d x}= V_{eff}(\r)\,.
\end{eqnarray}
We can now formally integrate the equations of motion 
\beqs
x(\r)=
\begin{cases}
\int_{\r}^{\infty}\frac{d\r}{V_{eff}(r)}&\quad x<\frac{L_{QQ}}{2}\,,\\
L_{QQ}-\int_{\r}^{\infty}\frac{d\r}{V_{eff}(r)}&\quad x>\frac{L_{QQ}}{2}\,.
\end{cases}
\eeqs

We need to specify the boundary conditions for  the string in 
Eq.~(\ref{ansa}). This is an open string, vibrating in the bulk of a closed 
string background. Following the ideas of 
\cite{Maldacena:1998im}, we add a D-brane at a very large radial distance 
where the open string will end. This string will then satisfy a Dirichlet 
boundary condition at $\r\to\infty$. This means that for large values of 
the radial coordinate
$\frac{dx}{d\sigma}$ must vanish. 
The only way of satisfying the equation of motion Eq.~(\ref{drdxfinal}) for $\rho\to\infty$, given that the left hand side has to be non-vanishing, is to have a divergent $V_{eff}(\r)$:
\beqs
\label{Eq:UVboundary}
\lim_{\r\rightarrow \infty} V_{eff}(\r)&=&\infty\,.
\eeqs
 This implies that there are restrictions on the asymptotic behavior 
 of the background functions [$f(\r), g(\r)$] in order for the string proposed in 
Eq.~(\ref{ansa}) to exist. This is ultimately the reason why we will perform the
calculations in backgrounds in class I in the next subsection.

Provided Eq.~(\ref{Eq:UVboundary}) is satisfied the string  moves to smaller values of the radial coordinate down to a turning point $\r_0$ where the quantity $\frac{d\r}{dx}(\r_0)=0$, i.~e. $V_{eff}=0$.
The turning point can be placed in any possible $\r_0$, with $\hat{\r}_0 < \r_0 < \infty$ (where $\hat{\r}_0$ is the end of the space),
 where the inversion point is given by imposing $C=f(\r_0)$.
It is  clear from Eq.~(\ref{drdxfinal}) 
that $V_{eff}(\r)$ 
controls not only the boundary condition at infinity, but also the 
possibility for the string to turn around and come back to the brane at 
infinity.


Given a  probe-string hanging from infinity, and that after turning
around at a point $\r_0$  goes back to the 
$D$-brane at infinity, we can 
 compute gauge theory quantities, like the  separation between the two 
ends of the string, which can be thought  of as  the separation between a quark-antiquark 
pair living on the $D$-brane and coupled to the end-points of the string.
And we can compute the Energy of the pair of quarks, that we  associate with 
the length of the string (computed along its path in the bulk).
 Both of these quantities will be functions of 
the turning point $\r_0$.

\subsubsection{The $D5$ on $S^2$ system.}

We now apply all of the above to the walking solutions of type I.
For our purposes, it is also convenient to fix $8e^{4\Phi_0}=1$, and $\hat{\r}_o=0$.
The functions we need for the probe string are:
\beqs
f^2(\r)&=&e^{2\Phi}\,=\,\sqrt{\frac{\sinh^2(2\r)}{(P^2-Q^2)P^{\prime}}}\,,\\
g^2(\r)&=&\frac{1}{2}P^{\prime}f^2(\r)\,,\\
\label{Eq:VeffMN}
V_{eff}^2(\r)&=&\frac{2}{C^2P^{\prime}}\left(\sqrt{\frac{\sinh^2(2\r)}{(P^2-Q^2)P^{\prime}}}\,-\,C^2\right)\,,
\eeqs
with $P$ and $Q$ defined earlier on, characterizing the background.

\begin{figure}[htpb]
\includegraphics[width=9cm]{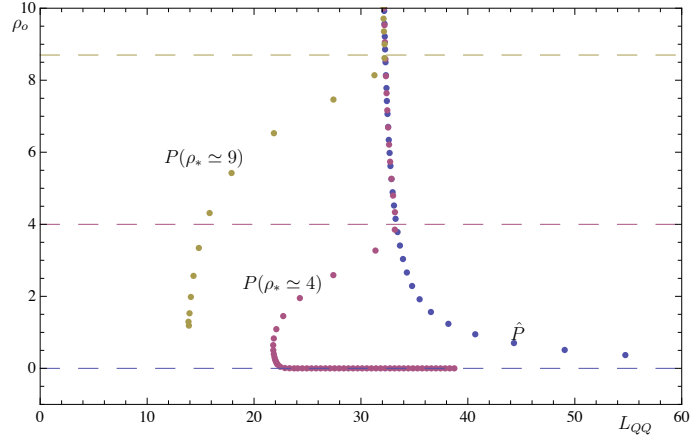}
\caption{The radial coordinate $\r_0$ of the turning point of the string
as a function of $L_{QQ}$, in the three backgrounds from Fig.~\ref{Fig:numericalP}, with same color-coding~\cite{NPR}.}
\label{Fig:study1}
\end{figure}

\begin{figure}[htpb]
\includegraphics[width=9.0cm]{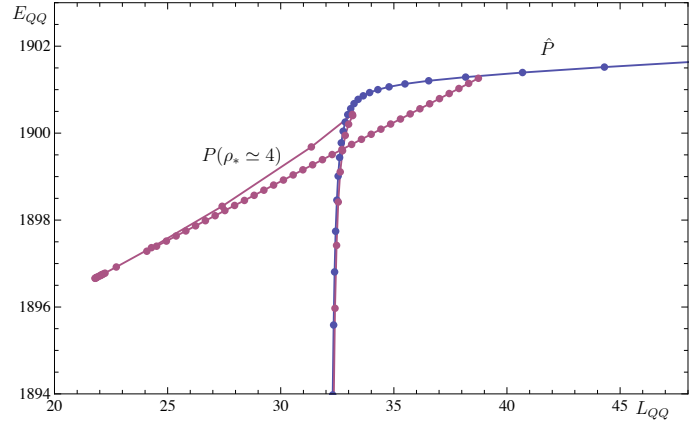}
\caption{The energy $E_{QQ}$ as a function of the quark-antiquark separation $E_{QQ}(L_{QQ})$.
The two  solutions in Fig.~\ref{Fig:numericalP} with $\r_{\ast}=0$ and $\r_{\ast}\simeq 4$ are used
for the background, and the results are shown using the same color-coding as in Fig.~\ref{Fig:numericalP}~\cite{NPR}. }
\label{Fig:study2}
\end{figure}

We set-up the configuration of the string by assuming that its extremes are
attached to the brane at $\r=\r_1\gg 1$, and treat this as a UV cut-off.
Because the background is known only numerically, and it is not asymptotically-AdS,
the calculations are performed at fixed cut-off.
The string is stretched in the Minkowski direction $x=x(\r)$, with $x(\r_1)=0$ for convenience.
We vary the integration constant $C^2 > f^2(0)$. For each choice of $C$ 
we  define $\r_0$ as $V^2_{eff}(\r_0)=0$. In this way, the coordinates of the string are
$(x(\r),\r)$,  where
\beqs
\label{Eq:distance}
x(\r)&=&\int_{\r}^{\r_1} \frac{\di r}{{V_{eff}(r)}}\,.  
\eeqs
The Minkowski distance between the end-points of the string is hence $L_{QQ}=2x(\r_0)$.
For the energy, because in the numerical study we do not remove the UV cut-off,
we use the unsubtracted 
 action (setting $T/(2\pi\alpha^{\prime})=1$)
evaluated up to the cut-off:
\beqs
\label{Eq:energy}
E_{QQ}&=&2 \int_{\r_0}^{\r_1}\di r \sqrt{\frac{f^2(r)g^2(r)}{f^2(r)-C^2}}\,.
\eeqs

The numerical results obtained for the three different solutions for  $P$ in Fig.~\ref{Fig:numericalP} 
are shown in Fig.~\ref{Fig:study1}, Fig.~\ref{Fig:study2}  and Fig.~\ref{Fig:strings}.

Let us first focus our attention on the background generated by $\hat{P}$ in
Eq.~(\ref{Eq:MN}), in which case a great deal of information can be extracted analytically.
The deeper the string probes the radial coordinate 
(smaller values of $\r_0$), the longer the separation $L_{QQ}$ between the 
end-points on the UV brane, in agreement with natural expectations.
In particular, setting $C^2=f^2(0)$ (equivalent to $\r_0\rightarrow 0$), and expanding around $\r\sim 0$
\beqs
V_{eff}^2(\r)&=&\frac{8\r^2}{9N_c}\,+\cdots\,,
\eeqs
which, by means Eq.~(\ref{Eq:distance}) and Eq.~(\ref{Eq:energy}),  implies that both 
$L_{QQ}$ and $E_{QQ}$ {\it diverge} for $C^2\rightarrow f^2(0)$.

Two regimes can be identified: as long as $\r_0 > \r_{IR}$, 
then $L_{QQ}$ varies very little with $\r_0$.
For small $\r_0<\r_{IR}$, further reductions of $\r_0$ imply much bigger increase in $L_{QQ}$.
The scale $\r_{IR}\sim {\cal O}(1)$ is the scale in which the function $Q$ changes from linear
to approximately quadratic in $\r$, and is also the scale below which the gaugino condensate
is appearing (the function $b(\r)$ in the background is non-zero).
This result is better visible in  Fig.~\ref{Fig:study1}.
The dependence of $L_{QQ}$ on $\r_0$ is monotonic, but shows two very different 
behaviors  for $\r_0<\r_{IR}$ and $\r_0>\r_{IR}$, respectively. The transition between the two
is completely smooth.

The physical meaning of this behavior is well illustrated by studying the total 
energy $E_{QQ}$ of the classical configurations, as a function of $L_{QQ}$
(see  Fig.~\ref{Fig:study2}).
For small $L_{QQ}$, the energy 
grows  very fast with $L_{QQ}$, until a critical value beyond which 
the dependence becomes linear. 
 Which can be interpreted in terms of the linear behavior of the
quark-antiquark potential obtained from the Wilson loop, in agreement with confinement.
The important conclusion is that the geometry and the end of the space
encodes the information about confinement.

\begin{figure}[htpb]
\begin{picture}(320,120)
\put(0,10){\includegraphics[width=5cm]{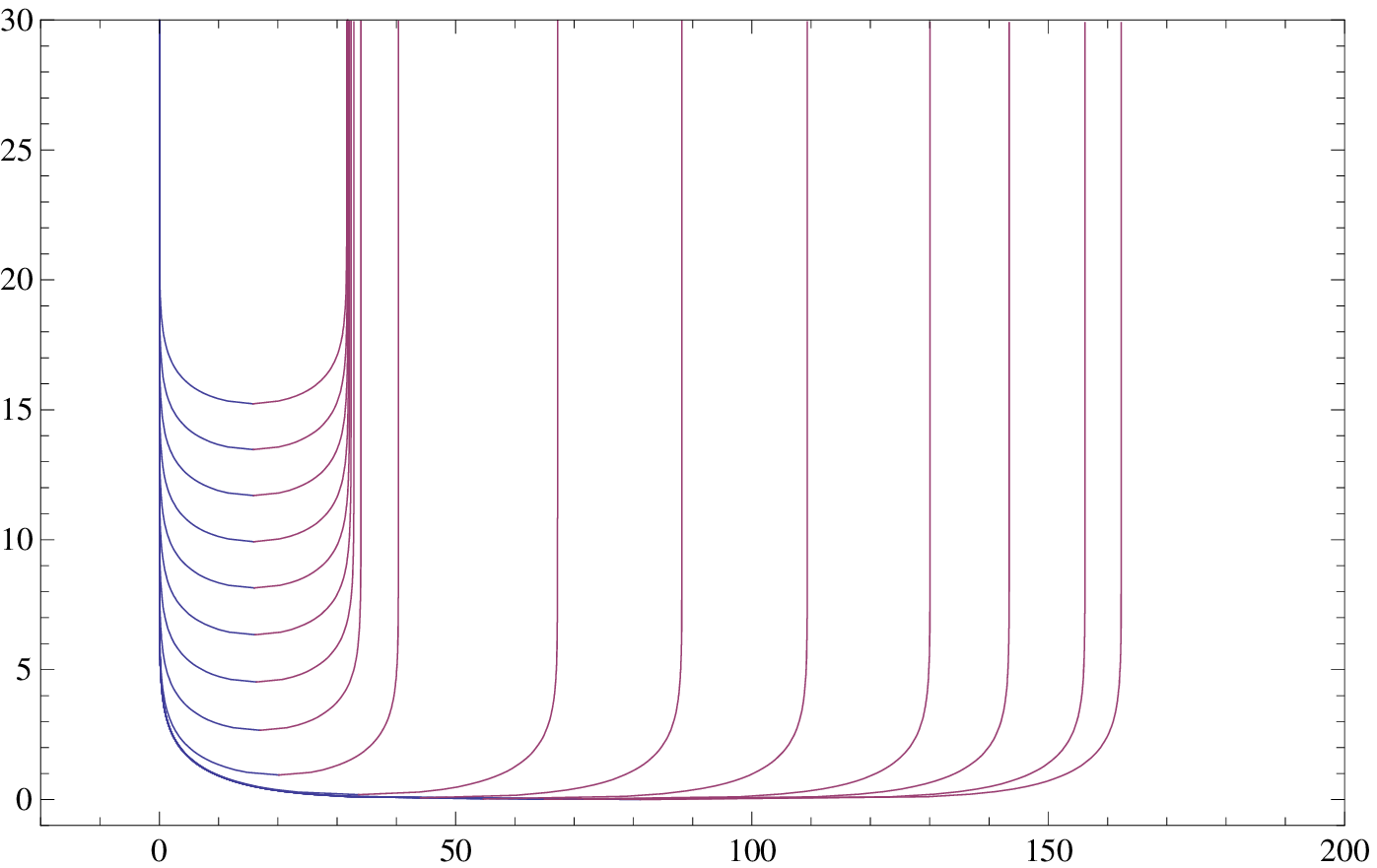}}
\put(150,10){\includegraphics[width=5cm]{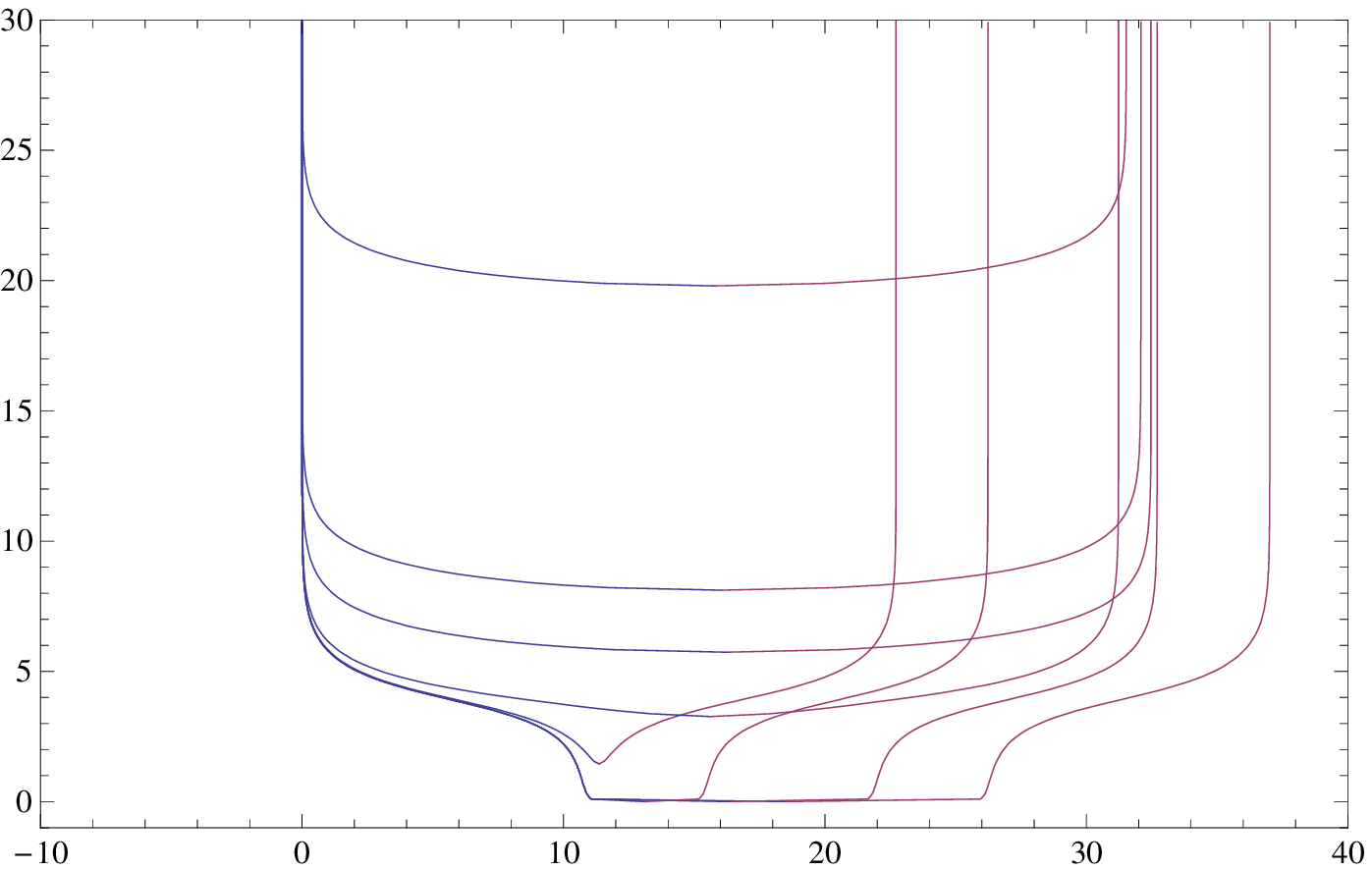}}
\put(-4,90){$\rho$}
\put(146,90){$\rho$}
\put(125,5){$x$}
\put(280,5){$x$}
\end{picture} 
\caption{The strings in  $(x,\r)$-plane, 
obtained with various choices of $C^2>f^2(0)$.
 Left to right, the background is described by  the  numerical solutions in Fig.~\ref{Fig:numericalP}, 
 characterized by   increasing values of $\r_{\ast}=0$ and $\r_{\ast}\simeq 4$, respectively. 
 All calculations performed with $\r_1=30$,
and other units set as explained in the text~\cite{NPR}.}
\label{Fig:strings}
\end{figure}

Comparing with the solutions that {\it walk} in the IR, shows a very 
different behavior.
Starting from  Fig.~\ref{Fig:study1}, one sees that 
as long as $\r_0>\r_{\ast}$, the dependence of $L_{QQ}$ from $\r_0$ reproduces the $\hat{P}$
case. Beginning from such large $\r_0$, we start pulling the string down to smaller values of $\r_0$,
and follow the classical evolution.
 Provided we do this
 adiabatically, we can describe the motion of the string as the set of 
 classical equilibrium solutions.
Going to smaller $\r_0$,  $L_{QQ}$ increases, 
 and nothing special happens
 until the tip of the string touches $\r_0\simeq \r_{\ast}$.
 At this point $L_{QQ}=L_{max}$ has a (local) maximum as a function of $\r_0$.
 From here on, the string can keep probing smaller values of $\r_0$ only at the price of becoming shorter in the Minkowski direction (smaller $L_{QQ}$).
Another change happens deep in the IR, when $L_{QQ}=L_{min}$ reaches a local minimum, at which point the 
  tip of the string 
entered the very bottom section  of the space, near its end
$\r_0<\r_{IR}$.
 From here on, further reducing $\r_0$
requires larger values of $L_{QQ}$. Asymptotically for 
$\r_0\rightarrow 0$, the separation between the end-points of the string is 
diverging, 
 $L_{QQ}\rightarrow \infty$.

Even more interesting is the behavior of the energy (see Fig.~\ref{Fig:study2}): 
for very short $L_{QQ}$, and again for very large-$L_{QQ}$,
 it is just a monotonic function, very similar to the one obtained from $\hat{P}$.
 But for a range $L_{min}<L_{QQ}<L_{max}$ there are three different configurations allowed
by the classical equations for the string we are studying.
One of the three solutions (smoothly connected to the small-$L_{QQ}$ configurations) 
is just the Coulombic potential already seen with $\hat{P}$.
The highest energy one is an unstable configuration, with much higher energy.
The third solution (smoothly connected 
to the unique solution with $L_{QQ}>L_{max}$) reproduces the linear potential typical of confinement.
 Notice (from  Fig.~\ref{Fig:study2})
that the solution at large-$L_{QQ}$ is linear, but has a slope much larger than what seen in 
the $\hat{P}$ case.

This co-existence of several disjoint classical solutions is expected in 
systems leading to phase transitions.
 The stability/metastability/instability of the 
 solutions can be illustrated by comparing Fig.~\ref{Fig:study1} and Fig.~\ref{Fig:study2} with Fig.~\ref{Fig:vdW} 
 in  Appendix~\ref{vanderwaals}. 
 This is just an analogy, and one should not push it too far. However, 
 identifying the pressure $P$, volume $V$ and Gibbs free energy $G$
as $L_{QQ}\leftrightarrow P$, $\r_0\leftrightarrow V$
 and $E_{QQ} \leftrightarrow G$, one sees obvious similarities.
In particular, there is a critical distance $L_{min}<L_c<L_{max}$
at which the minimum of $E_{QQ}$  is not differentiable, 
which suggests that this is a first-order (quantum) phase transition.

In order to better understand and characterize the solutions we find,
it is useful to look more in details at the shape of the string configurations,
focusing in particular on the right panel
in Fig.~\ref{Fig:strings}, in which we plot the string configuration 
that solves the equations of motion for various values of $\r_0$,
on the background with $\r_{\ast}\simeq 4$.
Consider those strings that penetrate below $\r_{\ast}$.
Besides having a shorter $L_{QQ}$, and higher $E_{QQ}$
than those for which $\r_0>\r_{\ast}$, these strings show another interesting feature.
They start developing a non trivial structure around their middle point, 
that becomes progressively more curved
the further the string falls at small $\r$.
Ultimately, this degenerates into a 
cusp-like configuration, which disappears once $\r_0$ approaches the end of the space.

Notice that, as a result, the three different solutions for $L_{min}<L_{QQ}<L_{max}$
have three very different geometric configurations. One (stable or metastable) configuration
is completely featureless, and practically indistinguishable from the solutions in the background
generated by $\hat{P}$.
The second (stable or metastable) configuration shows a funnel-like structure below $\rho_{\ast}$,
and most of  the string lies very close to the end of the space.
The third (unstable) solution presents a highly curved configuration around its middle point.
All of this seems to be consistent with the fact that the classical configurations prefer 
to lie either in the far-UV or deep-IR, outside the $\r_{IR}<\r<\r_{\ast}$ region.

Concluding, we learned two very important things from the study of the Wilson loop.
First of all, the end of the space in the IR can indeed be interpreted as the fact that
confinement is taking place. The (unphysical) singularity in the running of the gauge coupling 
yields the expected (physical) linear behavior of the quark-antiquark potential. 
Second, the existence of a plateau at intermediate scale has a very interesting physical meaning.
Because it is ultimately responsible for the arising of the first-order phase transition we found,
the hierarchy $\r_{\ast} > \r_{IR}$ cannot be undue by simply changing  renormalization scheme,
but is a physical effect.

\subsection{Summary.}

In this section we focused our attention on one very specific class of models,
in which the geometry is obtained by truncating type-IIB supergravity,
in the strongly-coupled limit of the system of $D5$-branes wrapping a 2-cycle inside
a CY3 manifold.

We found new classes of solutions to the BPS equations determining the background 
that exhibit some very peculiar properties, closely resembling those expected in a walking field theory.
Studying these backgrounds, we were able to show that
\begin{itemize}
\item the (non-perturbative) running of the four-dimensional gauge coupling exhibits a walking behavior, in the sense that there is a
finite interval of the radial direction within which the coupling is almost constant, but this behavior disappears both in the UV and in the IR,
\item in the IR, below the walking region, the theory confines, as signaled by the long-distance behavior of the Wilson loop,
and condensates emerge, spontaneously breaking the global symmetries of the theory,
\item the existence of the walking region is a physical effect, as signaled by the presence of a discontinuity of the derivative of the 
quark-antiquark energy as a function of the separation, 
\item the spectrum of the scalar glueballs contains a mass gap, and depending on the extension in the radial direction
of the walking region one scalar state becomes parametrically lighter than such gap. 
\end{itemize}

These are all very striking and important results, none of which can be obtained analytically 
by studying a four-dimensional, strongly-coupled gauge theory with traditional methods.
A long list of other things should be done.
I conclude with a sample of possible other open questions.
\begin{itemize}
\item The properties of the light scalar (its couplings in particular) are not known.
Holographic renormalization, at least in its simplest form, fails to provide a sensible treatment
 of the correlation functions, because  the 
background is not asymptotically AdS. Some clever strategy must be devised in order to be able to study the possible phenomenology of
such a  light scalar.
\item The rest of the spectrum (spin-1 fields and fermions in particular) has not been studied yet.
\item The spectrum of anomalous dimensions of the model is not known. This is a peculiarly difficult question, in
relation to the fact that the geometry is never really close to AdS, and hence the precise relation between radial direction and 
renormalization scale is not easily inferred.
\item It would be interesting to couple this model to the standard model, and compute the effect of the walking region 
on the oblique parameters.
\item The field and operator content of the dual field theory is known only in part, and a more systematic exploration would be useful.
\end{itemize}

Finally, it must be noted that there is no obvious reason why this specific set-up
is preferable, in relation to walking technicolor. To large extent, the reason why these studies have been performed in this specific context
has to do with the fact that backgrounds based on the conifold and its various deformations 
provide a very well studied and understood place from which to start.
But it would be very interesting to know if analogous results hold in completely different models,
possibly simpler, and maybe non-supersymmetric.

\newpage
\section*{Conclusions.}

Dynamical  electro-weak symmetry breaking relies
on the existence of a new strongly-coupled sector responsible for the 
spontaneous symmetry breaking that is at the core of the Standard Model.
There are two interconnected reasons why the ideas of gauge/string dualities
are useful in constructing and studying these models.
First of all, the non-perturbative nature of the underlying strong dynamics 
calls for a flexible and powerful set of tools which makes calculations doable.
But also, the huge amount of experimental data at our disposal
indicates quite clearly that a generic model is not going to pass
the precision tests. The new (strongly-coupled)
 physics sector has to have very special features
in order to be phenomenologically viable.
Walking technicolor is one such special possibility, 
which happens to be a 
natural candidate for phenomenological applications of
gauge/string dualities,
because of its quasi-conformal behavior,
its large anomalous dimensions, and its multi-scale nature.

In these notes, besides reviewing the phenomenology of
strongly-coupled extensions of the standard model
(at a very pedagogical level, by comparing to the minimal version of the standard model
itself and to  its weakly-coupled extensions), 
I provided a few examples illustrating how the ideas 
taken from gauge/string dualities can indeed yield very
useful and interesting results, taking them both from the
context of five-dimensional effective theories (bottom-up approach)
and from  more rigorous ten-dimensional string-theory constructions (top-down approach).

Summarizing, the main messages I hope I was able to convey to the reader are that
\begin{itemize}
\item the unhealthy behavior of the minimal version of the Standard Model, when extrapolated at arbitrary high scale
(where Landau poles and fine-tuning appear), 
requires that it be completed, particularly in the sector responsible for electro-weak symmetry breaking, and 
many proposals exist that do so,
\item at low energy the Standard Model provides such an efficient and precise description of 
the vast amount of data at our disposal, that only very peculiar, special types of 
new physics models can be  phenomenologically viable,
\item  walking technicolor, supplemented by tumbling ETC,
 is one such special,  phenomenologically  viable possibility, in which new strongly-coupled interactions 
 and non-perturbative phenomena are present,
\item  with the incoming of LHC data, actual calculations of physical observable quantities within this strongly coupled
framework are urgently needed,
\item gauge/gravity dualities can be used to perform these calculations, and several examples 
of useful results exist,
\item the technology of gauge/gravity duality is at such a stage of development  that,
provided a sensible model is identified, the calculation of actual physical quantities 
can be done, and is often comparatively simple,
\item while a completely satisfactory proposal for the dual of a  walking technicolor theory does not exist at present,
yet,  circumstantial evidence from 
the examples we have, together with the existing vast literature on the subject 
and the number of open possibilities for model-building
 in the gauge/gravity context, strongly suggests that there is a lot that can be learned in this direction,
  which represents a very promising avenue for present and future research.
\end{itemize}

\newpage
\appendix
\section{Van der Waals gas.\label{vanderwaals}}
Here I summarize some aspect of first-order phase transitions that
plays an important conceptual role in the body of the paper. 
This appendix is taken from~\cite{NPR}, and extended discussions can be found in 
textbooks on  statistical mechanics.
As an example, I review the
classical treatment of the van 
der Waals gas,  in terms of the pressure $P$, temperature $T$ and volume 
$V$ of $n$ moles of particles, by means of the  equation of state 
\beq
P= \frac{nRT}{V- b n} - \frac{n^2 a}{V^2}\,,
\label{vdw}
\eeq
where $R, b, a$ are constants.

Fig.~\ref{Fig:vdW} shows one isotherm. 
The condition 
for stability of the equilibrium $\left(\partial^2 F /\partial V^2\right)_T = - \left(\frac{\partial P}{\partial V}\right)_{T} >0$ 
is not satisfied in some region. This implies that a phase transition is taking place.

\begin{figure}[htpb]
\includegraphics[width=7cm]{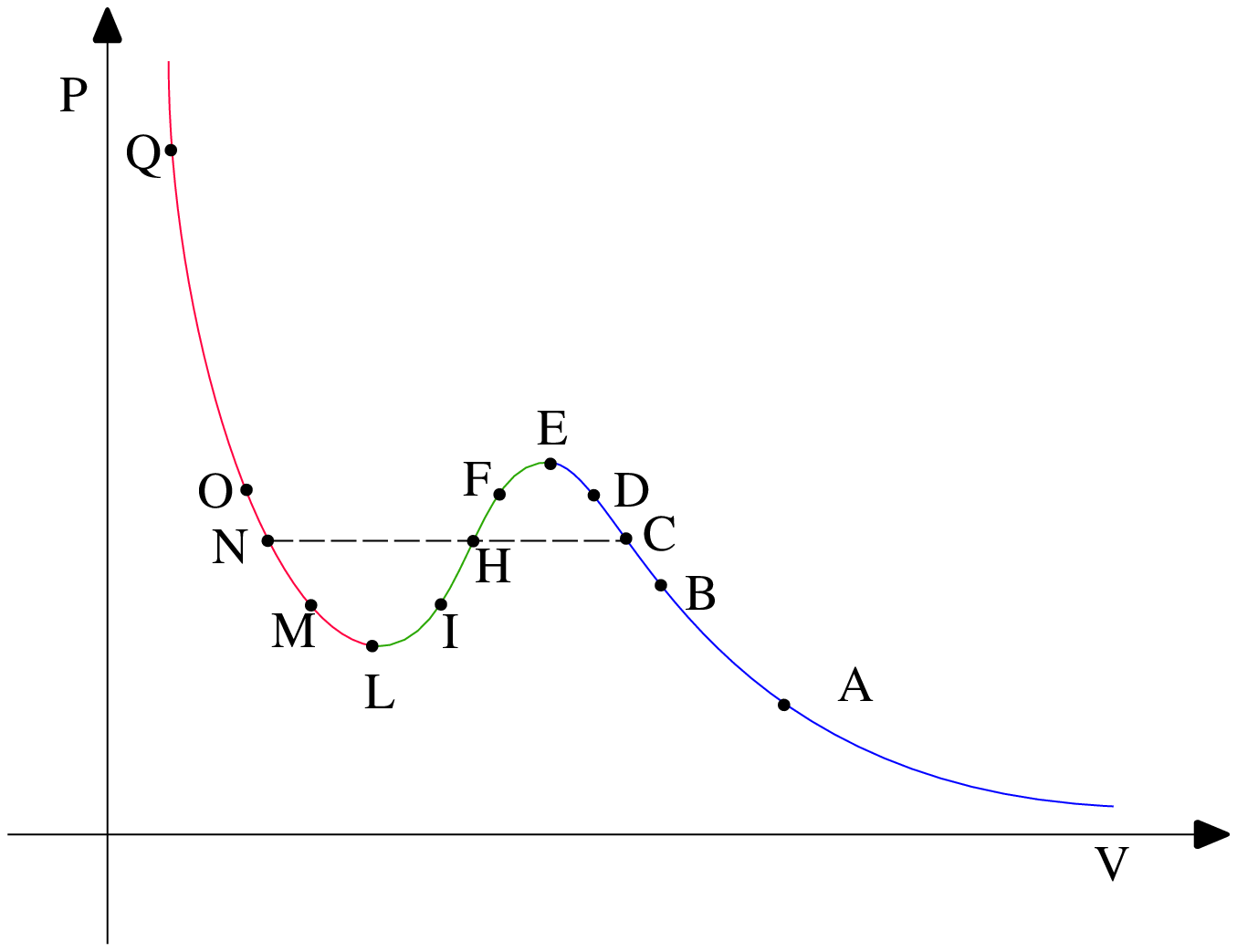}
\includegraphics[width=7cm]{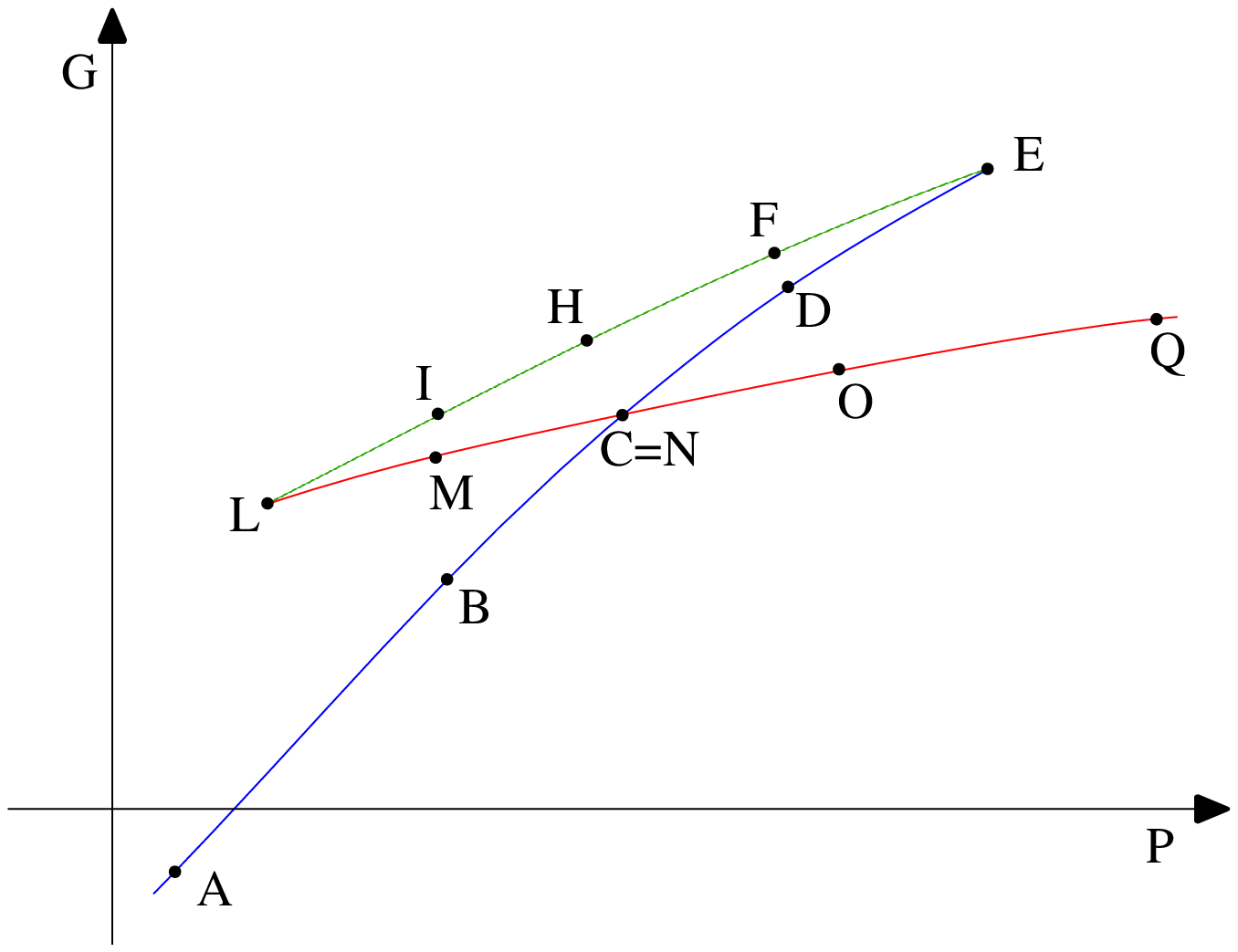}
\caption{The pressure $P$ as a function of the volume $V$ (left panel)
and the Gibbs free energy $G$ as a function of the pressure $P$(right panel) for
the same isotherm curve~\cite{NPR}.}
\label{Fig:vdW}
\end{figure}

In order to understand what the physical trajectory followed by the
 system at equilibrium is,  Fig.~\ref{Fig:vdW} shows the  
 Gibbs free energy $G=G(T,P)=G(P)$ for the same isotherm. 
 From this plot, one sees that  the system evolves on the 
path $ABCOQ$, where $C=N$, in such a way that for every choice of $P$
the Gibbs free energy $G$ is at its minimum.
 The evolution is smooth along $ABC$ (gas phase: $|\partial P/\partial V|$ is small), 
 but at $C=N$ the free energy is not
 differentiable, signaling that a first-order phase transition is taking place.
 In the $(P,V)$ plane the system runs along the horizontal line 
 (constant $P$) joining $C$ and $N$.
 This explains  the {\it Maxwell rule} introducing a curve of constant 
pressure that separates two regions of  equal areas above and below it, delimited by the original isotherm.
Afterwards, 
the evolution follows smoothly the curve $NOQ$ (liquid phase: $|\partial P/\partial V|$ is large). 

While the trajectory $ABCNOQ$ follows the stable equilibrium configurations,
it is possible to 
have the system evolving along the path $CDE$ or $LMN$. Both these 
paths represent metastable configurations, because $\left(\partial P/\partial V\right)_T<0$. 
Indeed, these metastable 
states can be realized in laboratory experiments. For example, in a  bubble chamber or in supercooled water,
the metastability is exploited as a detector device, because 
small perturbations induced by passing-by charged particles  are sufficient to
drive the system out of the state and into the stable minimum.
The evolution along the path $EFHIL$ is completely unstable and not realized 
physically (it is a local maximum, as clear from the right panel of Fig.~\ref{Fig:vdW},
and by the fact that $\left(\partial P/\partial V\right)_T>0$).

The analogy  with the examples in the main body 
of the paper is apparent.
Notice for instance that the pressure $P$ in this system as a function of the 
volume $V$ behaves as a non monotonic function, hence there are {\it inversion 
points} in the curve (the points $E,L$ in figure \ref{Fig:vdW}).

\vspace{1.0cm}
\begin{acknowledgments}
I  thank Daniel Elander, Biagio Lucini,  Ioannis Papadimitriou and Agostino Patella for useful comments on the manuscript,
and Carlos Nunez, Antonio Rago and Mark Round for the collaborations
part of these lectures are based upon.
I would like to thank the hospitality of University of Barcelona, where
lectures based on parts of these notes have been presented.
The work of MP is supported in part  by
the Wales Institute of
Mathematical and Computational Sciences and
by the STFC Grant
ST/G000506/1.

\end{acknowledgments}


\end{document}